\documentclass[oldversion]{aa}
\usepackage{graphicx}
\usepackage{txfonts}
\usepackage{natbib}
\usepackage{url}
\usepackage{longtable}
\usepackage{caption} 
\bibpunct{(}{)}{;}{a}{}{,} 

\def\ecs{erg~cm$^{-2}$s$^{-1}$}
\def\lum{erg~s$^{-1}$}

\usepackage{subfig}
\usepackage{wasysym}          
\usepackage{fixltx2e}         
\usepackage{adjustbox}        
\usepackage{enumitem}         

\newcommand{\ql}{\textquoteleft}
\newcommand{\qr}{\textquoteright}

\begin{document}

\title{Searching for the most powerful thermonuclear X-ray bursts \\
with the Neil Gehrels Swift Observatory}
\titlerunning{Searching for the most powerful thermonuclear X-ray bursts with Swift}

\author{J.J.M. in 't Zand\inst{1}, M.J.W. Kries\inst{1,2}, D.M. Palmer\inst{3}
  \& N. Degenaar\inst{4}}
\authorrunning{in 't Zand, Kries, Palmer \& Degenaar}

\institute{
 SRON Netherlands Institute for Space Research, Sorbonnelaan 2, 3584 CA Utrecht, the Netherlands
\and Dep. of Physics, Utrecht University, PO Box 80000, 3508 TA Utrecht, Netherlands
\and Los Alamos National Laboratory, B244, Los Alamos, NM 87545, U.S.A.
\and Anton Pannekoek Instituut for Astronomy, University of Amsterdam, Science Park 904, 1098 XH Amsterdam, the Netherlands }

\date{}

\abstract{We searched for thermonuclear X-ray bursts from Galactic
  neutron stars in all event mode data of the Neil Gehrels Swift
  Observatory collected until March 31, 2018. In particular, we are
  interested in the intermediate-duration bursts (shell flashes fueled
  by thick helium piles) with the ill-understood phenomenon of strong
  flux fluctuations. Nine such bursts have been discussed in the
  literature to date. Swift is particularly suitable for finding
  additional examples. We find and list a total of 134 X-ray bursts;
  44 are detected with BAT only, 41 with XRT only, and 49 with
  both. Twenty-eight bursts involve automatic slews. We find 12
  intermediate-duration bursts, all detected in observations involving
  automatic slews. Five show remarkably long Eddington-limited phases
  in excess of 200 s. Five show fluctuations during the decay phase;
  four of which are first discussed in the present study. We discuss
  the general properties of the fluctuations, considering also 7
  additional literature cases. In general two types of fluctuations
  are observed: fast ones, with a typical timescale of 1 s and up and
  downward fluctuations of up to 70\%, and slow ones, with a typical
  timescale of 1 min and only downward fluctuations of up to 90\%. The
  latter look like partial eclipses because the burst decay remains
  visible in the residual emission. We revisit the interpretation of
  this phenomenon in the context of the new data set and find that it
  has not changed fundamentally despite the expanded data set. It is
  thought to be due to a disturbance of the accretion disk by
  outflowing matter and photons, causing obscuration and reflection
  due to Thompson scattering in an orbiting highly ionized cloud or
  structure above or below the disk.  We discuss in detail the most
  pronounced burster SAX J1712.6-3739. One of the bursts from this
  source is unusual in that it lasts longer than 5600 s, but does not
  appear to be a superburst.}


\maketitle

\section{Introduction}

One of the brightest phenomena in the keV X-ray sky is the Type I
X-ray burst phenomenon \citep[e.g.,][]{grindlay1976,swank1977} with
peak fluxes sometimes reaching 10$^{-6}$~erg~s$^{-1}$cm$^{-2}$ (1--20
keV). We refer to these simply as X-ray bursts or bursts. They are due
to thermonuclear shell flashes on neutron stars with a relatively low
magnetic field, during which the hydrogen and helium accreted from a
Roche-lobe overflowing companion star in a compact low-mass X-ray
binary is burned
\citep[e.g.,][]{hansen1975,maraschi1977,joss1977}. The thermonuclear
runaway commences within seconds, after which the neutron star cools
on a timescale of 10~s to 10 hr, depending on the ignition
depth. During the flashes, the photosphere reaches a temperature on
the order of 10 MK and radiates predominantly in X-rays \citep[for
  reviews, see, e.g.,][]{lewin1993, stroh2006,galloway2017}. The
nuclear power generated in such a flash can be excessive, reaching a
level where the radiation pressure gradient overcomes the strong
gravitational pull by the neutron star. In other words, the Eddington
limit is reached. A wind and shell may be blown from the neutron
star. The outflowing photon and matter flux interacts with the
accretion flow/disk and can seriously disturb it, depending on its
geometry \citep{zand2011}.

As of March 31, 2018, 111 bursting neutron stars were known in the
Galaxy (see Table~\ref{table:xrtdata})\footnote{See also URL
  www.sron.nl/$\sim$jeanz/bursterslist.html}. The number of bursts
detected per neutron star varies widely. Since the burst process is
fueled by accretion, the momentary accretion rate determines to a
large extent the recurrence time. Transiently accreting neutron stars
can only burst when there is accretion, which on average occurs a few
percent of the time \citep[cf.,][]{yan2015} or maybe somewhat more
\citep{degenaar2010b}. Neutron stars that accrete persistently on a
timescale of a human life can burst at a rate anywhere between roughly
once per hour to once per month.

Three major kinds of X-ray bursts can be identified:
\begin{enumerate}
\item common X-ray bursts \citep[e.g.,][]{lewin1993}, with recurrence
  times of a few hours and durations of tens of seconds. The accretion
  rate is larger than a few percent of the Eddington limit. The
  bursts may be fueled by a mixture of hydrogen and helium or only
  helium. In the latter case, the luminosity often reaches the
  Eddington limit and the photosphere is seen to expand;
\item intermediate-duration bursts
  \citep[e.g.,][]{Zand20052S,cumming2006,zand2007,Falanga2008}, with
  recurrence times of many weeks and durations of tens of minutes. The
  accretion rate is smaller than 1\% of the Eddington limit, except in
  GX 17+2 \citep{kuulkers2002GX}. They are thought to be fueled by a
  large pile of helium. The combination of helium as a fuel, which
  burns relatively quickly, and the amount of fuel provides a
  circumstance where the nuclear power is many times higher than the
  Eddington limit;
\item superbursts with recurrence times of about one year and durations
  of about one day. These bursts are seen to occur at a wide range of
  accretion rates \citep[for a recent review, see][]{zand2017b}. They
  are thought to be fueled by carbon at a column depth that is about
  10$^3$ larger than for common He/H bursts
  \citep{cumming01,stroh02}. Superbursts are usually sub-Eddington,
  due to the smaller energy output per nucleon, the smaller mass
  fraction of the carbon in the ignited layer, and the longer heat
  diffusion time.
\end{enumerate}

Some long bursts have very strong flux fluctuations with respect to
the typical power-law decay of the burst \citep{zand2014,zand2017},
with amplitudes above and below the cooling curve of order 50\% on
timescales of 1 s to 1 min. So far, nine such bursts have been
reported \citep{jvpm15,stroh02,Zand20052S,
  Molkov2005SLX,zand2008,zand2011,palmer2011,
  Degenaar2013IGR,barriere2015} as measured with the instruments
Ginga-LAC, INTEGRAL-JEM-X, BeppoSAX-WFC, RXTE-PCA, Swift-XRT, and
NuSTAR \citep[for a recent overview, see][]{degenaar2018}. The common
characteristics are that the fluctuations are at least 10\% and that
they occur with some delay (order one minute) after the
Eddington-limited phase. The nature of this phenomenon is not fully
understood, but it seems clear that the accretion disk plays a
role. In order to get a more complete understanding of the phenomenon,
more data would be useful. Swift is an efficient tracker of these
bursts. Therefore, we set out to investigate all Swift XRT and BAT
data on X-ray bursts.

To detect rare kinds of bursts it is crucial to use wide-field imaging
devices. There are two kinds of wide-field devices. There are those
with a limited field of view (FOV) that scan the sky. Examples are
RXTE/ASM and ISS/MAXI that scan $>80$\% of the sky every 1.5 hr. This
revisit time is sufficiently short to identify superbursts
\citep{Kuulkers2002,serino2017,iwakiri2018}, but not
intermediate-duration bursts. Second, there are devices with truly
wide FOVs, such as the BeppoSAX Wide Field Camera instrument
\citep[WFC;][]{jager1997}, that are efficient in detecting bursts of
all durations.

The spectra of X-ray bursts have a peak temperature of k$T$=2.0--2.5
keV, which implies that the photon spectrum peaks at 3--4 keV and the
energy spectra at 6--7.5 keV. Thus, the detector of a wide-field device
should preferably be able to detect photons in the 2--10 keV range,
such as BeppoSAX/WFC (FOV 8\% of the sky) or INTEGRAL/JEM-X
(0.3\%).

There are other wide-field devices in orbit, but they are active at
somewhat unfavorable bandpasses. They include the non-imaging Fermi
Gamma-ray Burst Monitor (GBM; 8 keV -- 1 MeV), which has 100\% sky
coverage (being limited only by the  Earth obscuring a third of the
sky), and the imaging Burst Alert telescope on the Neil Gehrels Swift
Observatory (BAT; 15--$\sim$150 keV) with a FOV of 18\% of the sky and
mostly pointed away from Earth. The sensitivities of these devices on
a 5 sec timescale to a 2.5 keV blackbody  are about 1 Crab
\citep{jenke2016} and 100 mCrab, respectively. The better sensitivity
of the BAT is due to a larger detector and the imaging capability. The
GBM detects about 250 bursts per year during the phases when the
bursts are hotter than k$T$=2.5 keV \citep{jenke2016}. This implies
that superbursts can hardly  be detected because the temperature peaks
at lower values.

While the detection capabilities of GBM and BAT are good, the
diagnostic capabilities are limited. Both instruments are only able to
detect the hottest phases of X-ray bursts, losing the signal quickly
during the burst decay where most of the fluence is emitted. The
BAT has one important advantage: it can trigger automatic pointings of
the Swift X-Ray Telescope (XRT) at the burst. The XRT is very
sensitive because it is a focusing telescope. Automatic slews typically
happen within about 100 s. This implies that the BAT+XRT combination
is particularly efficient in detecting intermediate-duration
bursts. These bursts are super-Eddington and therefore reach the
highest possible temperatures. Also, they last a long time so that they are
easily picked up with XRT despite the response delay.

We estimate that 12 to 14 thousand X-ray bursts have been detected
since the advent of X-ray astronomy in the 1960s. Judged from a subset
of one thousand well-observed cases \citep{galloway2008}, it appears
that some 20\% show evidence of the radiation-pressure effect
mentioned above. This evidence pertains to a temporary increased
emission region size as derived from the spectrum. This is called
photospheric radius expansion (PRE). In addition, there are a few
($\sim$1\%) PRE X-ray bursts that show superexpansion, where the
expansion is so large (with a radius growing from $\sim$10 km to
$\ga10^3$~km) that the temperature drops out of the 1--10 keV bandpass
and the illusion of a precursor event is created
\citep{Zand2010}. Many intermediate-duration bursts seem to have
superexpansion if the sensitivity is sufficient to detect it given the
short nature.

We set out to obtain an overview of burst detections with Swift-BAT
and XRT for which event data is available so that light curves can be
followed to small fluxes and intermediate-duration bursts can be
distinguished, and to concentrate on those that show fluctuations.  In
\S~\ref{ch:Instr} we present the relevant details of the Swift
instrumentation and operations, in \S~\ref{sec:search} the approach
used to search for bursts, in \S~\ref{ch:data} our analysis of all
relevant new data, in \S~\ref{sec:literature} an overview of relevant
data from the literature, in \S~\ref{sec:comparison} a comparative
analysis of all intermediate-duration bursts and fluctuations, and in
\S~\ref{ch:Discussion} a discussion of the results and the
implications for the model of the fluctuations.

\section{Instrumentation and follow-up algorithm}
\label{ch:Instr}
The Neil Gehrels Swift Observatory is a NASA Medium Explorer mission
launched on November 20, 2004.  The main goal of the observatory is to
search and study gamma-ray bursts \citep[GRBs;][]{swift2004}, but it
is also very successful in the study of X-ray bursts (e.g., the
discovery of 13 new X-ray burst sources, see
Appendix~\ref{appendix:1}).  To observe GRBs and their afterglows,
Swift has three instruments: the wide-field BAT
\citep[][]{Barthelmy2005BAT}, and the narrow-field X-ray
\citep[XRT;][]{Burrows2005XRT} and Ultraviolet/Optical Telescopes
\citep[UVOT;][]{uvot2005}.

\renewcommand{\tabcolsep}{0.7mm}
\begin{table*}[t]
\caption{Overview of the 28 BAT-triggered bursts from 24 sources with
  coverage by the XRT, ordered by observation date. The bold typeset
  entries are bursts with fluctuations. All times are with a typical
  uncertainty of 1--5 seconds.}
\label{tab:table10}
\centering
\tiny
\begin{tabular}{lllrrrrrccl}
    \hline
    Object                 &Trigger& Trigger time (UTC)          & Peak     & Start XRT   & Duration   & Duration     & touch & Pre- & Inter- & Reference \\
                           &number &                             & flux     & data after  & $t_{5\%}$   & $t_{\rm BAT}$ & down  & cur- & mediate? & \\
                           &       &                             & [$10^{-1}$& burst onset &   [s]      &  [s]         & [s]   & sor? &        & \\
                           &       &                             & c~s$^{-1}$& [s]         &            &             &       &      &   & \\
                           &       &                             & cm$^{-2}$]&             &            &             &       &      &   & \\
    \hline                        
    4U 1812-12             &106799   & 2005-02-24   12:40:45     & 0.64(7)  & 160  & $<160$&  21 &  15 & y &   & \\
    Swift J1749.4-2807     &213190   & 2006-06-02   23:54:34     & 3.1(2)   &  90  & $<90$ &  20 &   7 &   &   & \cite{wijnands2009} \\
    1A 1246-588            &223918   & 2006-08-11   02:59:56     & 2.0(2)   & 230  &  290  &  70 &  40 &   & y & \cite{zand2008} \\    
    SAX J1810.8-2609       &287042   & 2007-08-05   11:27:26     & 2.1(2)   &  80  &  100  &  23 &  14 & y &   & \cite{degenaar2013}  \\
    XTE J1810-189          &306737   & 2008-03-18   22:32:52     & 0.65(14) & 100  &$<100$ &  10 &   0 &   &   &  \\
    IGR J17473-2721        &308196   & 2008-03-31   09:03:33     & 1.7(2)   &$<$0  &   75  &  25 &  10 & y &   & \\
    1RXH J173523.7-35401   &311603   & 2008-05-14   10:32:37     & 1.0(1)   & 250  &$>250$ & 240 & 200 &   & y & \cite{degenaar2010} \\
    XTE J1701-407          &317205   & 2008-07-17   13:29:59     & 3.1(2)   & 140  &  190  & 125 &  55 &   & y & \cite{linares2009} \\
    XTE J1701-407          &318166   & 2008-07-27   22:31:20     & 2.0(2)   & 120  &$<110$ &  15 &  10 &   &   & \cite{linares2009} \\
    IGR J17511-3057        &371210   & 2009-09-30   18:31:57     & 1.6(2)   &  50  & $<50$ &  20 &  10 &   &   & \cite{bozzo2009} \\
    SAX J1712.6-3739       &426405   & 2010-07-01   14:55:41     & 1.7(2)   & 100  &  140  &  30 &  20 &   &   & \cite{strohmayer2010} \\
    XTE J1810-189          &455640   & 2011-06-19   00:59:37     & 0.75(12) & 300  &  750  & 400 & 200 &   & y & \\
    Swift J185003.2-0056   &456014   & 2011-06-25   00:06:08     & 2.4(3)   & 100  &$<100$ &  40 &  10 &   &   & \cite{degenaar2012b} \\
    {\bf SAX J1712.6-3739} &{\bf 504101}&{\bf 2011-09-26 20:11:29}&{\bf 1.6(2)}& {\bf 180}  & {\bf  1050} & {\bf 315} & {\bf 190} &   & {\bf y} & {\bf \cite{palmer2011}} \\
    Swift J1922.7-1716     &506913   & 2011-11-03   14:12:13     & 2.2(3)   & 120  &$<120$ &  50 &  20 & y &   & \cite{degenaar2012b} \\
    {\bf IGR J17062-6143}  &{\bf 525148}&{\bf 2012-06-25 22:42:32}&{\bf 1.6(3)}& {\bf 260}  & {\bf  1300} & {\bf 420} & {\bf 220} &   & {\bf y} & {\bf \cite{Degenaar2013IGR}} \\
    XMM J174457-2850.3     &530588   & 2012-08-11   04:43:54     & 2.1(2)   &  95  &  130  &  75 &  55 & y &   & \cite{degenaar2014} \\
    Swift J174805.3-244637 &530808   & 2012-08-13   09:13:34     & 1.5(2)   &  90  & $<90$ &  35 &  15 & y &   & \\
    {\bf IGR J18245-2452}  &{\bf 552369}&{\bf 2013-03-30 15:10:38}&{\bf 0.57(17)}&{\bf 120}  & {\bf rise!} & {\bf 130} & {\bf 125} &   & {\bf y} & {\bf \cite{barthelmy2013}} \\
    {\bf Swift J1734.5-3027}&{\bf 569022}&{\bf 2013-09-01 09:13:17}&{\bf 1.8(2)}& {\bf 145}  & {\bf  600}  & {\bf 155} & {\bf  75} &   & {\bf y} & {\bf \cite{Bozzo2015Swift}} \\
    MAXI J1421-613         &584155   & 2014-01-18   08:39:20     & 0.89(16) &$<$0  &   30  &  15 &   5 & y &   & \cite{serino2015} \\
    {\bf 4U 1850-087}      &{\bf 591237}&{\bf 2014-03-10 21:05:00}&{\bf 2.0(2)}&{\bf  610}  & {\bf  1400} & {\bf 840} & {\bf 530} &   & {\bf y} & {\bf \cite{zand2014}} \\
   {\bf SAX J1712.6-3739}  &{\bf 609878}&{\bf 2014-08-18 17:10:04}&{\bf 1.1(1)}&{\bf  270}  & {\bf  $>$5530} &{\bf $>$800}&{\bf 540} & {\bf y} & {\bf y} & {\bf \cite{cummings2014}} \\
    4U 0614+09             &631747   & 2015-02-19   16:42:24     & 8.2(4)   &  85  &$<120$ &   6 &   3 &   &   & \cite{breeveld2015} \\
    SAX J1808.4-3658       &637765   & 2015-04-11   19:36:25     & 6.3(5)   &  40  &   50  &  25 &   7 &   &   & \cite{malesani2015} \\
    IGR J00291+5934        &650221   & 2015-07-25   02:12:05     & 3.2(6)   & 110  &  270  &  20 &  10 &   & y & \cite{falco2017} \\
    SAX J1806.5-2215       &745022   & 2017-04-01   19:00:53     & 0.81(14) & 205  &  550  & 250 & 130 &   & y & \cite{barthelmy2017} \\
    Swift J181723.1-164300 &765081   & 2017-07-28   16:57:58     & 2.9(5)   &  80  & $<80$ &  15 &   7 & y &   & \cite{barthelmy2017b} \\
   \hline
\end{tabular}
\end{table*}
\renewcommand{\tabcolsep}{2mm}

The BAT \citep{Barthelmy2005BAT} is a coded aperture camera with a
large FOV of 2.3 sr (for a response larger than 5\% of the maximum),
which is 18\% of the entire sky, an angular resolution of 17 arcmin
(on-axis), and a localization accuracy of 1 to 4 arcmin (depending on
the strength of the signal above the background). The CdZnTe detector
array provides a 15--150 keV bandpass with a resolution of about 7
keV. The BAT is used to detect GRBs, automatically localizing them on
board and triggering the observatory to automatically slew the optical
axis of the satellite with the co-aligned narrow-field instruments to
the GRB.

The XRT \citep{Burrows2005XRT} is an X-ray telescope active between
0.2 and 10 keV with an angular resolution of 18--22\arcsec\ (HPD; 1.5--8
keV), a typical location accuracy of 3\arcsec, a $600\times600$ pixels
CCD camera with a spectral resolution of 140 eV at 5.9 keV and an
effective area of 125 cm$^2$ at 1.5 keV and 20 cm$^2$ at 8 keV. The
sensitivity is about $3\times10^{-11}$~\ecs\ in 10 s. The CCD
full-image readout time is 2.5 s, but the CCD can also be read out in
the Windowed Timing mode, when the columns are collapsed and limited
to a width of 8\arcmin. The readout time then reduces to 1.7 ms,
allowing the measurement of photon intensities of a few hundred counts
per second with a pile-up fraction of less than 1\%.

The onboard software uses two algorithms to identify a trigger. First,
it checks for temporary excesses in the photon rates above the
background and persistent cosmic sources in (sections of) the detector
array in four channel ranges (15--25, 15--50, 25--100, 50--$\sim$150 keV) and
on timescales between 4 ms and 32 s. When a photon rate increase is
detected above a certain significance (from 5 to $12\sigma$ depending
on energy and timescale) the excess counts in each detector pixel are
deconvolved to give a sky image using an FFT-based algorithm that
takes 6 s.  Each significant peak in the sky image is then reanalyzed
with a photon back-projection algorithm that provides a higher
accuracy determination of the source location, flux, and significance.
In addition, the software analyzes images on 64 s, 320 s, and full
pointing timescales to search for sources that do not cause rapid rate
variations on the shorter timescales.

Each peak location is compared to an onboard catalog of 1300 known
sources (including most known bursters).  If the location of a
significant image peak is not found in the catalog, it is declared to
be a GRB.  If the location is found in the catalog, a threshold and a
merit parameter are extracted from that source's catalog entry.  The
image flux (or in some cases the fluence) is compared to the threshold
to determine whether the detection is interesting.  An interesting
detection results in immediate ground notification, extended
event-data recording, and possible pointing follow-up.  Most source
detections are not interesting in this sense; the Crab and other
bright sources are detected whenever they are in the FOV, and so their
thresholds are set so that typical emission is not considered
interesting.  Likewise, large numbers of bursts from known sources are
detected and recorded, but usually do not trigger a further response.
After a detection, the catalog threshold for that source is
automatically set to double the measured value so that further
activity at the same level is not considered interesting until the
threshold is manually commanded back to a lower level.  This prevents
excess resource use when a transient source appears and persists above
its original interesting threshold.

\begin{figure*}[p]
\includegraphics[width=0.325\textwidth, trim=0.0cm 1cm 3cm 0.5cm,clip=true]{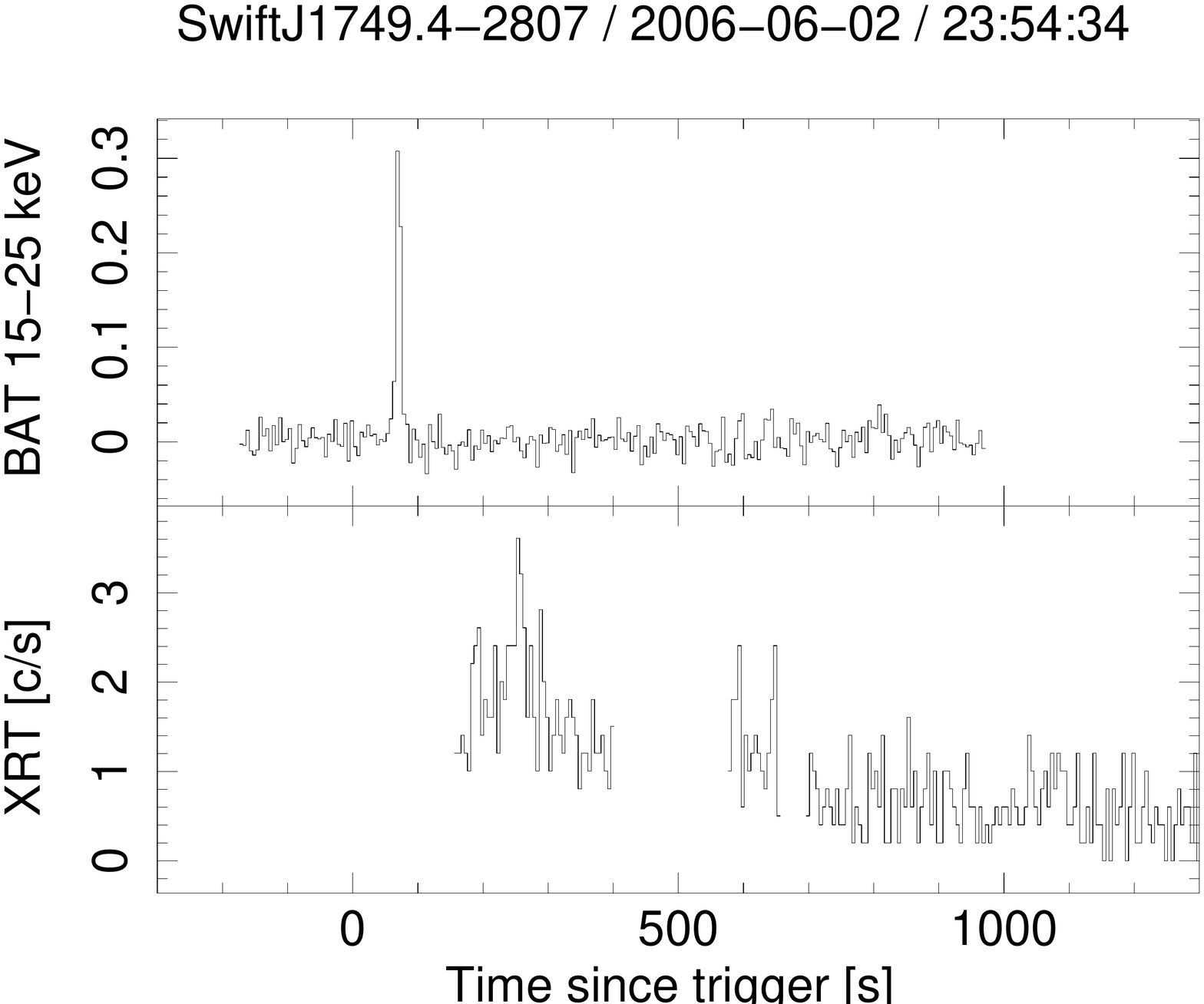}
\includegraphics[width=0.31\textwidth, trim=1.1cm 1cm 3cm 0.5cm,clip=true]{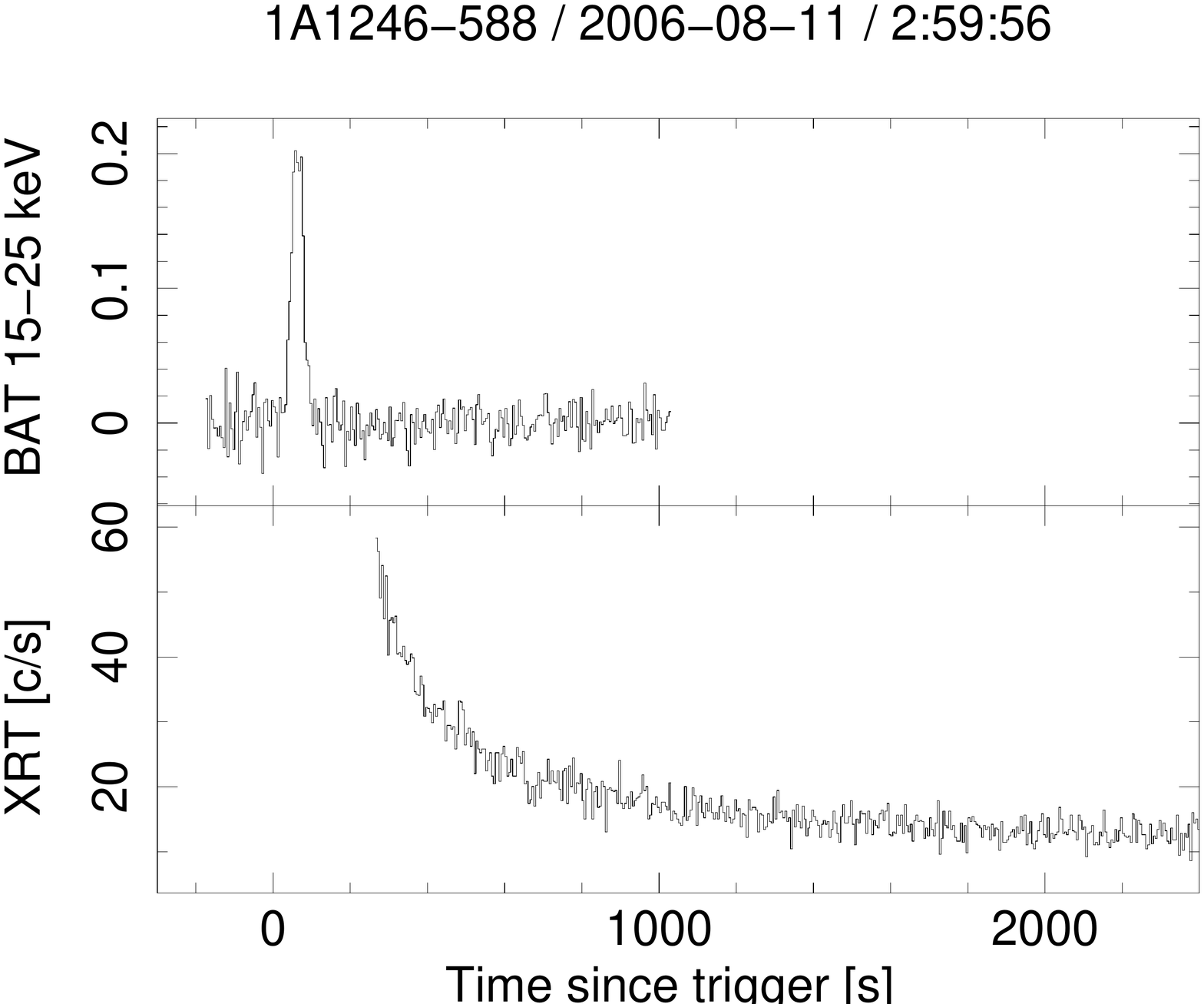}
\includegraphics[width=0.31\textwidth, trim=1.1cm 1cm 3cm 0.5cm,clip=true]{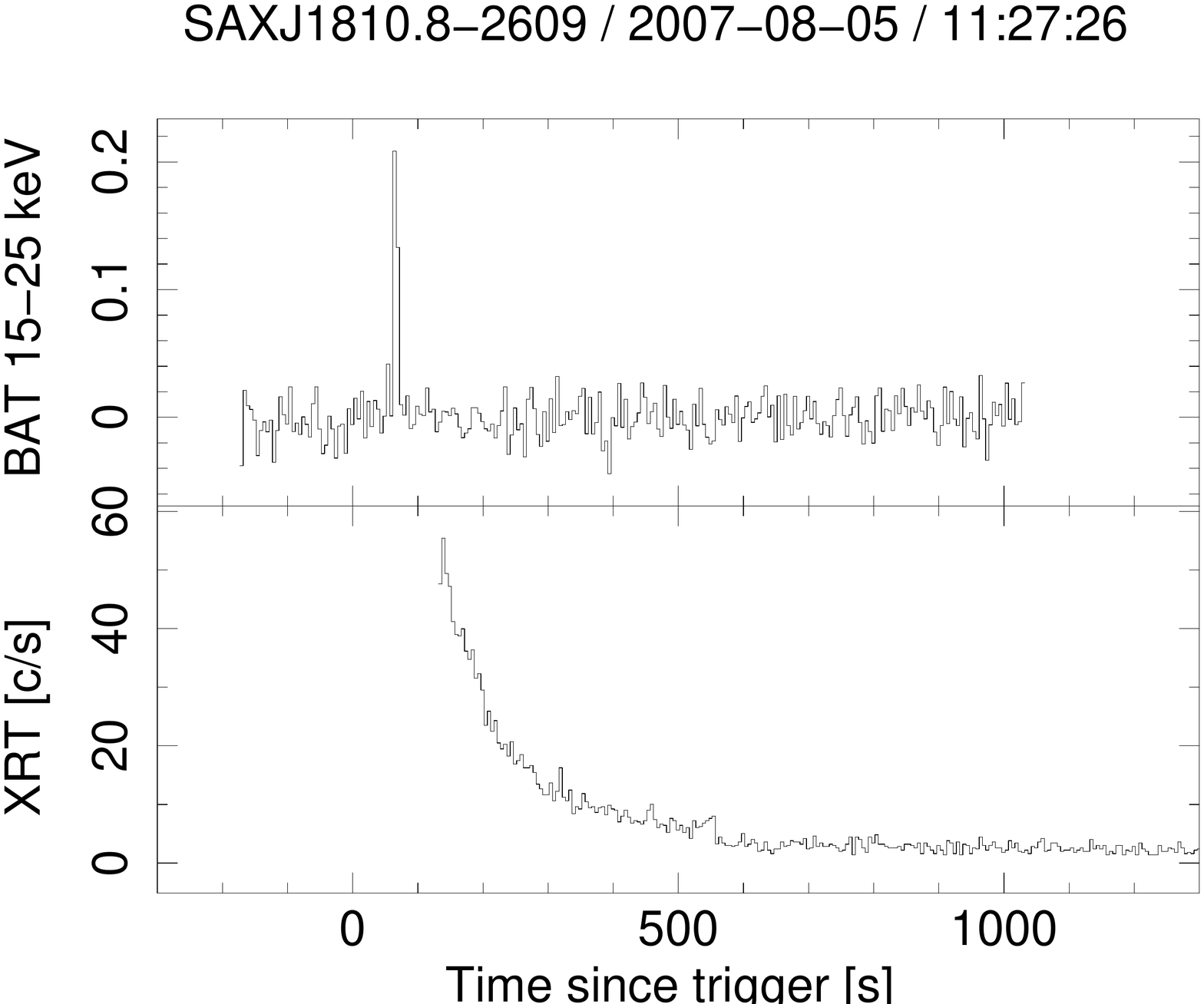}\vspace{3mm}
\includegraphics[width=0.325\textwidth, trim=0.0cm 1cm 3cm 0.5cm,clip=true]{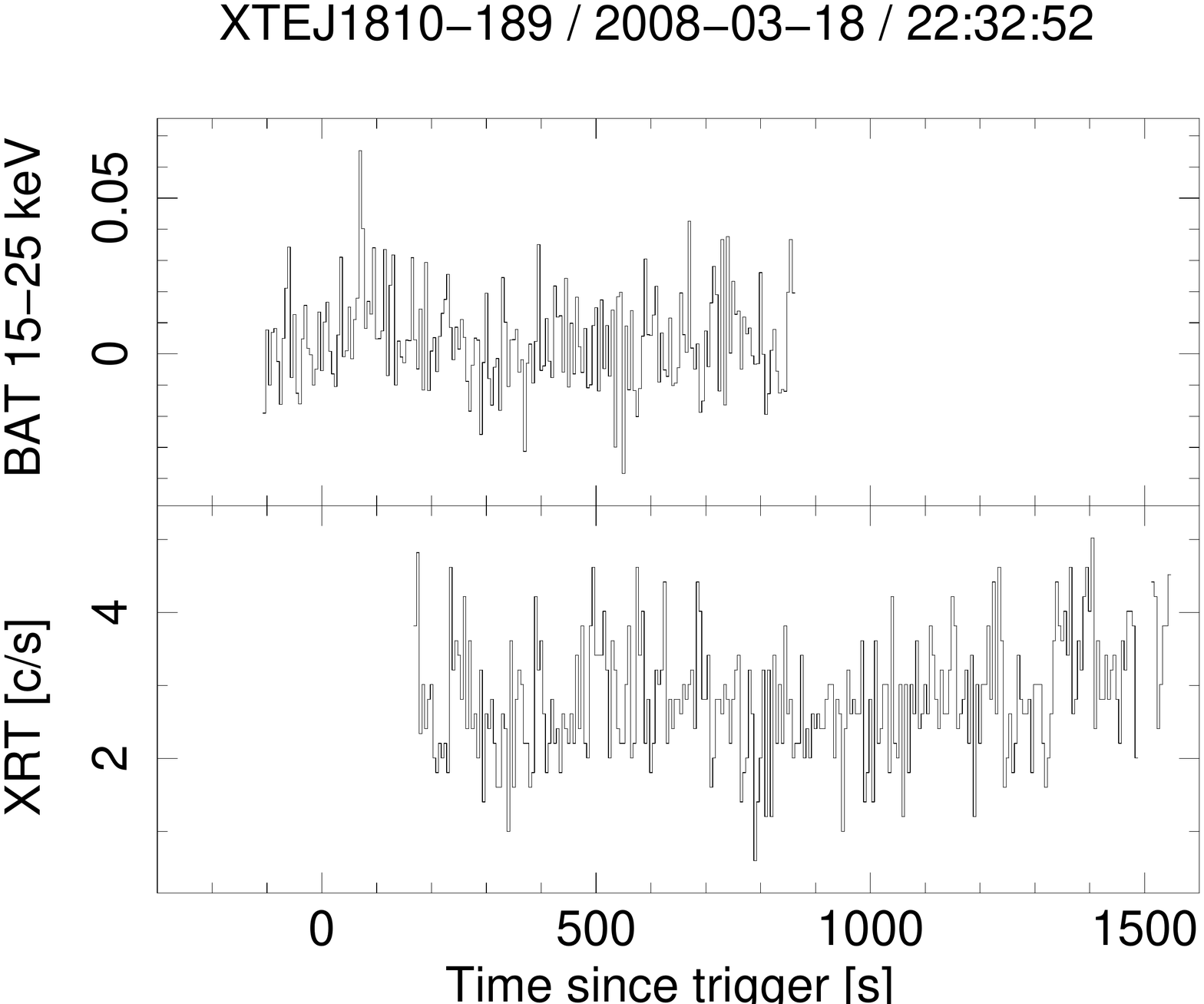}
\includegraphics[width=0.31\textwidth, trim=1.1cm 1cm 3cm 0.5cm,clip=true]{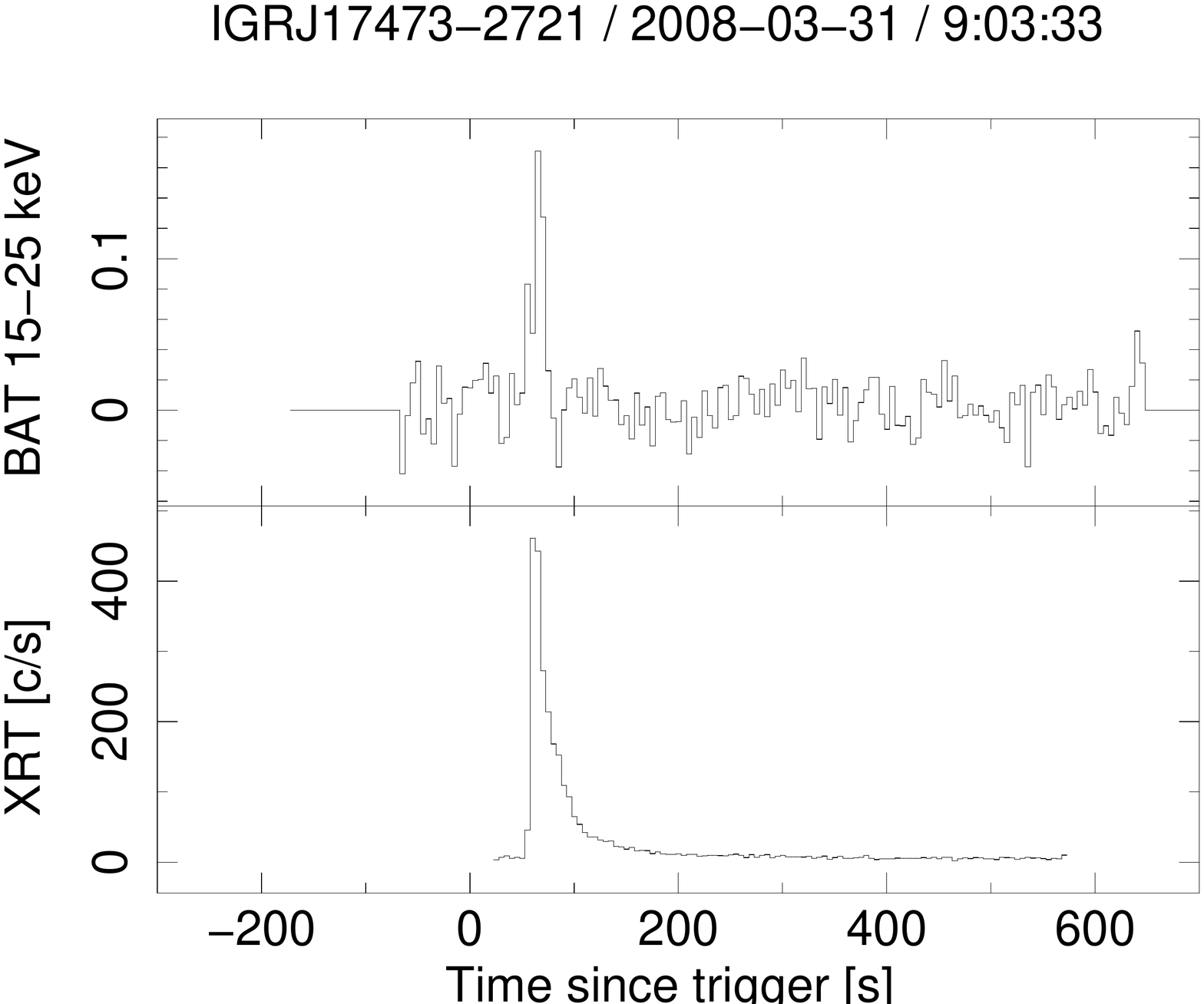}
\includegraphics[width=0.31\textwidth, trim=1.1cm 1cm 3cm 0.5cm,clip=true]{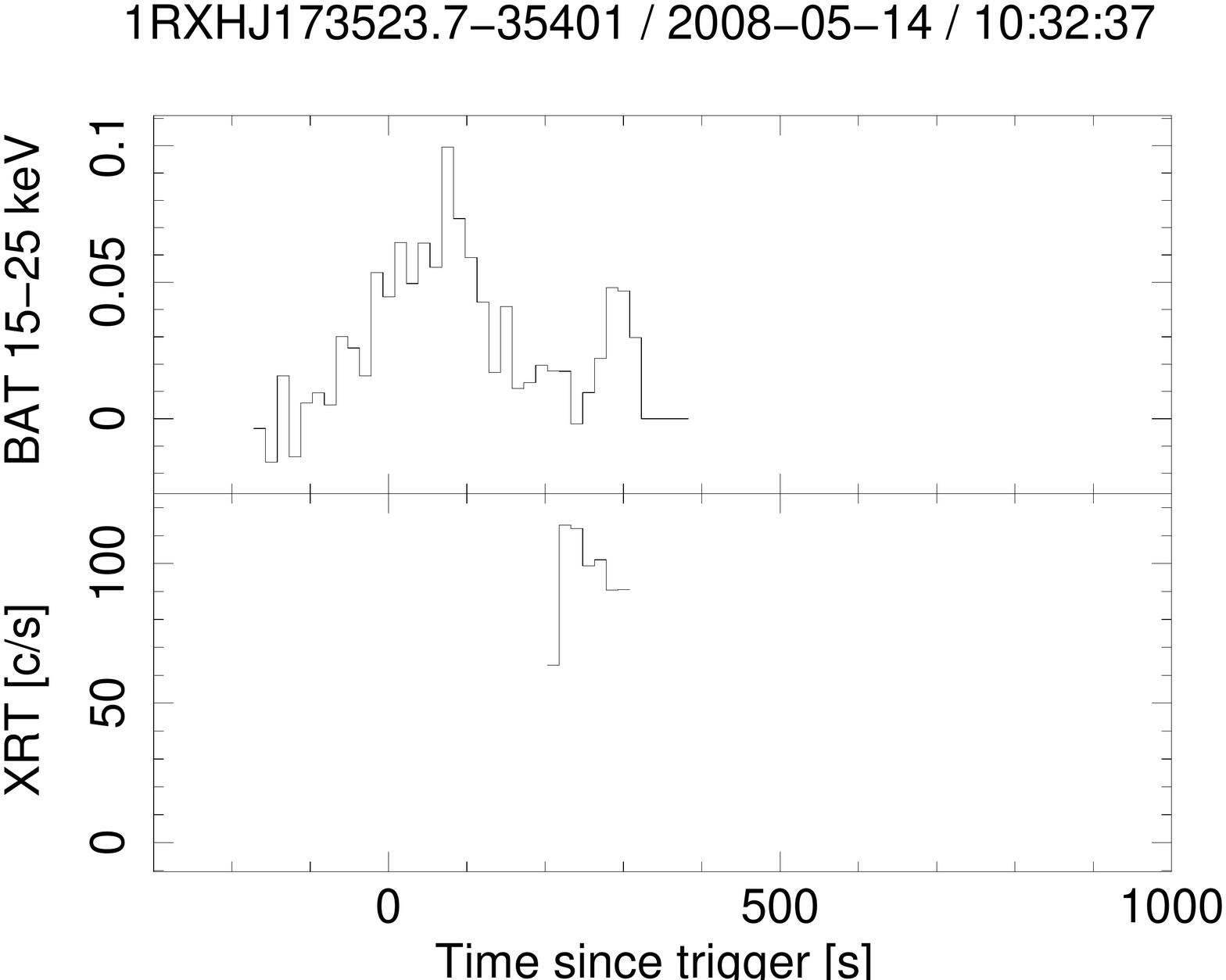}\vspace{3mm}
\includegraphics[width=0.325\textwidth, trim=0.0cm 1cm 3cm 0.5cm,clip=true]{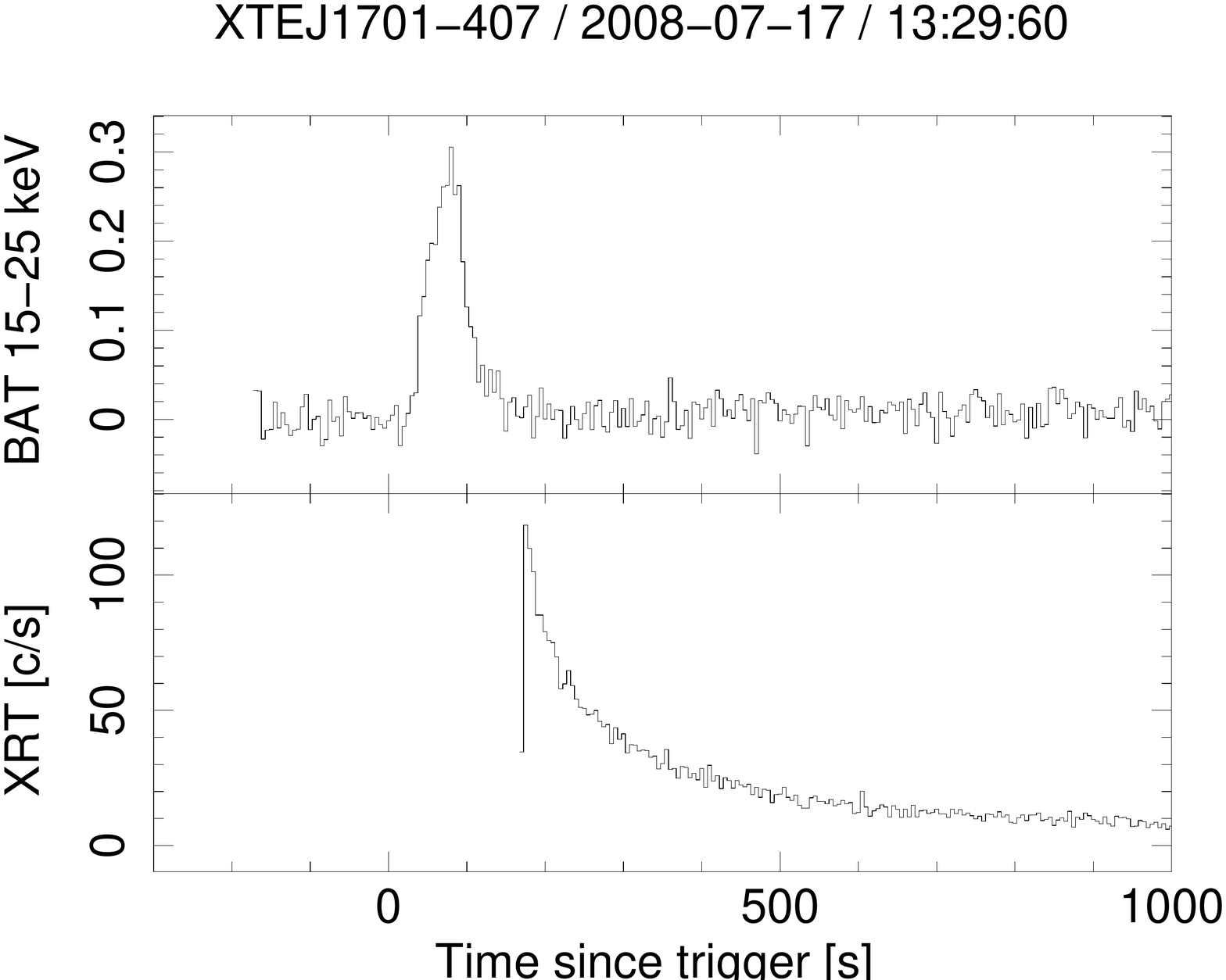}
\includegraphics[width=0.31\textwidth, trim=1.1cm 1cm 3cm 0.5cm,clip=true]{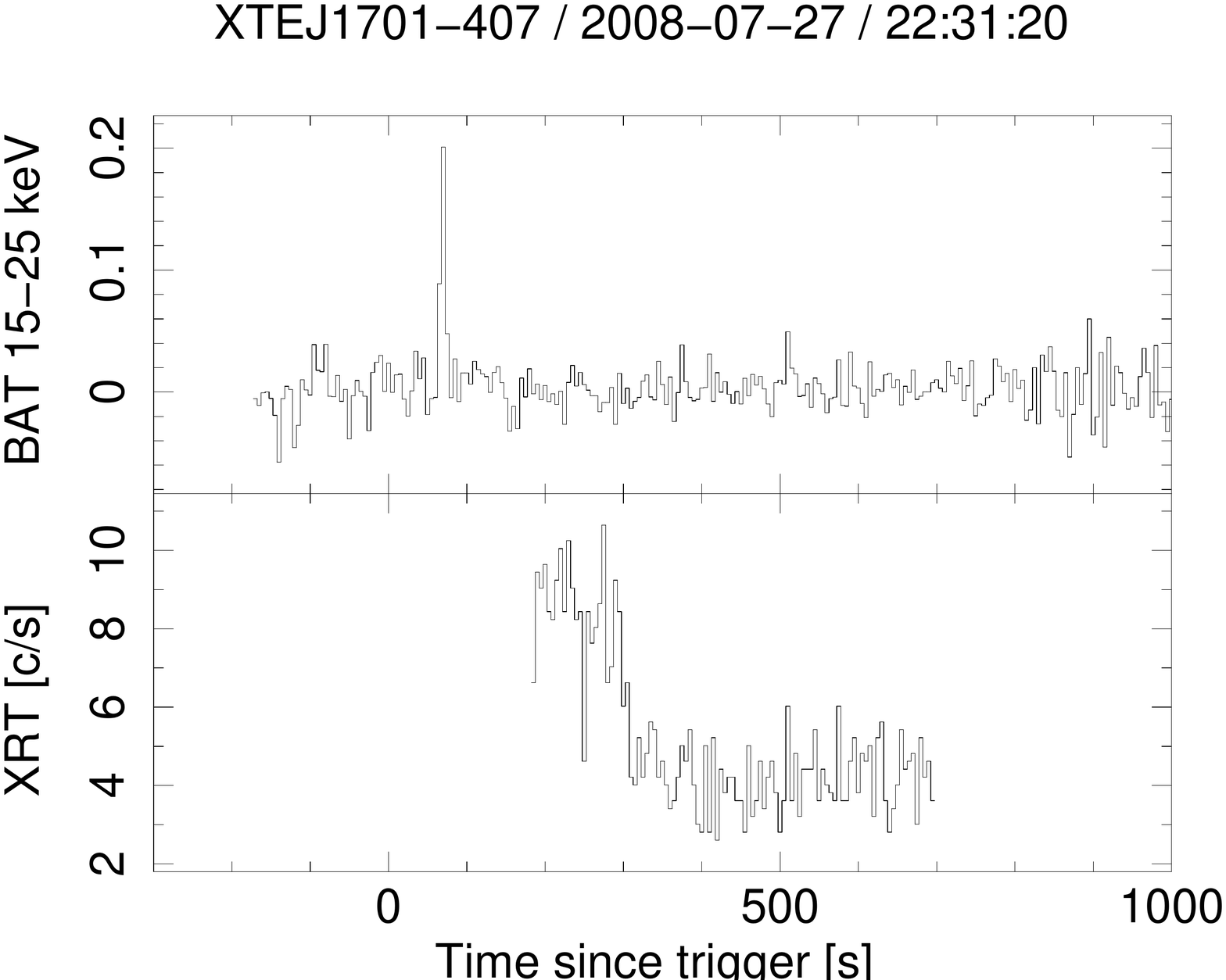}
\includegraphics[width=0.31\textwidth, trim=1.1cm 1cm 3cm 0.5cm,clip=true]{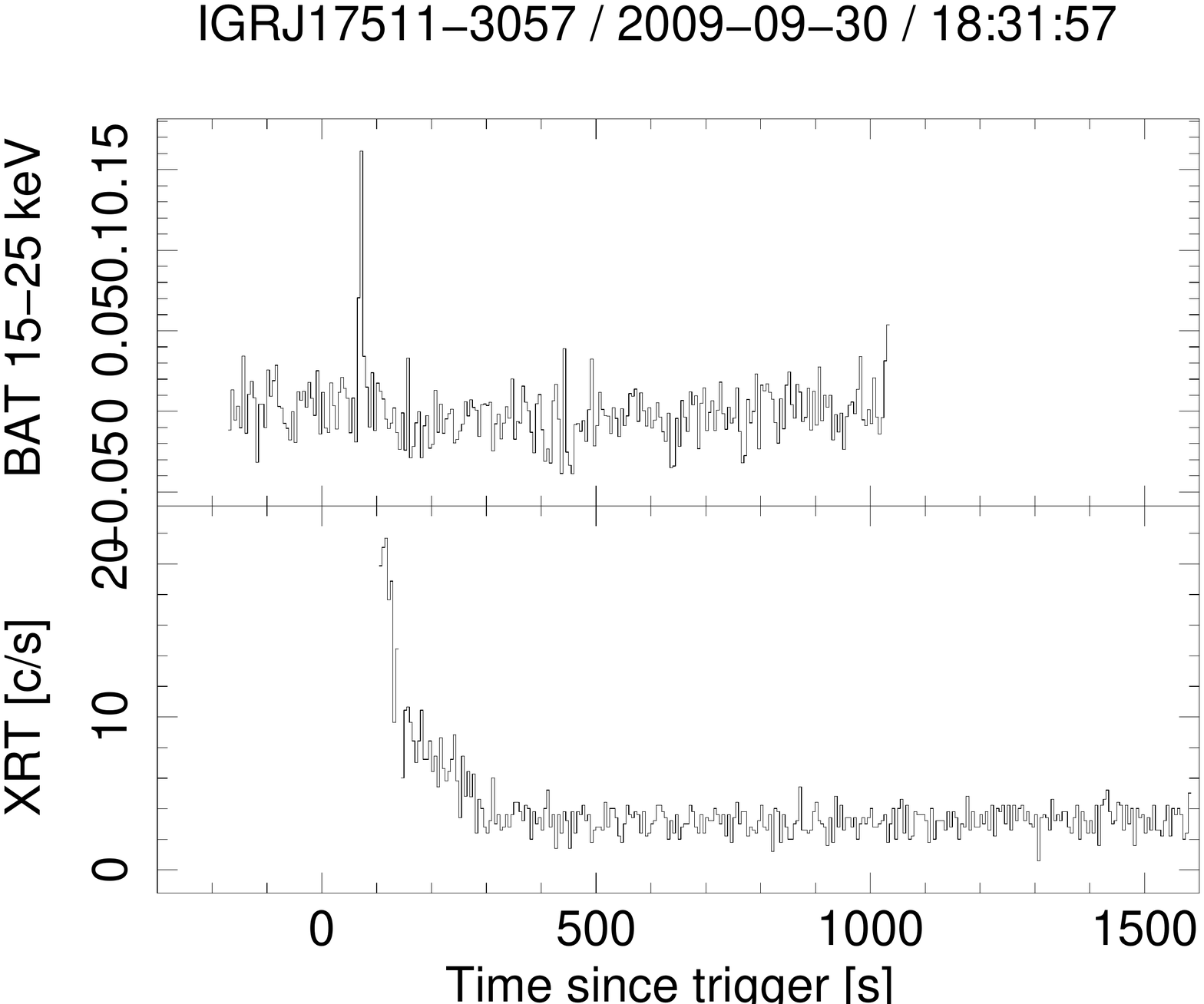}\vspace{3mm}
\includegraphics[width=0.325\textwidth, trim=0.0cm 1cm 3cm 0.5cm,clip=true]{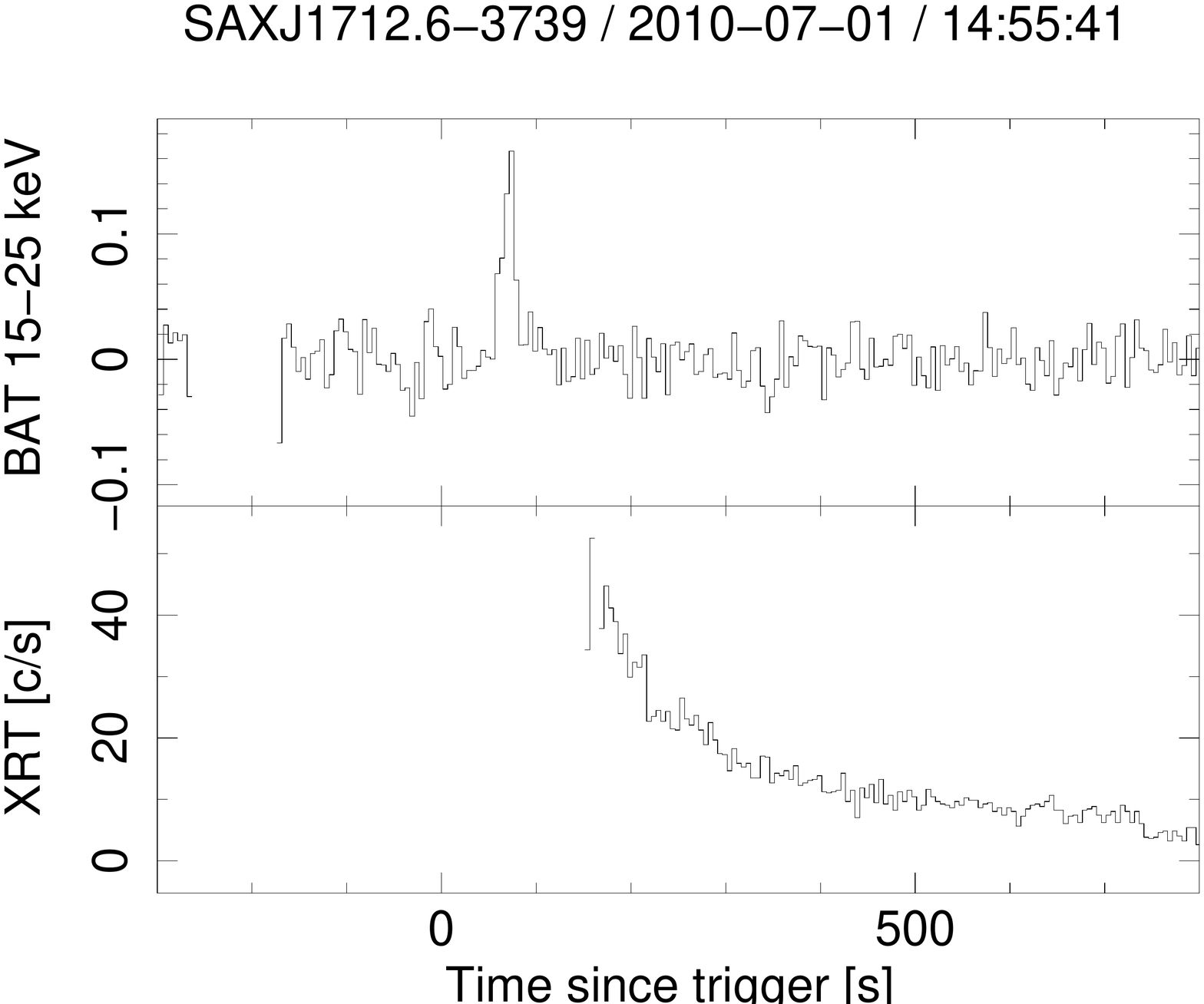}
\includegraphics[width=0.31\textwidth, trim=1.1cm 1cm 3cm 0.5cm,clip=true]{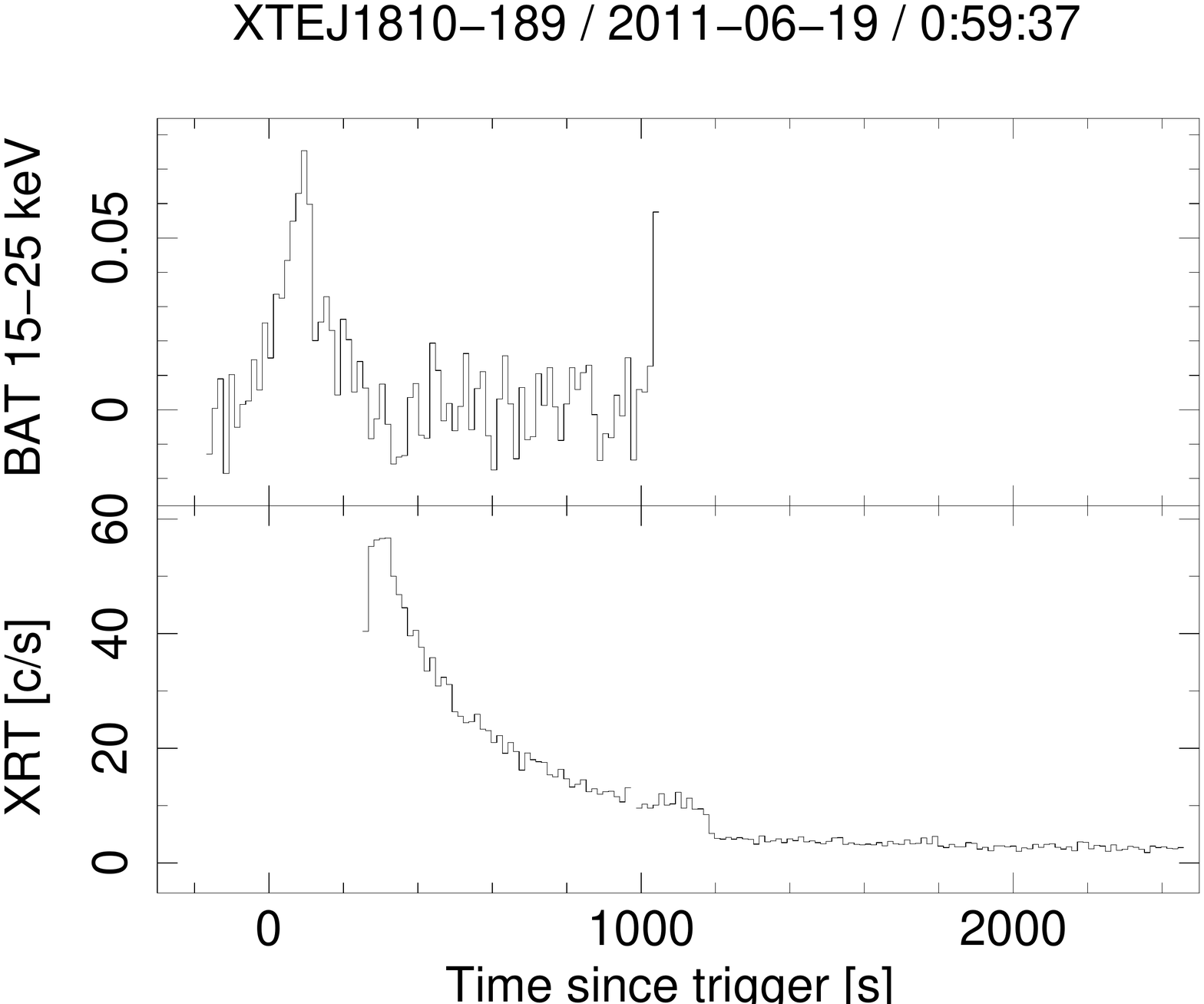}
\includegraphics[width=0.31\textwidth, trim=1.1cm 1cm 3cm 0.5cm,clip=true]{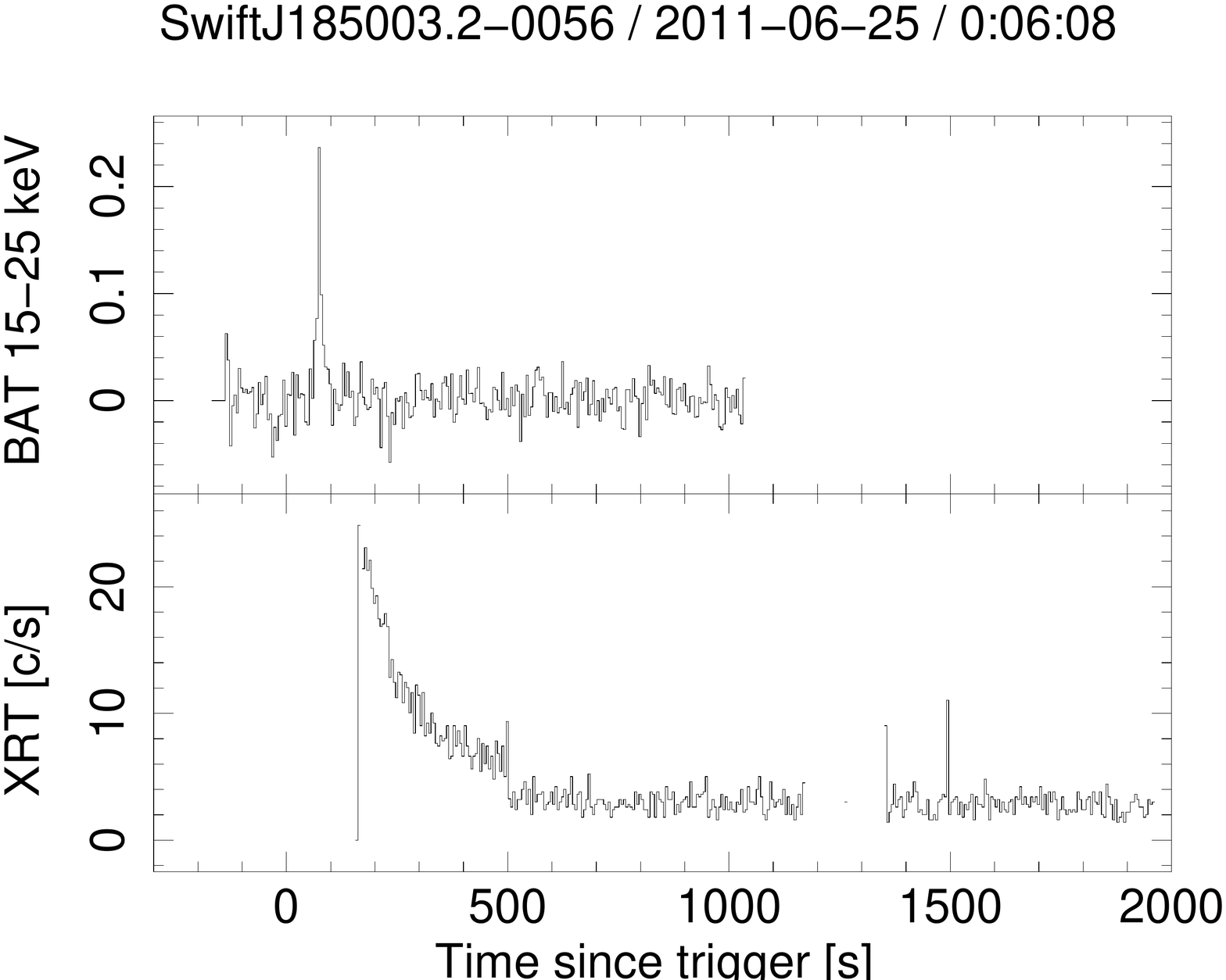}\vspace{3mm}
\includegraphics[width=0.325\textwidth, trim=0.0cm 0cm 3cm 0.5cm,clip=true]{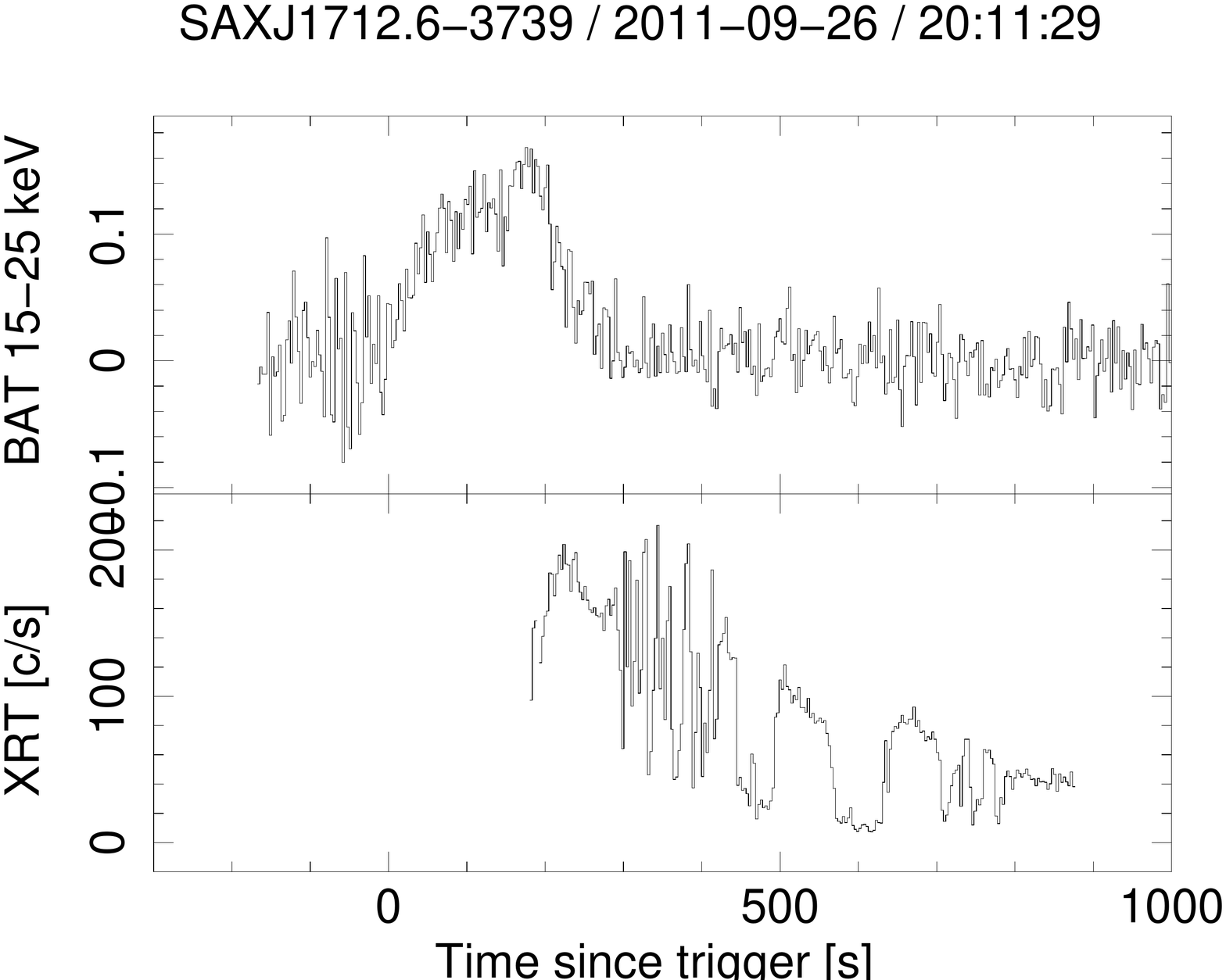}\hspace{2mm}
\includegraphics[width=0.325\textwidth, trim=1.1cm 0cm 3cm 0.5cm,clip=true]{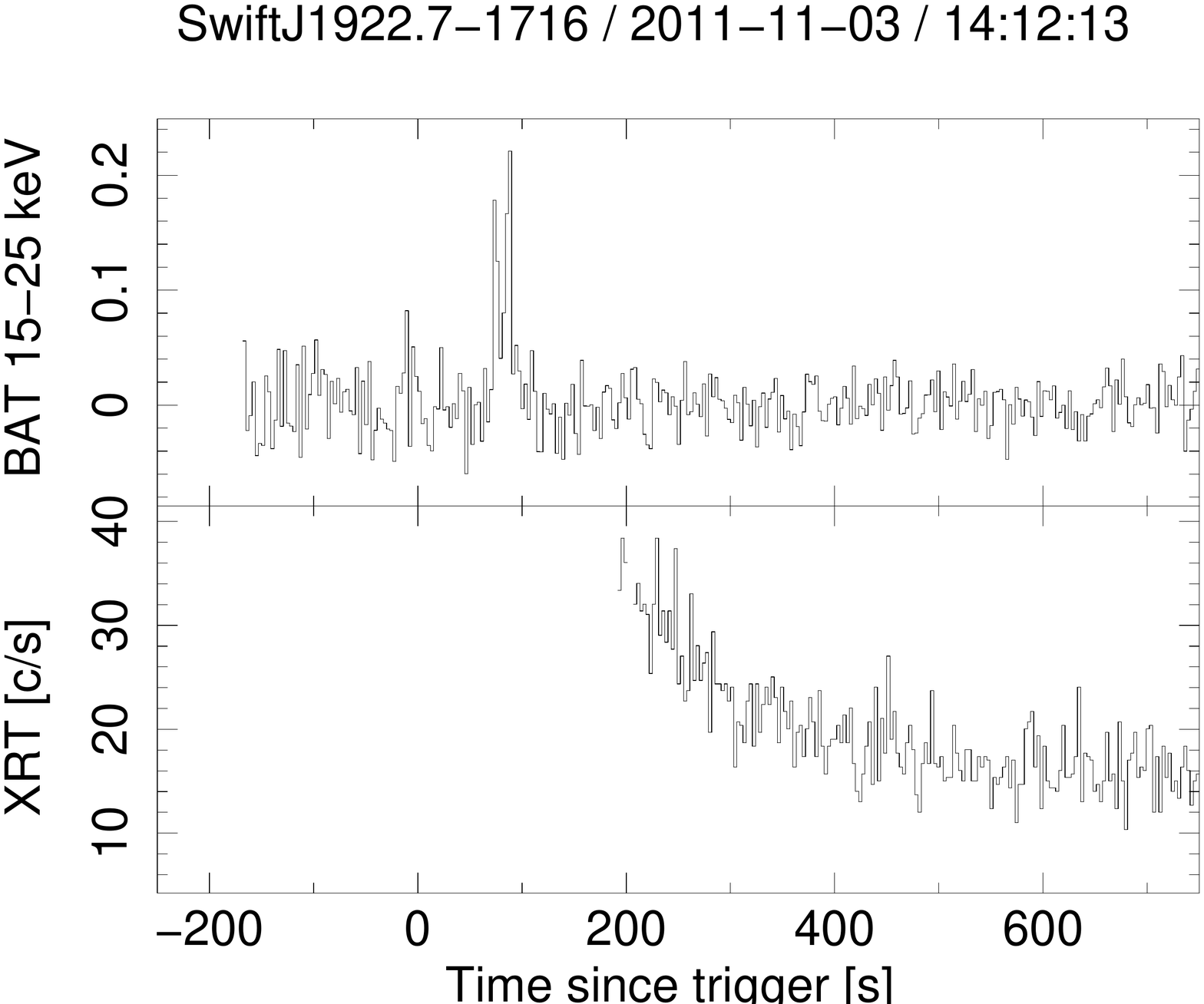}\hspace{2mm}
\includegraphics[width=0.325\textwidth, trim=1.1cm 0cm 3cm 0.5cm,clip=true]{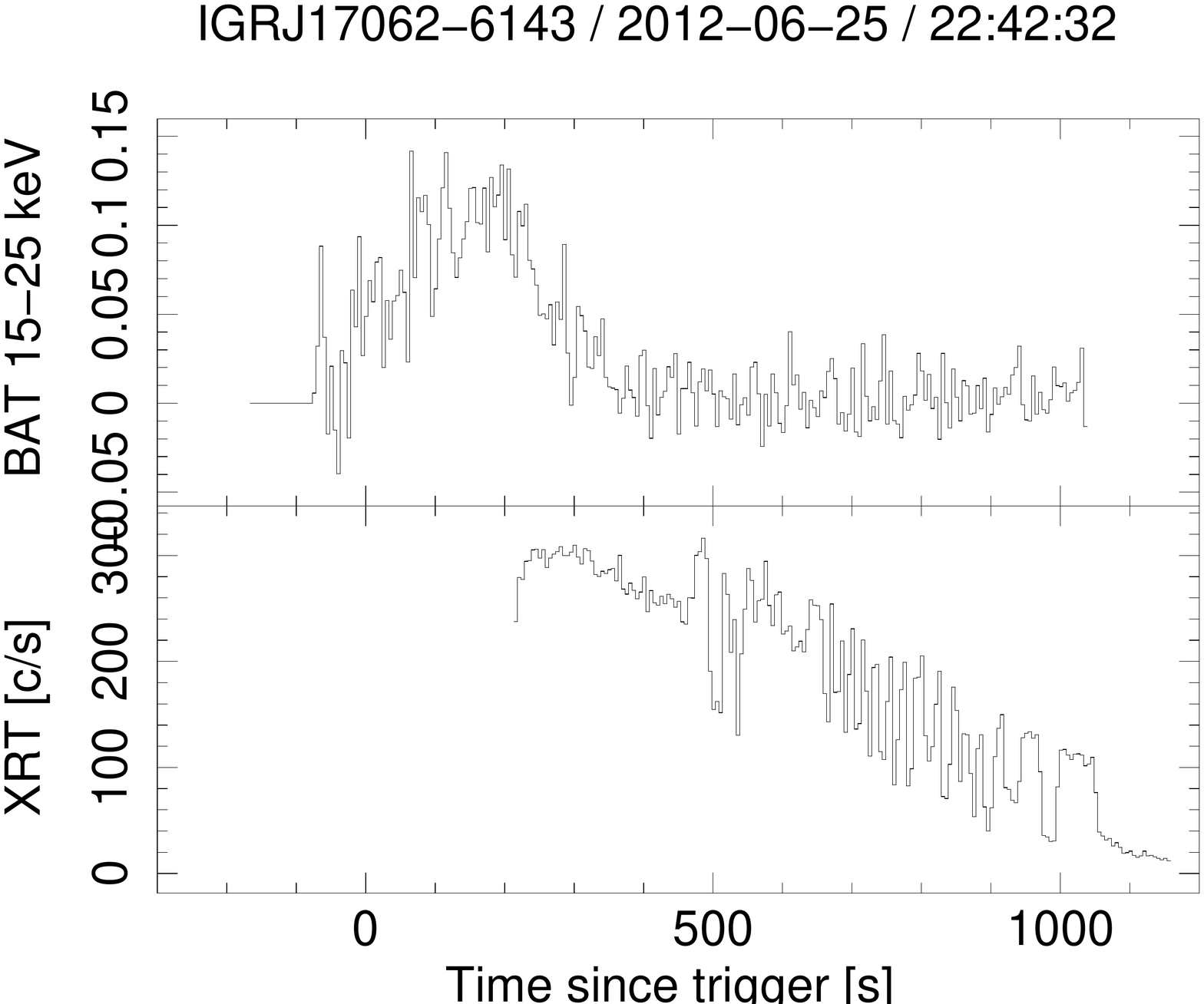}
    \caption{BAT and XRT light curves of all 28 X-ray bursts with BAT
      event files and followed up with XRT, except
      4U~1812-12/2005-02-24 because its XRT coverage is sparse.  The
      BAT light curves are normalized and corrected for a changing
      pointing due to the slews. The XRT data are for the complete
      detector and without corrections for pile-up or background. The
      steps that can sometimes be seen in XRT data are due to a switch
      in CCD data collection mode with different pile-up fractions.}
    \label{fig:lc1}
\end{figure*}

\setcounter{figure}{0}
\begin{figure*}[!t]
\includegraphics[width=0.3325\textwidth, trim=0.0cm 1cm 3cm 0.5cm,clip=true]{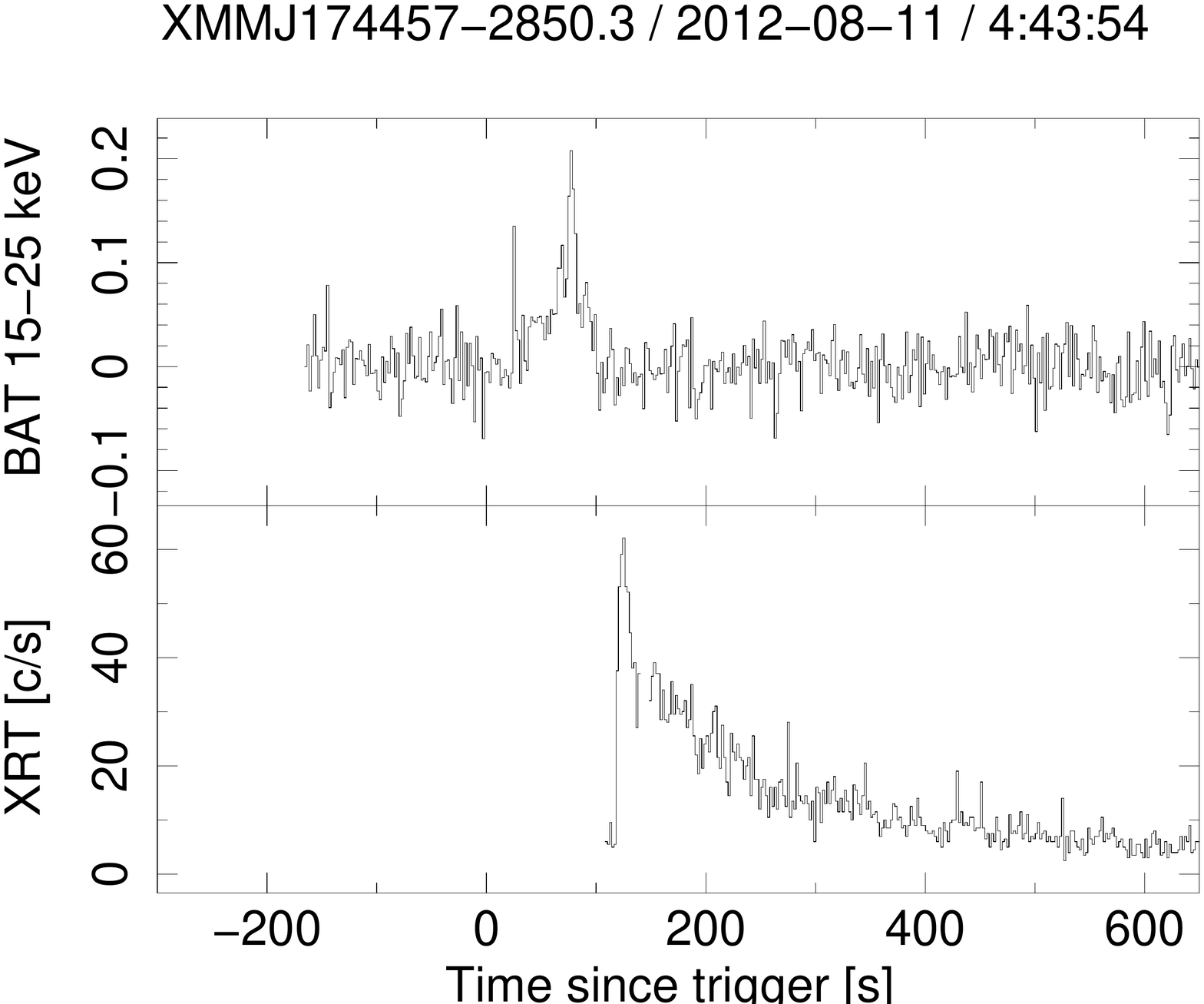}
\includegraphics[width=0.31\textwidth, trim=1.1cm 1cm 3cm 0.5cm,clip=true]{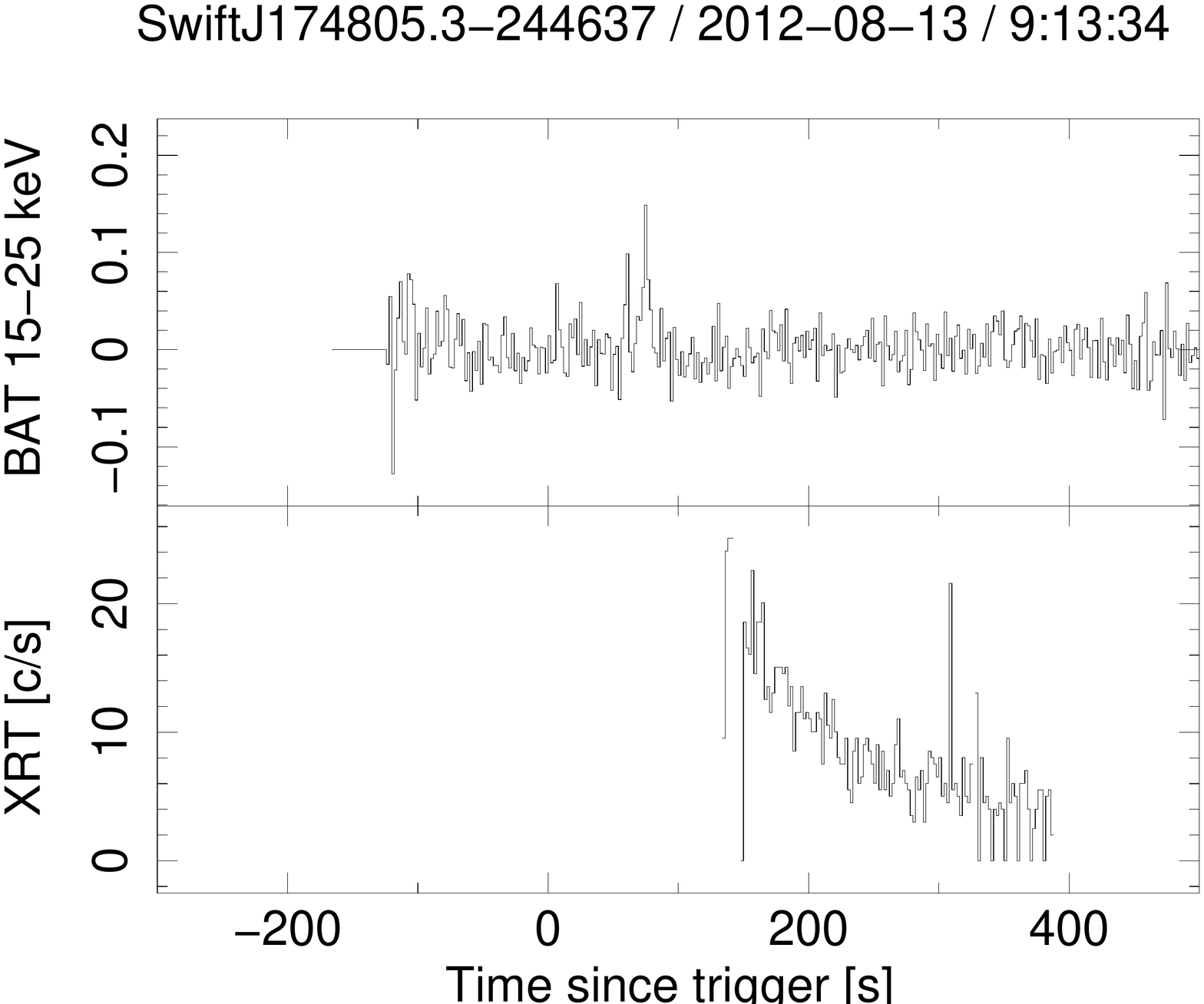}
\includegraphics[width=0.31\textwidth, trim=1.1cm 1cm 3cm 0.5cm,clip=true]{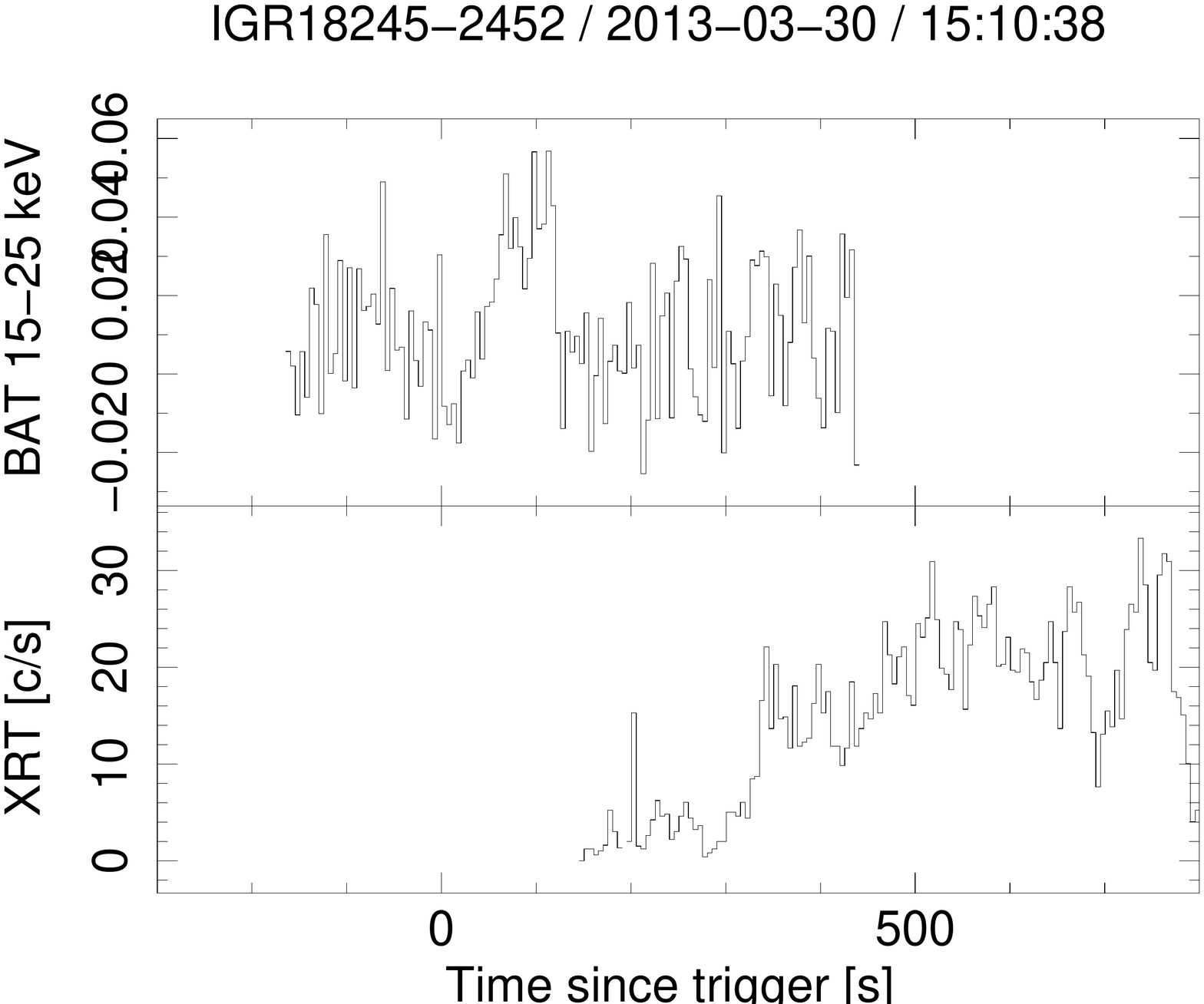}\vspace{3mm}
\includegraphics[width=0.325\textwidth, trim=0.0cm 1cm 3cm 0.5cm,clip=true]{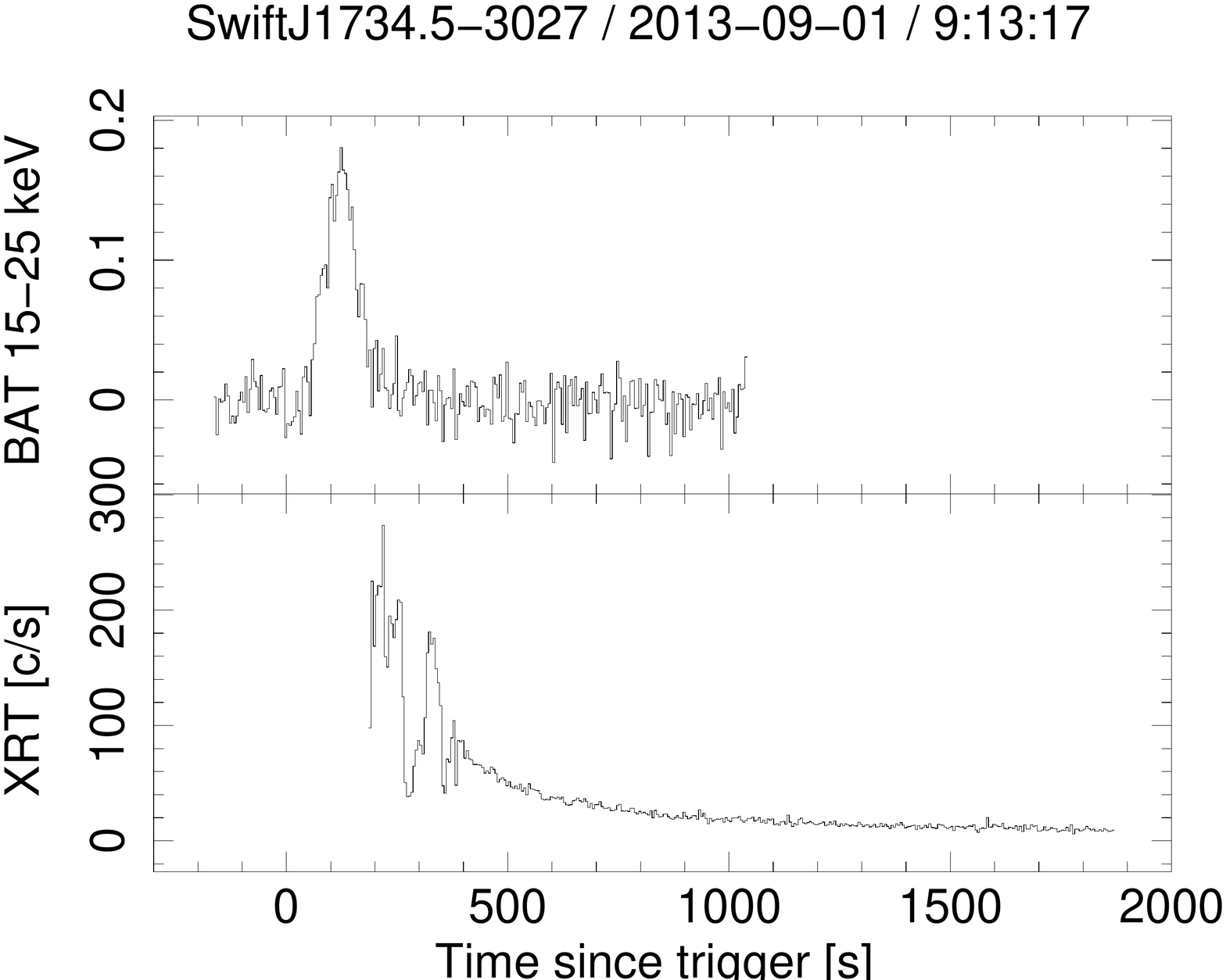}
\includegraphics[width=0.31\textwidth, trim=1.1cm 1cm 3cm 0.5cm,clip=true]{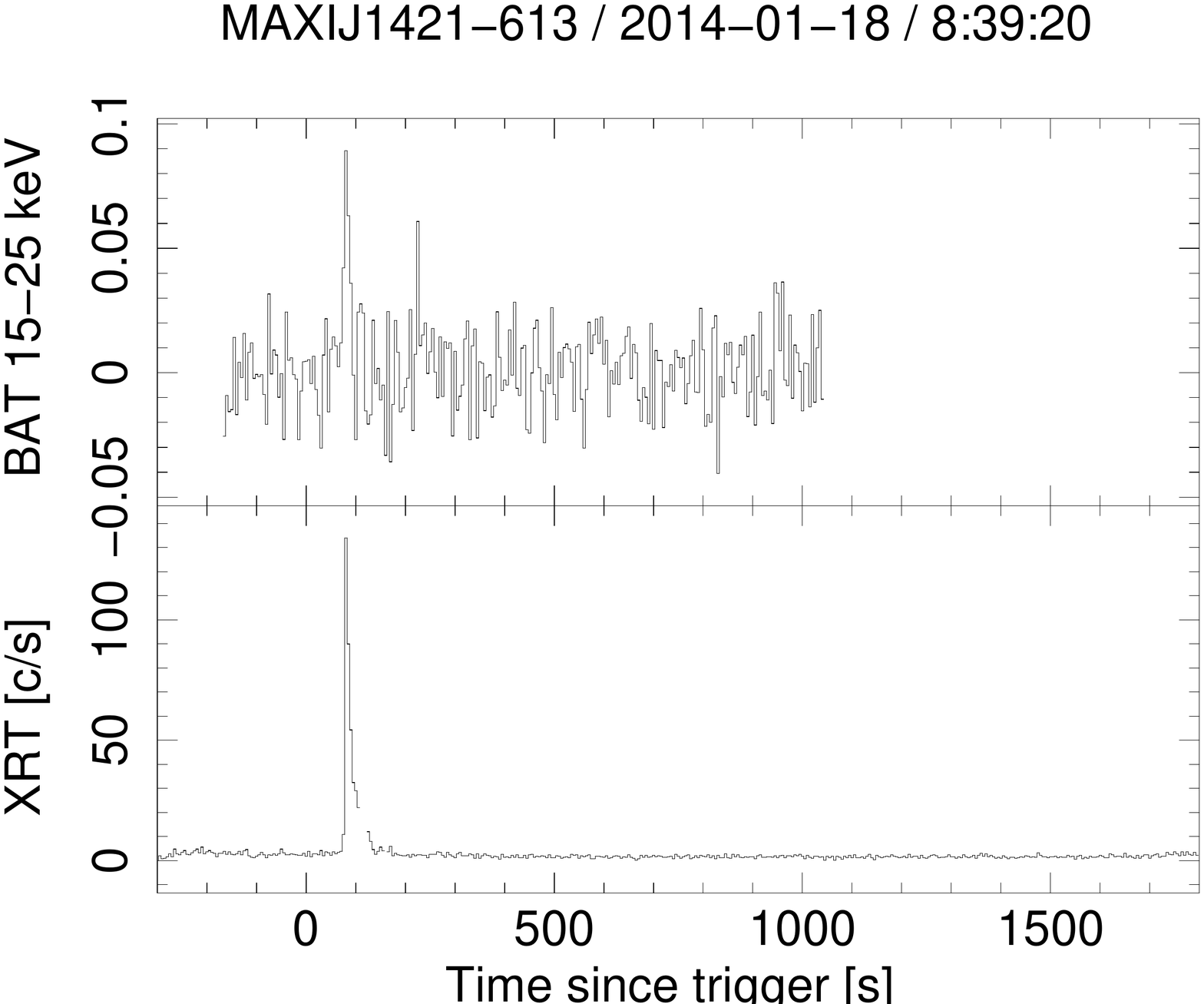}
\includegraphics[width=0.31\textwidth, trim=1.1cm 1cm 3cm 0.5cm,clip=true]{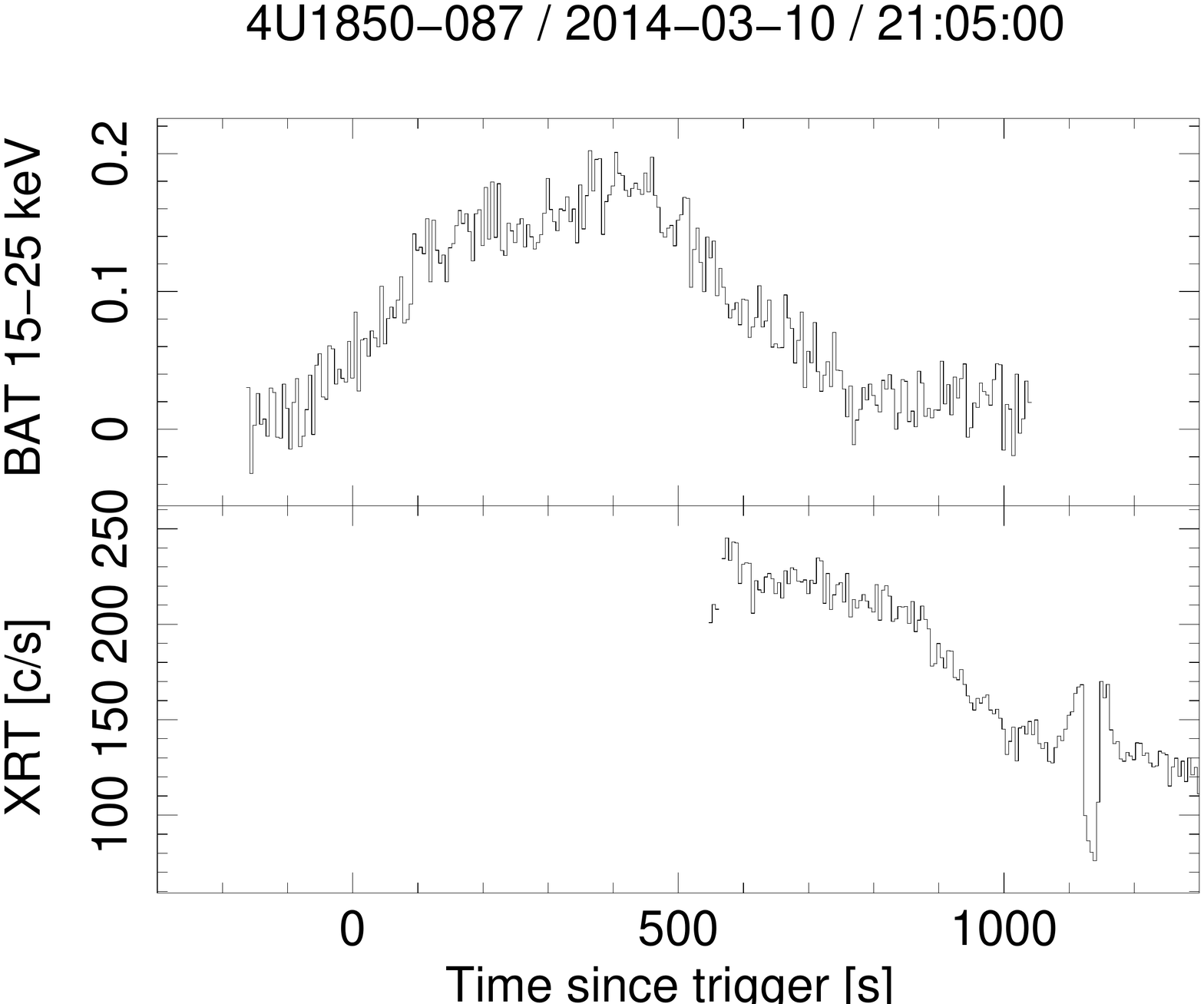}\vspace{3mm}
\includegraphics[width=0.325\textwidth, trim=0.0cm 1cm 3cm 0.5cm,clip=true]{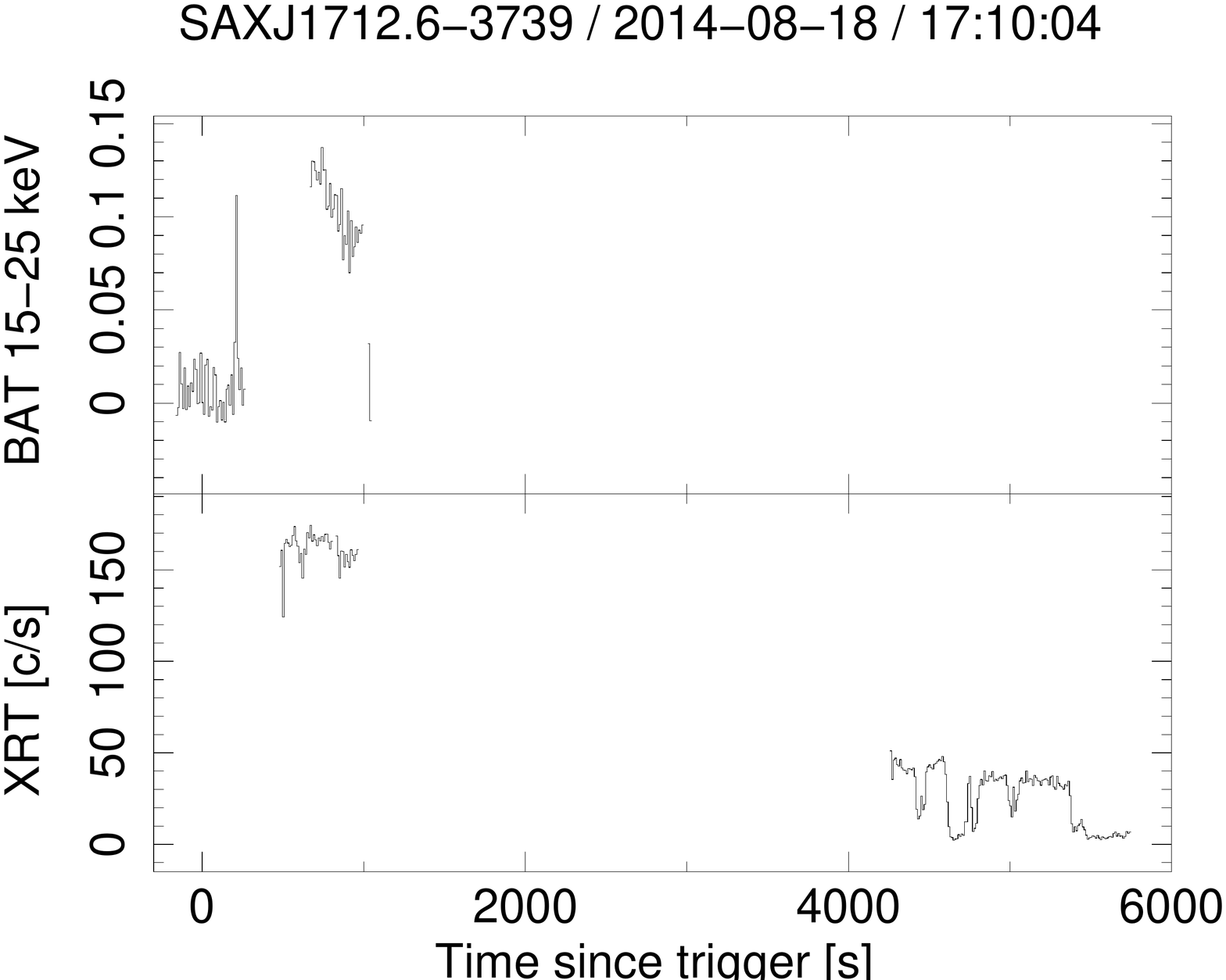}
\includegraphics[width=0.31\textwidth, trim=1.1cm 1cm 3cm 0.5cm,clip=true]{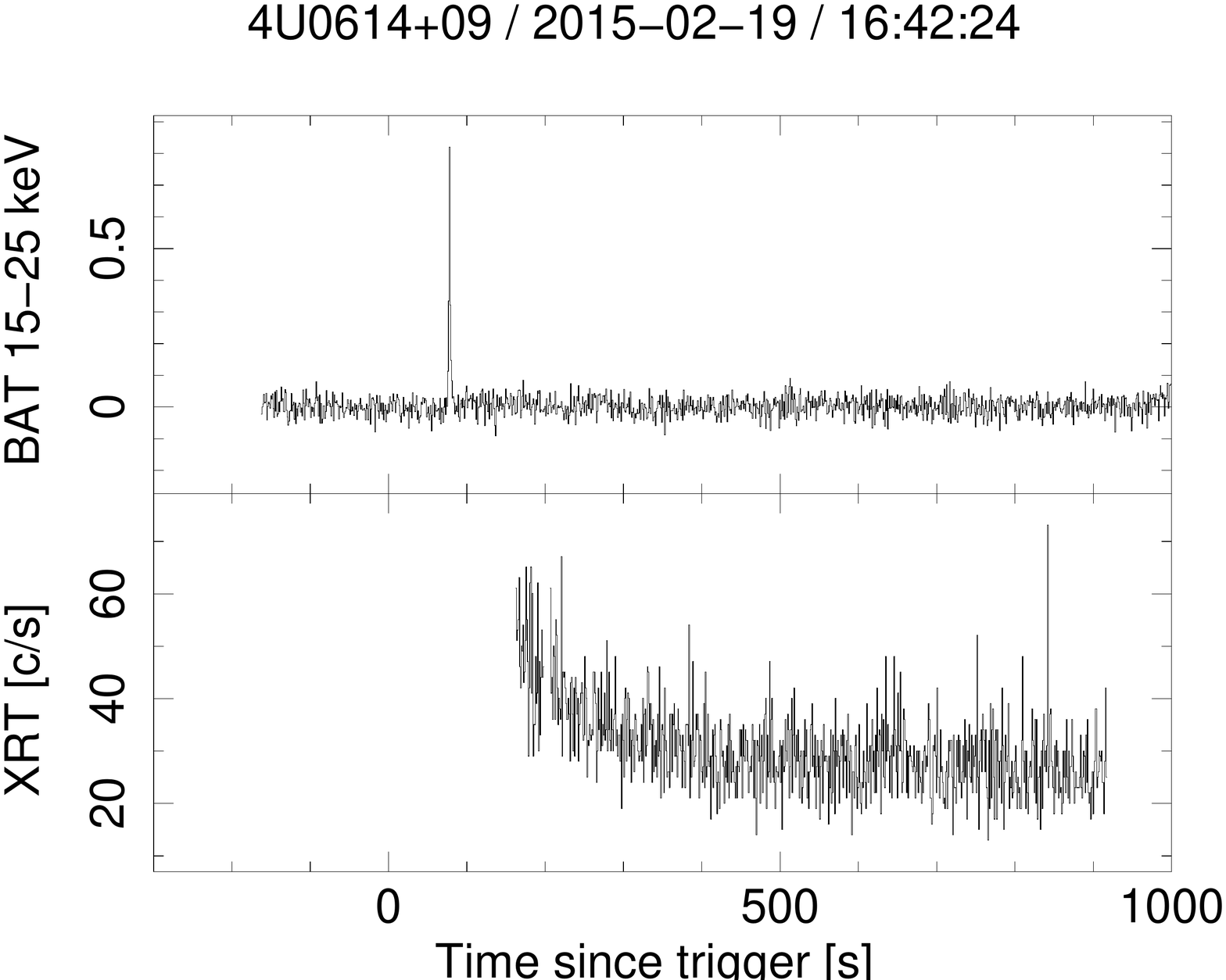}
\includegraphics[width=0.31\textwidth, trim=1.1cm 1cm 3cm 0.5cm,clip=true]{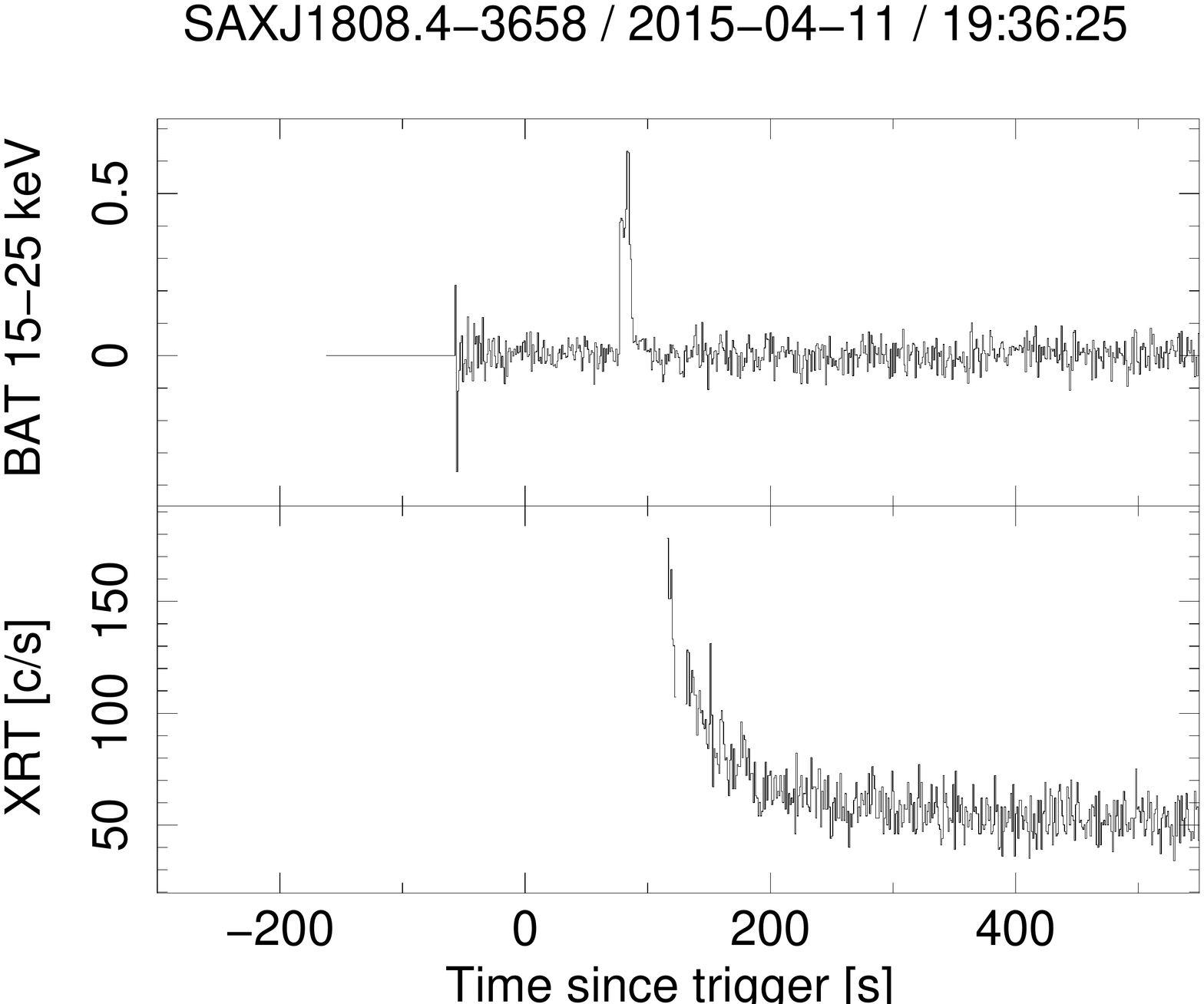}\vspace{3mm}
\includegraphics[width=0.325\textwidth, trim=0.0cm 0cm 3cm 0.5cm,clip=true]{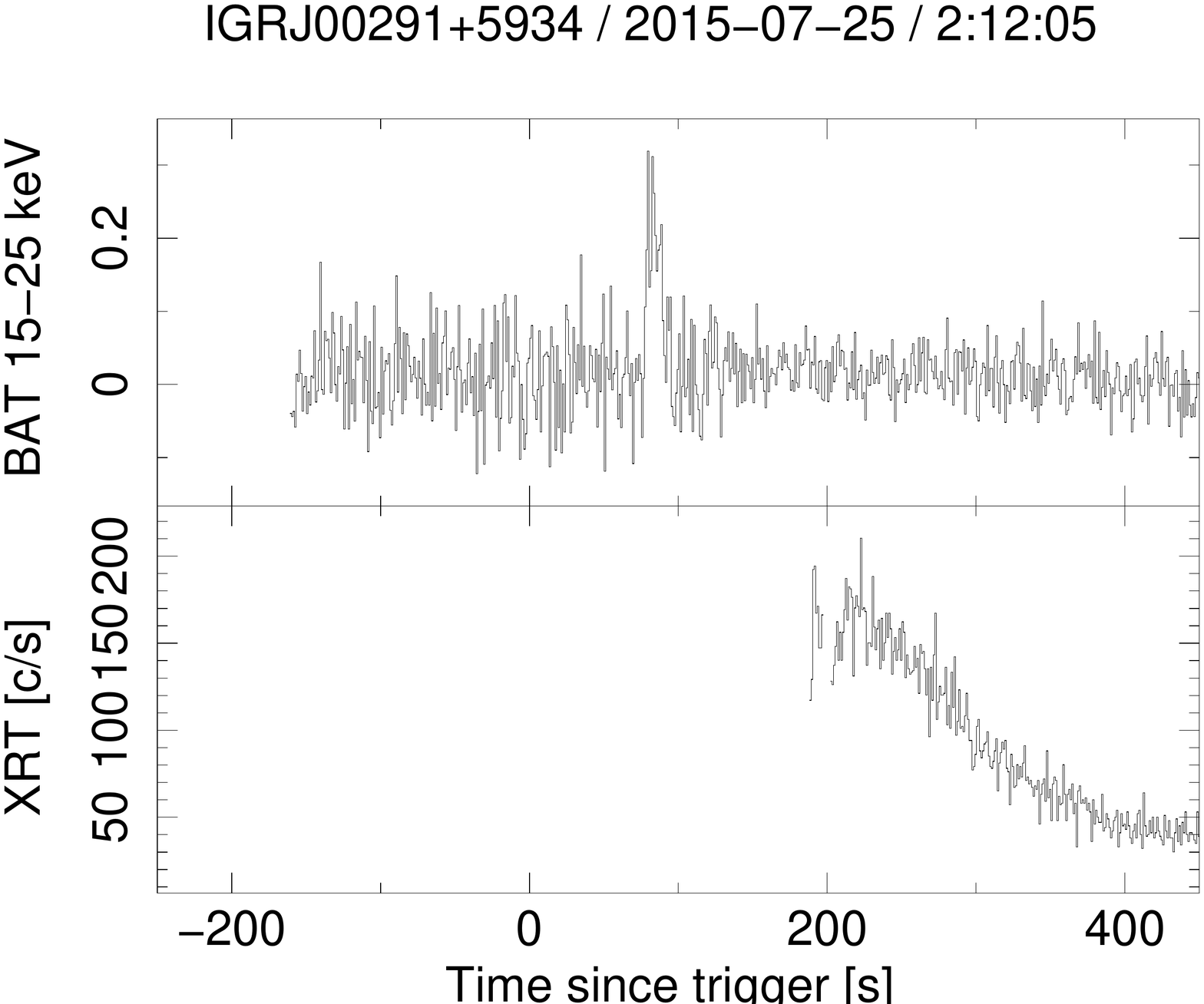}\hspace{2mm}
\includegraphics[width=0.325\textwidth, trim=1.1cm 0cm 3cm 0.5cm,clip=true]{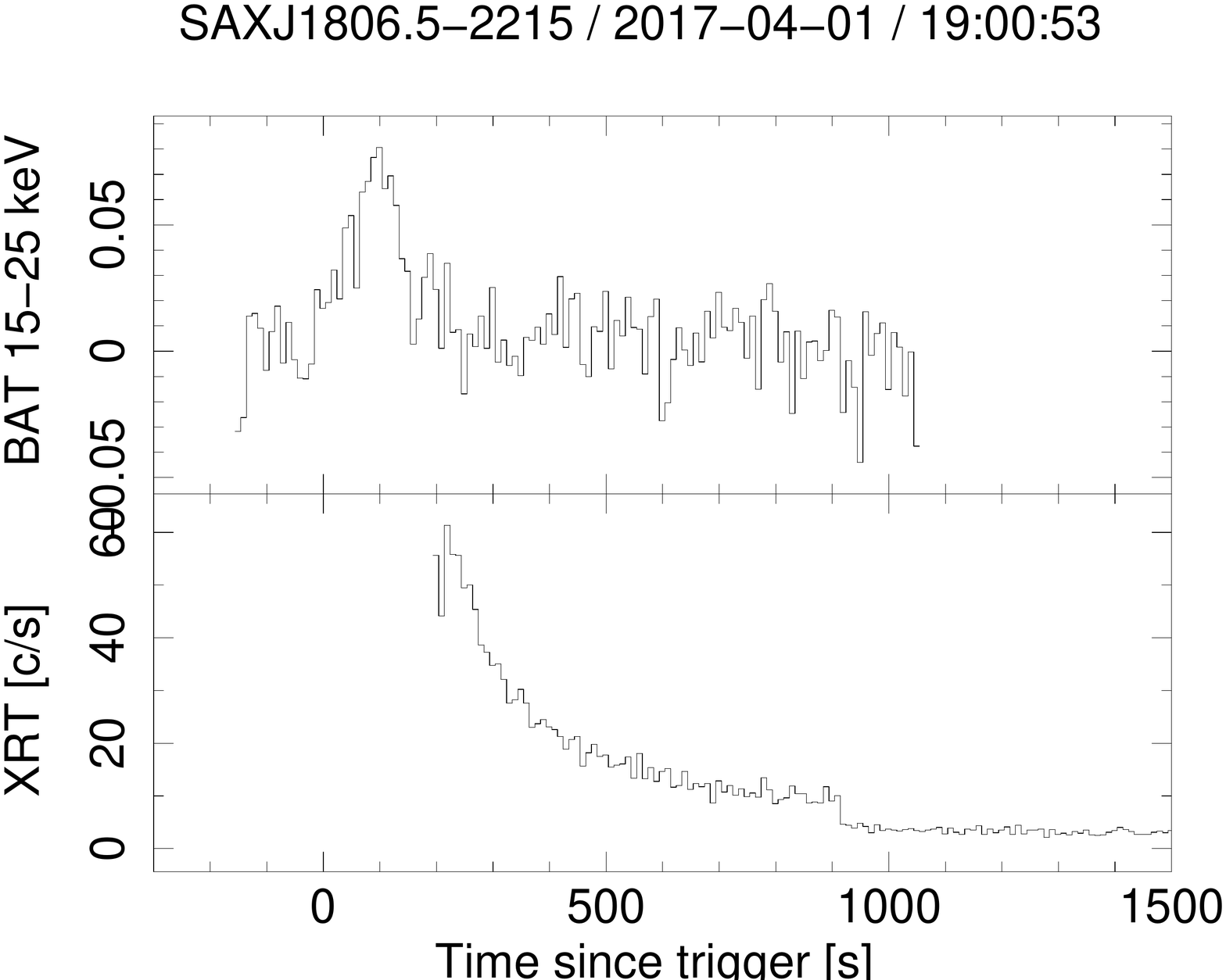}\hspace{2mm}
\includegraphics[width=0.325\textwidth, trim=1.1cm 0cm 3cm 0.5cm,clip=true]{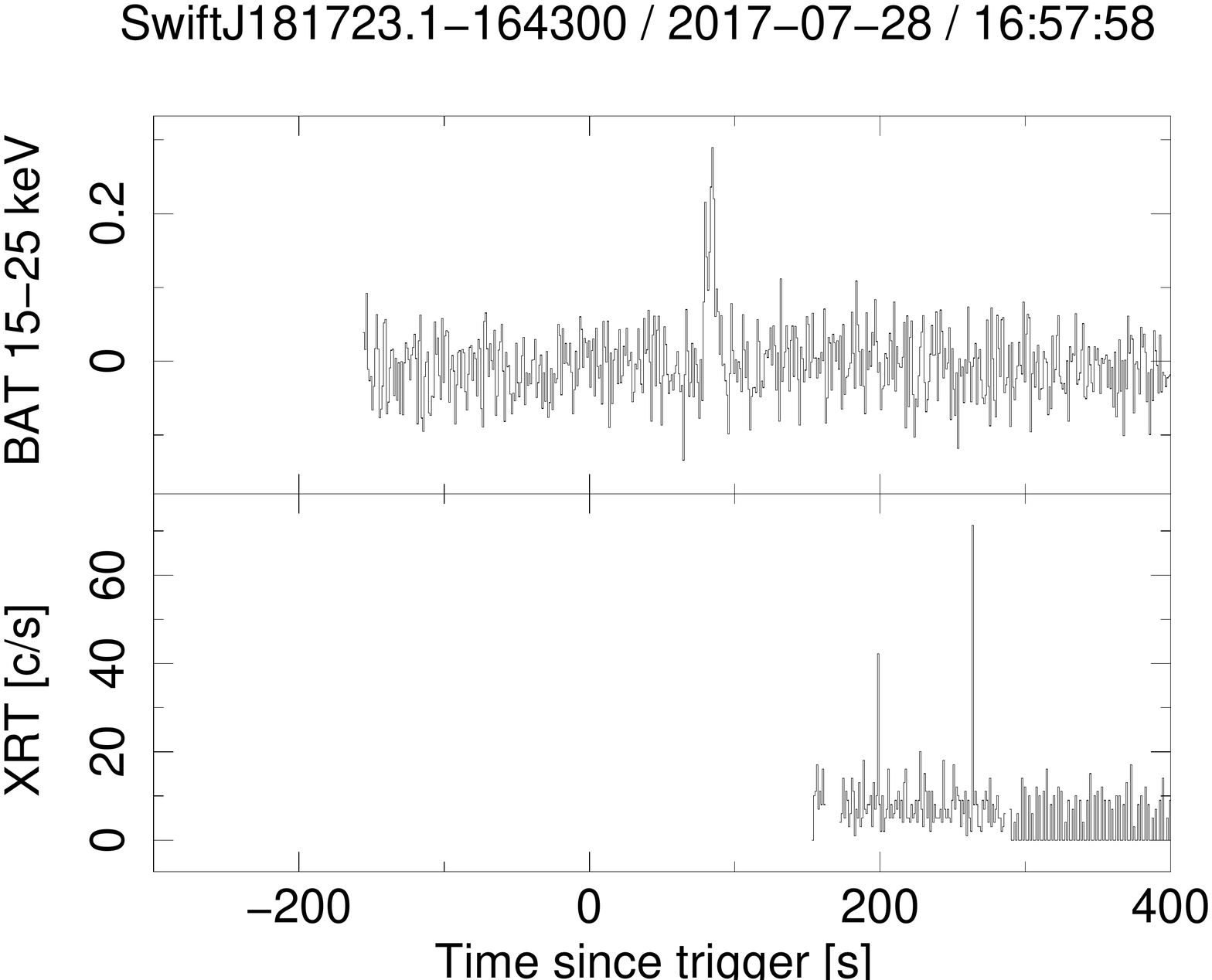}
    \caption{{\it (continued)}}
    \label{fig:lc2}
\end{figure*}

If a source detection is considered interesting, that source's merit
parameter is combined with other parameters (such as its location
relative to the Sun and Moon exclusion zones) to give an overall
Figure of Merit (FoM) value, and the source becomes Swift's new
Automated Target (AT) unless there is already an active AT with a
higher FoM.  Swift continuously tests whether the current AT is in the
observable part of the sky and, if so, compares its FoM to that of the
Pre-Planned Science Target (PPST) scheduled for that time.  If the AT
has a higher FoM, Swift slews to the AT and continues pointed
observations until it is no longer observable, a higher-FoM PPST takes
precedence, or the AT completes a pre-determined total observation
time.  These pointed observations allow XRT and UVOT follow-up
observation within ~100 s after detection (Swift's typical slew time)
if the source's merit is sufficient.

In the first five years of the mission, the FoM for known sources
(including X-ray bursters, unless not yet known as such) were set to
very low values and they would not be observed with the XRT, unless
accidentally because the onboard catalog was not up to date. In 2009
this approach changed. A few other targets received higher
priority. Bursters with very low burst rates were allowed to be
followed up (prior to May 2009 M15 X-2, 1RXS J170854.4-321857, 1RXS
J171824.2-402934, 4U 1722-30, SLX 1737-282 and SAX J1752.3-3138;
C. Markwardt, priv. comm.) next to of course new bursters.

\section{Search}
\label{sec:search}

We are looking for rare kinds of X-ray bursts that have coverage in
the sub-10 keV band. Such coverage is necessary because fluctuations
have only been seen in the tails of bursts when the temperature may be
too low for useful data in the $>15$ keV band. We are most likely to
find these bursts among BAT triggers that resulted in automatic slews
and have positions coincident with known X-ray bursters.

We first searched for detections of X-ray bursts with BAT that
triggered the instrument into downloading event data, by querying the
Swift Master Catalog at HEASARC for all observations with BAT event
data for more than 5 s. This image-encoded data, sampled at 0.1 ms,
enables the verification of a true burst signal from the location of a
bonafide X-ray burster. BAT provides event mode data whenever it is
triggered by one of the onboard rate or image trigger
algorithms. Depending on the follow-up script as specified per source,
either 21 mins of such event mode data are downloaded or 40 s. Bursts
that result in an AT will always have 21 mins of event mode data
available.

Our query on March 31, 2018, yielded 2319 observations, including 985
gamma-ray bursts. We downloaded the mask-tagged light curves for the
triggers and compared the R.A. and Decl. coordinates of the trigger
object with those of all known X-ray bursters (see
Table~\ref{table:xrtdata}). With a match within 0.2 deg, this brought
the list down to 96 events. We investigated the light curves to search
for a clear burst signal and found 70.

28 of these bursts with BAT event data were slewed to and 25 have XRT
data showing the later phases of the burst. Those lacking an XRT
detection are 2005-02-24, 2008-03-18 and 2017-07-28. In two cases, XRT
was already observing the source and the burst is completely covered
with XRT. The 28 bursts are listed in Table \ref{tab:table10} and the
BAT and XRT light curves are presented in Fig. \ref{fig:lc1}. Five of
these 28 bursts have not been reported elsewhere (2005-02-24,
2008-03-18, 2008-03-31, 2011-06-19 and 2012-08-13). 12 bursts have
been discussed in some detail in papers. 11 have merely been reported
in GCN circulars or Astronomical Telegrams. The 28 bursts are from 24
sources. Three bursts are from a single source: SAX J1712.6-3739.

Additionally, we searched all XRT data of the 111 currently known
Galactic X-ray bursters for X-ray bursts and found 65 additional
bursts for a total of 90 bursts detected with XRT (see
Table~\ref{table:xrtdata} in Appendix~\ref{appendix:1}). Therefore, we
count 134 detections of X-ray bursts with event data from BAT or XRT;
49 of these have both BAT and XRT detections (25 as a result of a
slew) while 41 have only XRT detections and 44 have only BAT
detections. A total 69 out of the 134 have triggered BAT, yielding
event mode data, while the remaining 65 bursts were first found in XRT
data for which 21 times a signal was detected in non-event-mode data
from BAT.

It should be noted that Swift detects many more X-ray bursts with BAT,
but most are below the threshold for triggering BAT into a mode that
provides imaged light curves for more than a few seconds to the ground
at sufficient time resolution of $\la$1~s or for triggering an
automatic slew by Swift. These bursts do not allow to distinguish
intermediate-duration bursts. They will be discussed in a forthcoming
paper.

BAT light curves of thermonuclear X-ray bursts have a faster decay
than in the classical 1--10 keV bandpass. Due to the 15 keV low-energy
threshold and the 3 to $<1$ keV cooling of the burst, BAT loses the
signal fast. BAT light curves can be divided in three classes: (1)
fast rise, slow decay, (2) slow rise, fast decay and (3) precursor
followed by slow rise and fast decay. Classes 2 and 3 are PRE
bursts. Class (2) is more common among PRE bursts than in the
classical band, again because of the BAT bandpass. Sometimes PRE
bursts show a rise lasting up to minutes. PRE is characterized by a
relatively short phase of expansion (a few seconds) followed by a slow
phase of contraction until the photosphere shows a minimum in the
emission region size as measured through the blackbody radius and a
maximum in the spectral hardness as measured through the blackbody
temperature \citep[e.g.,][]{lewin1993}. This is known as the touchdown
point.  Due to the BAT spectral response, the PRE phase is observed as
a slow rise in the BAT-measured flux. The touchdown point is followed
by a decay because of a decreasing bolometric flux and spectral
hardness.

The  BAT trigger times are the start times of time intervals over which
the trigger criterion is assessed, ranging from 4 ms to the duration
of a full pointing. They are not necessarily equal to the start time
of the bursts. To estimate the burst start time we qualitatively
determine where the slope of the BAT light curve rise intersects the
pre-burst flat level. Start times are not determined very accurately
and are estimated to have a 5 s uncertainty.

Of interest here are intermediate-duration bursts. The definition of
these bursts naturally depends on the duration. On the one hand, it is
a somewhat difficult parameter to measure; often there is only partial
coverage in X-rays with XRT. On the other hand, there is no generally
accepted threshold for duration above which we speak of
intermediate-duration bursts. We define a duration as the length of
time $t_{5\%}$ that the flux stays above 5\% of the peak flux
\cite[see also][]{zand2017}. To be able to determine this with BAT and
partial XRT data, we assume that the 15--25 keV BAT-measured peak flux
happens at touchdown when the blackbody temperature peaks at about 3
keV \citep[e.g.,][]{galloway2008}. For such a spectrum, a flux of 0.1
c~s$^{-1}$cm$^{-2}$ in BAT (15-25 keV) is equivalent to roughly 400
c~s$^{-1}$ in XRT (full bandpass and $N_{\rm
  H}<6\times10^{21}$~cm$^{-2}$). From the actual BAT peak flux we
calculate what the 5\% level is for the XRT, determine or estimate
through extrapolation, if sensible, at what time the flux decays
through that level, determine when the burst starts to rise above that
level, and measure how much time passes between the two points in
time. For a time profile consisting of a fast rise and an exponential
decay, $t_{5\%}$ is 3 times the exponential decay time (the factor of
3 is equal to -ln(0.05)). We define intermediate-duration bursts as
$t_{5\%}>$150 s, or an exponential decay time that is longer than 50
s. This is commensurate with durations longer than expected for common
X-ray bursts \citep[e.g., Fig. 1 in ][]{zand2017}. We find 12 such
cases in our BAT+XRT sample of 28 (see Table~\ref{tab:table10}). This
shows that BAT, thanks to the high temperatures of
intermediate-duration bursts, is quite efficient in detecting them. In
the 44 BAT-only bursts, there is one additional case of an
intermediate-duration burst because the rise time is very long: 80
s. This is a burst from XTE J1701-407 on 2018-03-09. In the 90 XRT
bursts, there are two more bursts with an e-folding time in excess of
50 s: one burst from GS 1826-24 on October 21, 2009, and one burst
from 1RXS J180408.9-342058 on March 31, 2015. However, we regard these
as examples of prolonged H-burning due to rapid proton captures
\citep[e.g.,][]{zand2017} instead of cooling after a deep ignition as
per intermediate-duration burst.

There are some exceptional cases among the intermediate-duration
bursts.  Five are longer than $t_{5\%}=750$ s: two (out of three)
bursts from SAX J1712.6-3739 ($1050$ and $>5530$ s), one from 4U
1850-087 ($1400$ s), XTE J1810-189 (750 s), and IGR J17062-6143
($1300$ s). These values approach those of bursts from SLX 1735-269
\citep[$\sim1600$ s][]{Molkov2005SLX} and SLX 1737-282
\citep[$\sim1800$~s][]{zand2002}. The longest burst from SAX
J1712.6-3739 is extremely long ($>5530$ s). It looks somewhat like a
superburst, but we argue that it is not (see \S \ref{sec:super}). If
we interpret the peak of the BAT light curve as the touchdown point at
the end of the PRE phase, very long PRE phases are noticeable in some
X-ray bursts: 530 s for 4U 1850-087, 220 s for IGR J17062-6143, and
200 s for XTE J1810-189 and 1RXH J173523.7-35401. For the literature
burst from SLX 1735-260 the time is 450 s \citep{Molkov2005SLX}.

\begin{figure*}
\centering \includegraphics[width=0.33\textwidth]{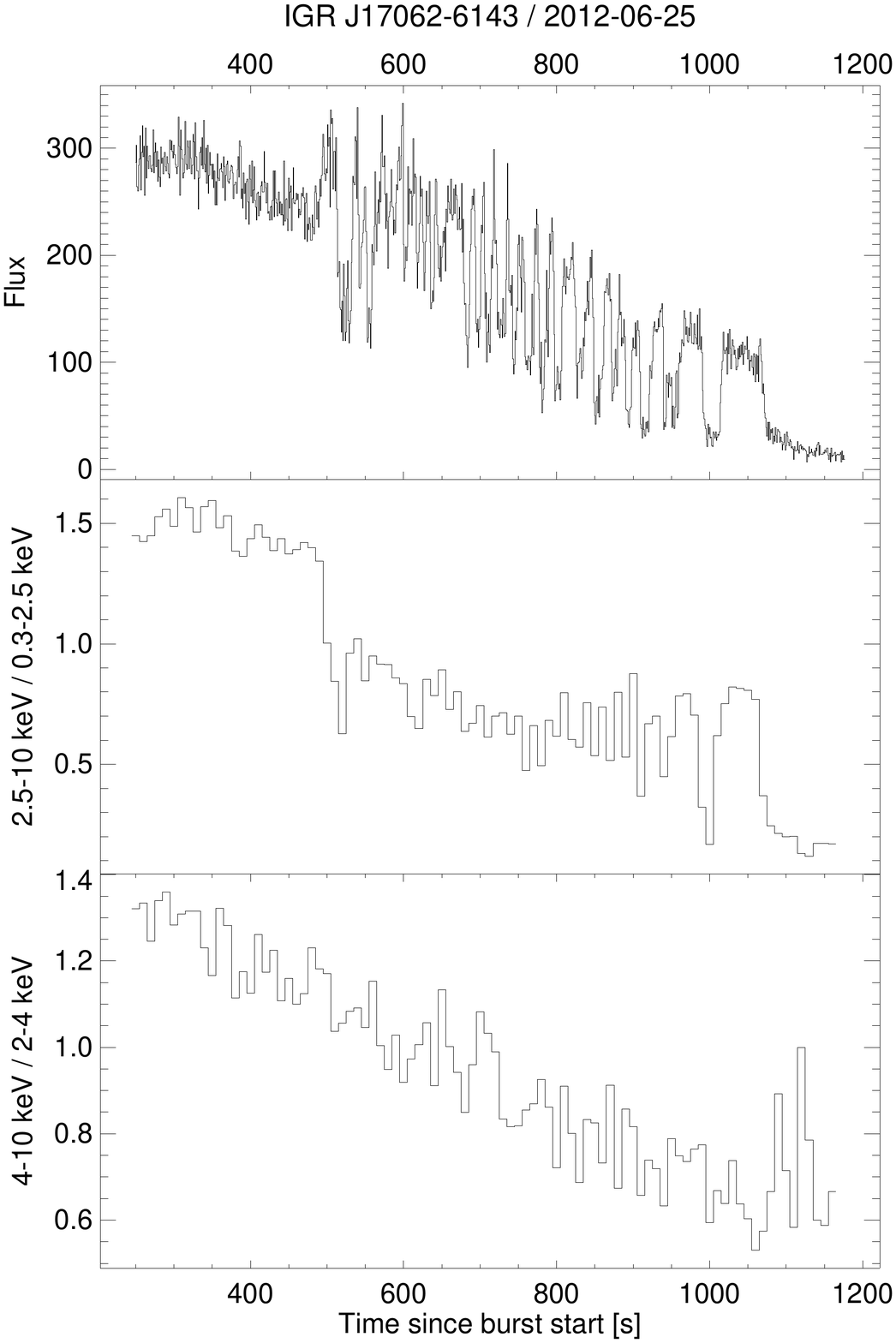}
\centering \includegraphics[width=0.33\textwidth]{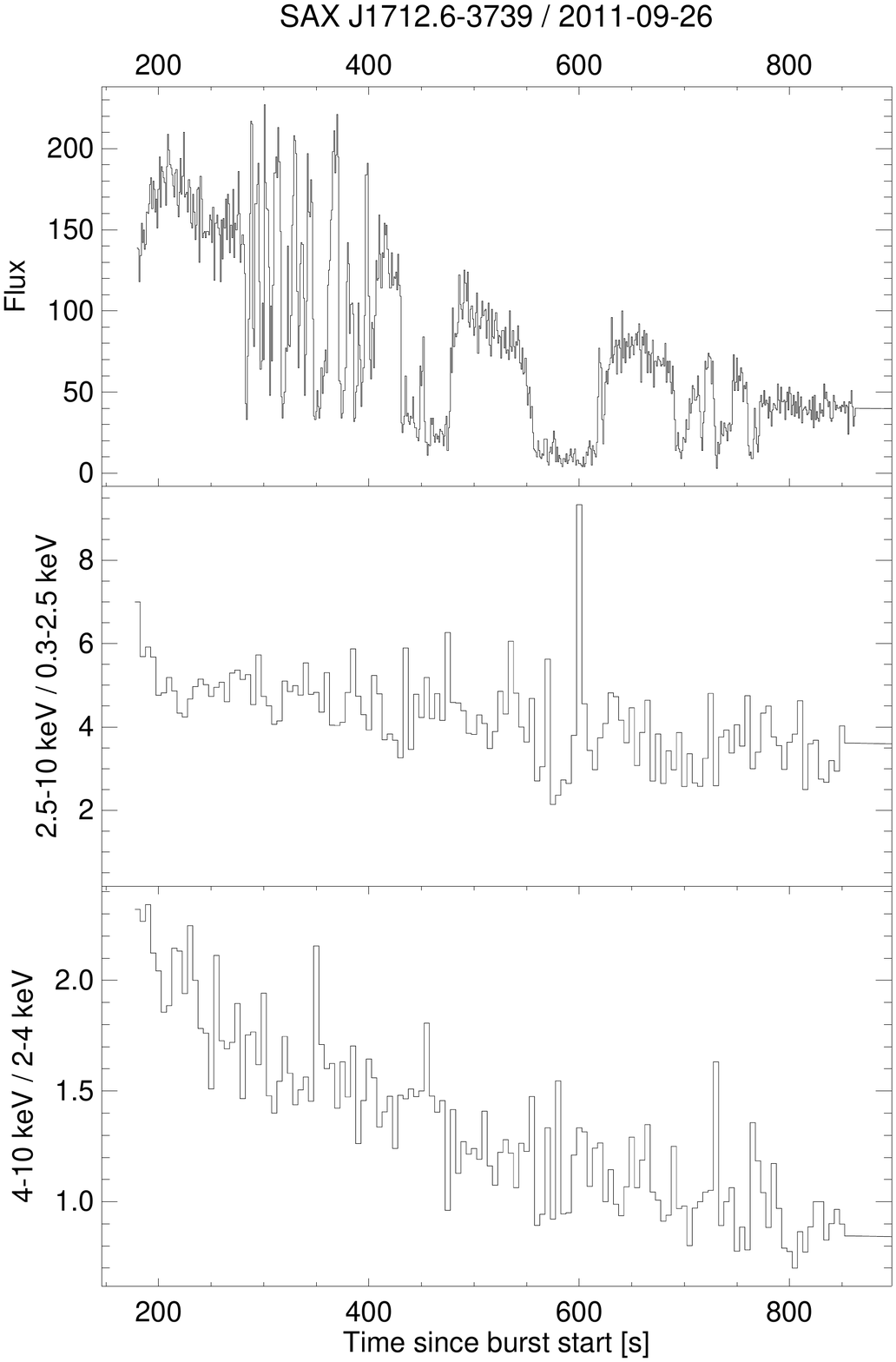}
\centering \includegraphics[width=0.33\textwidth]{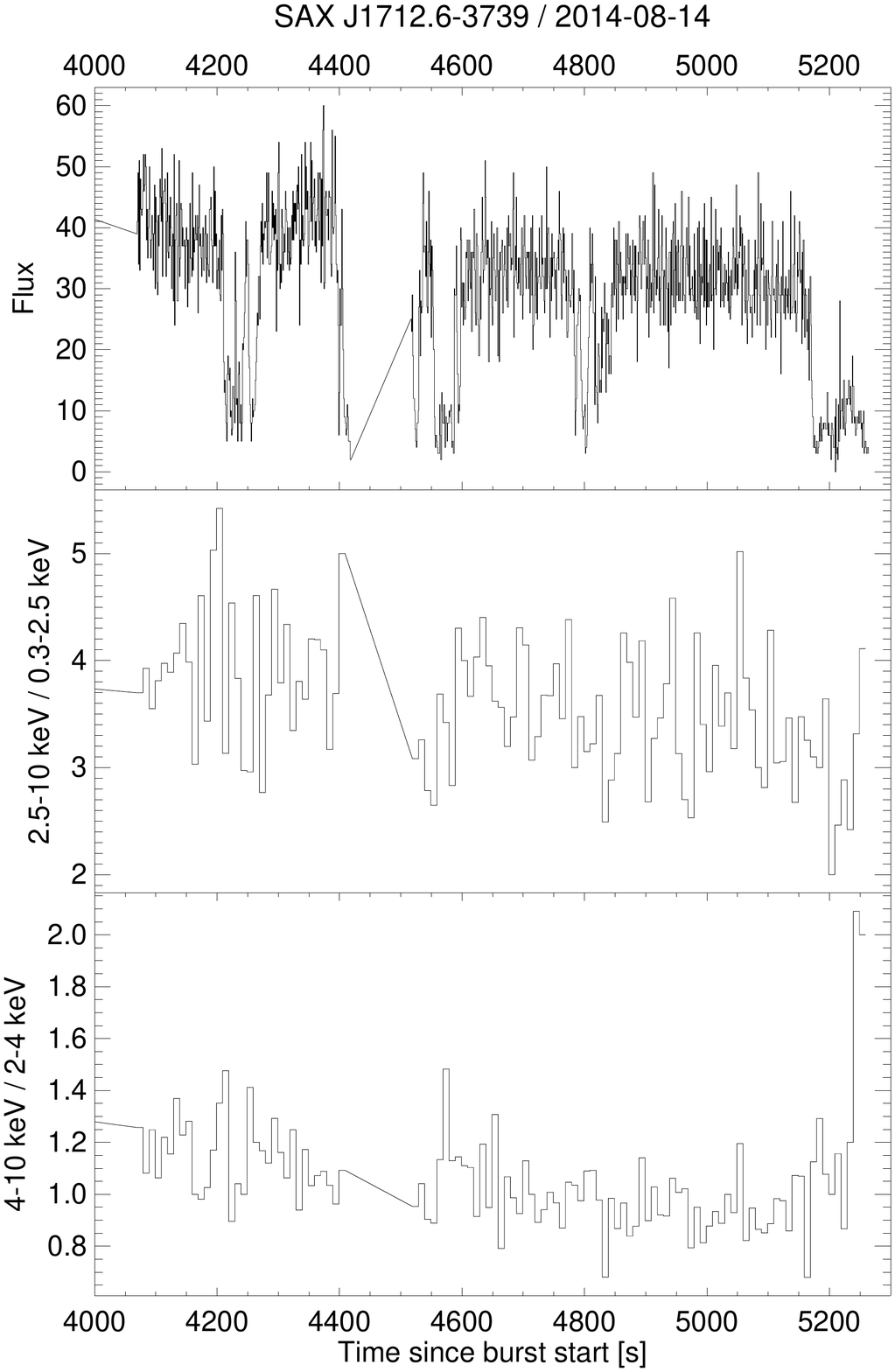}
\centering \includegraphics[width=0.33\textwidth]{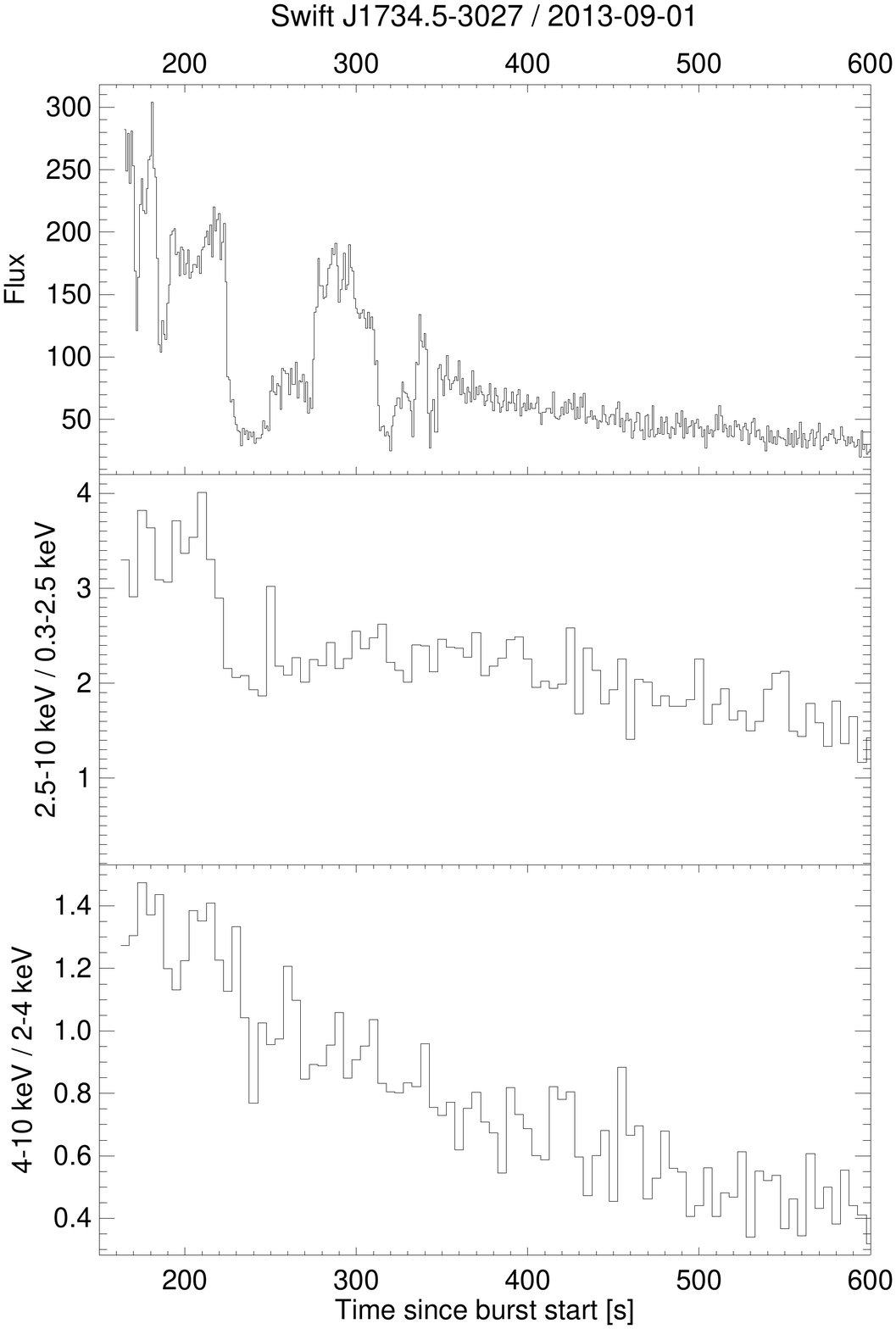}
\centering \includegraphics[width=0.33\textwidth]{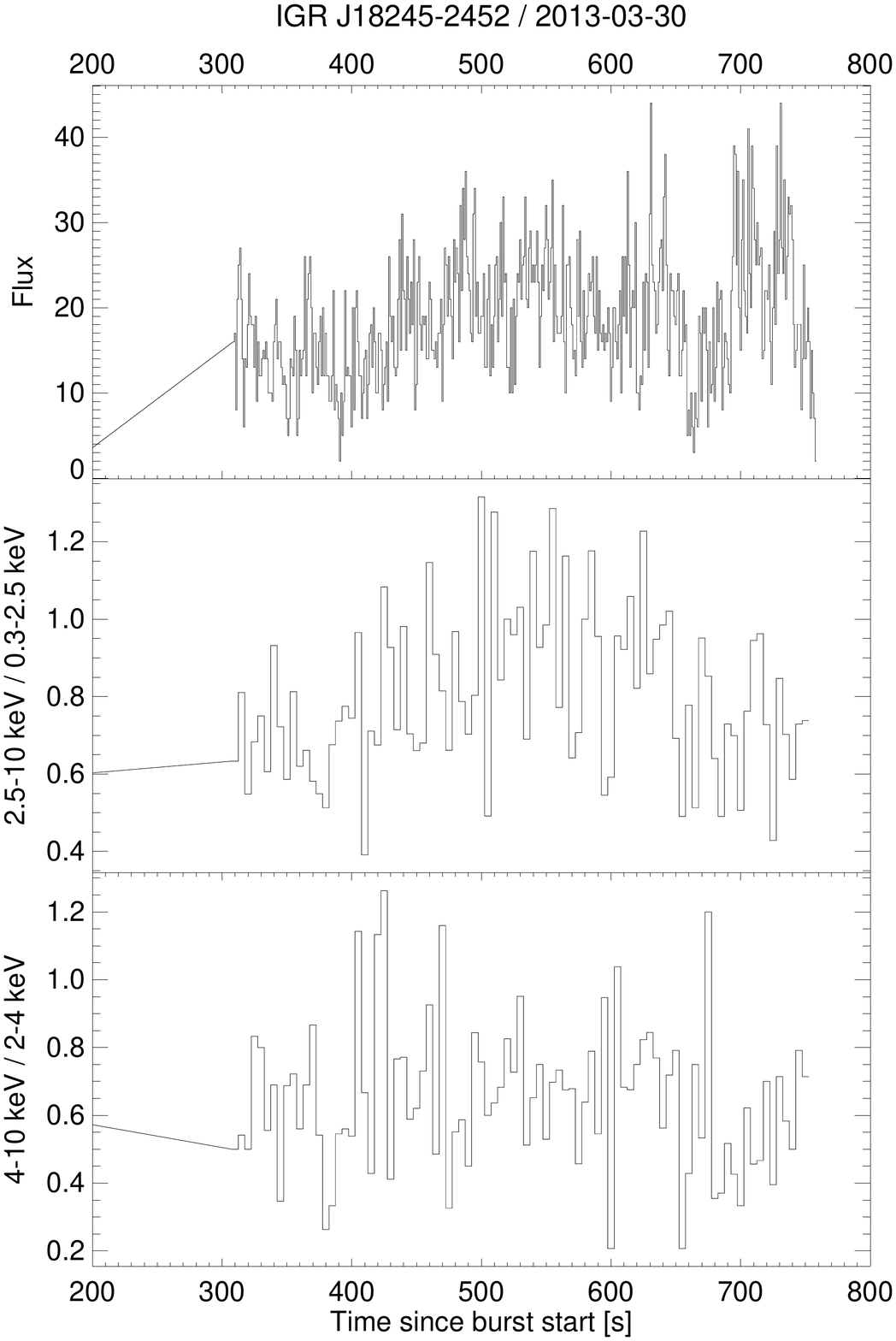}
\centering \includegraphics[width=0.33\textwidth]{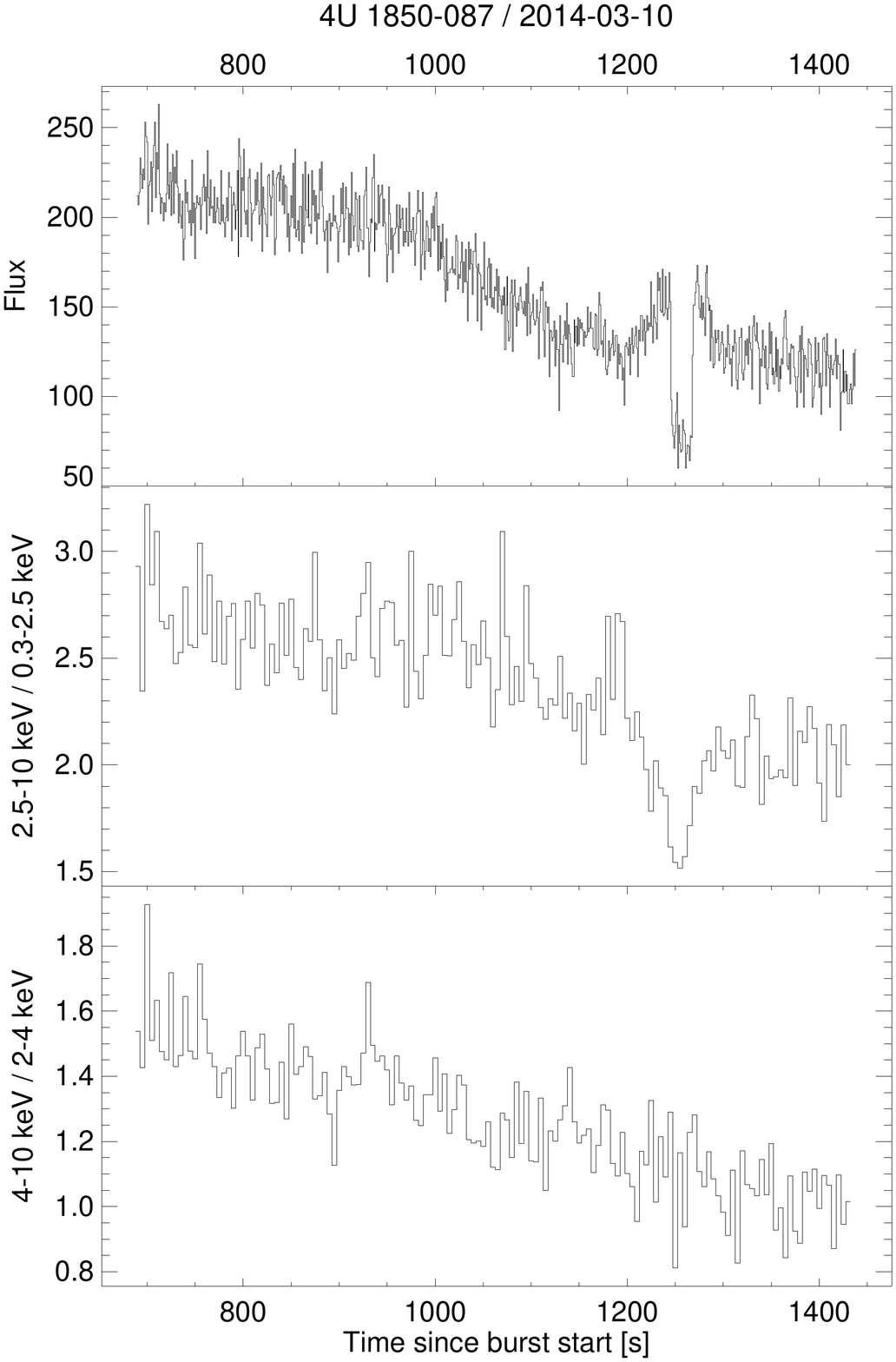}
\caption[]{Time profiles of the XRT data of the six selected bursts:
  of the intensity (top panels), the 2.5--10 keV / 0.25--2.5 keV
  hardness ratio (middle panels) and the 4--10 keV / 2--4 keV hardness
  ratio (bottom panels). Times are since burst onset as determined from BAT
  data. \label{fig:lcs}}
\end{figure*}

There are six bursts, marked in bold in Table~\ref{tab:table10}, that
show strong flux fluctuations in the XRT data during the decay,
reminiscent of such features seen earlier in WFC and PCA data
\citep{Zand20052S,zand2008,zand2011}. In the following we focus on
these bursts, all of which are intermediate-duration bursts.

\section{Data analysis and results}
\label{ch:data}
We present an analysis of the XRT data of the six bursts with
fluctuations (marked in bold in Table~\ref{tab:table10}). The
analysis pertains to the light curve and spectrum.  We start by
making some general notes about how we extracted the data and what the
results of a standard analysis show, and continue with a discussion of
the six bursts individually in order of right ascension. It is noted
that to date only the fluctuations in the XRT data from IGR J17062-6143 have
 been completely analyzed and published
\citep{Degenaar2013IGR}. This is also the burst with the highest
quality XRT data.

\begin{figure*}
\centering \includegraphics[height=0.33\textwidth,angle=270]{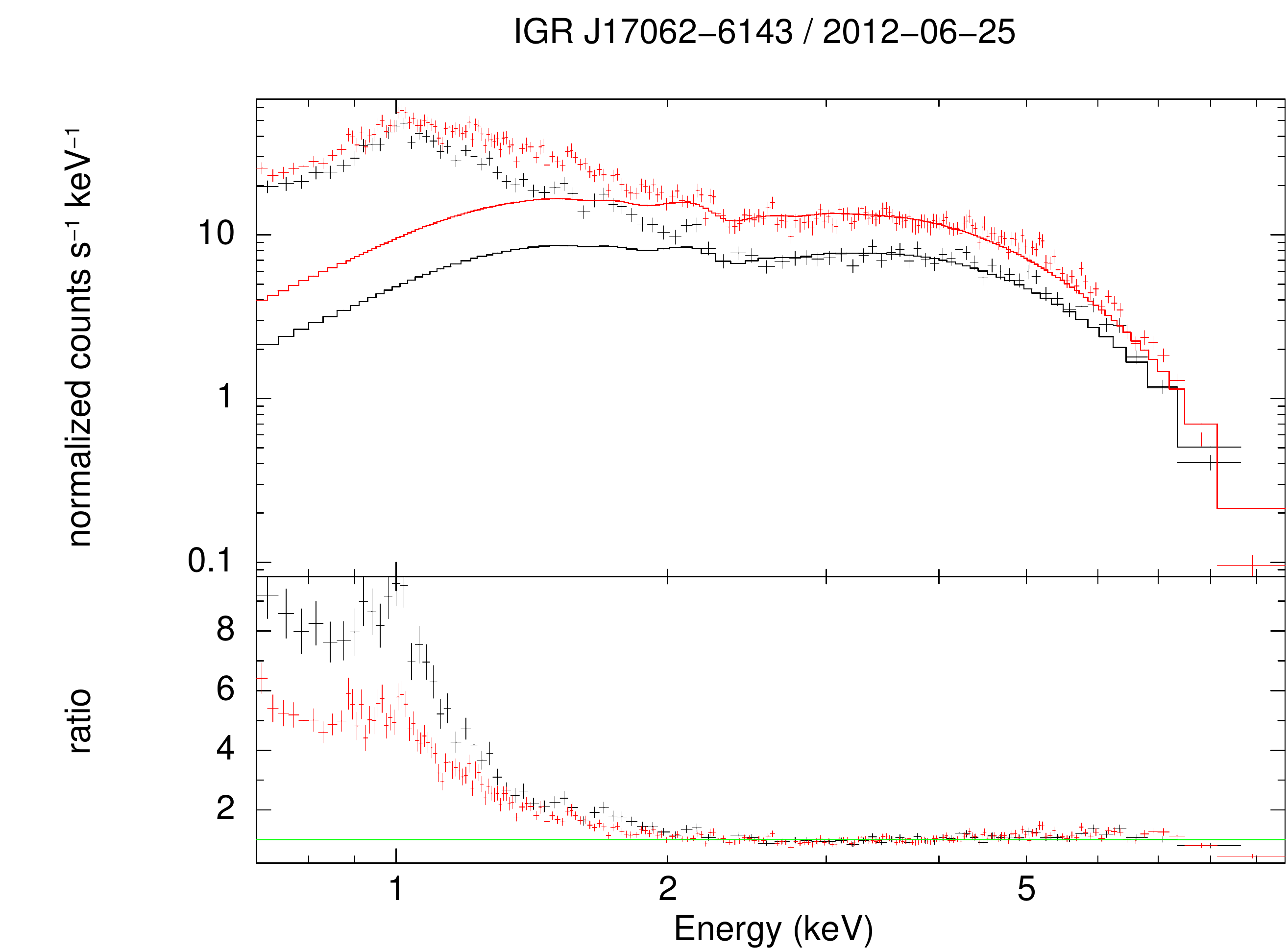}
\centering \includegraphics[height=0.33\textwidth,angle=270]{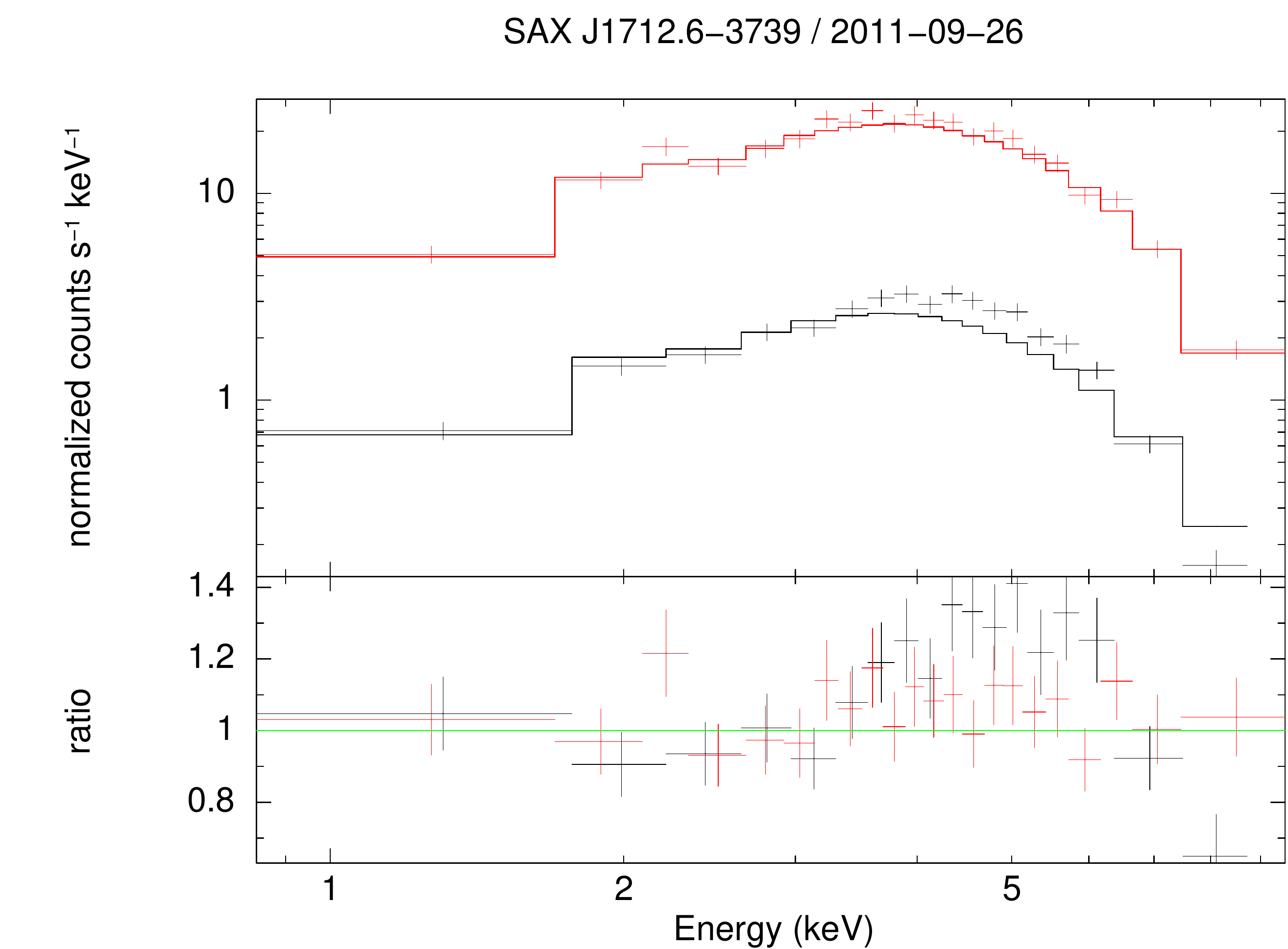}
\centering \includegraphics[height=0.33\textwidth,angle=270]{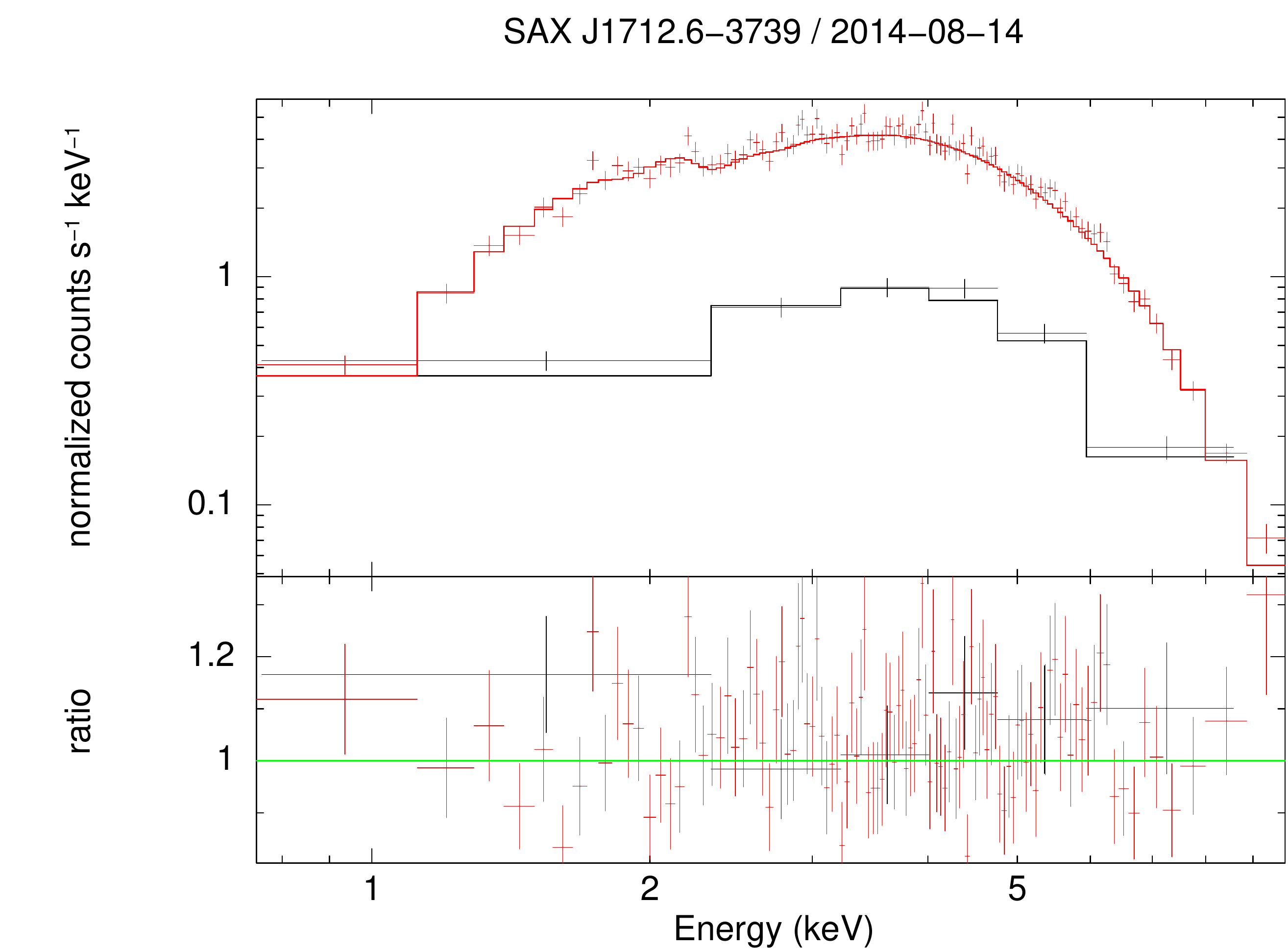}
\centering \includegraphics[height=0.33\textwidth,angle=270]{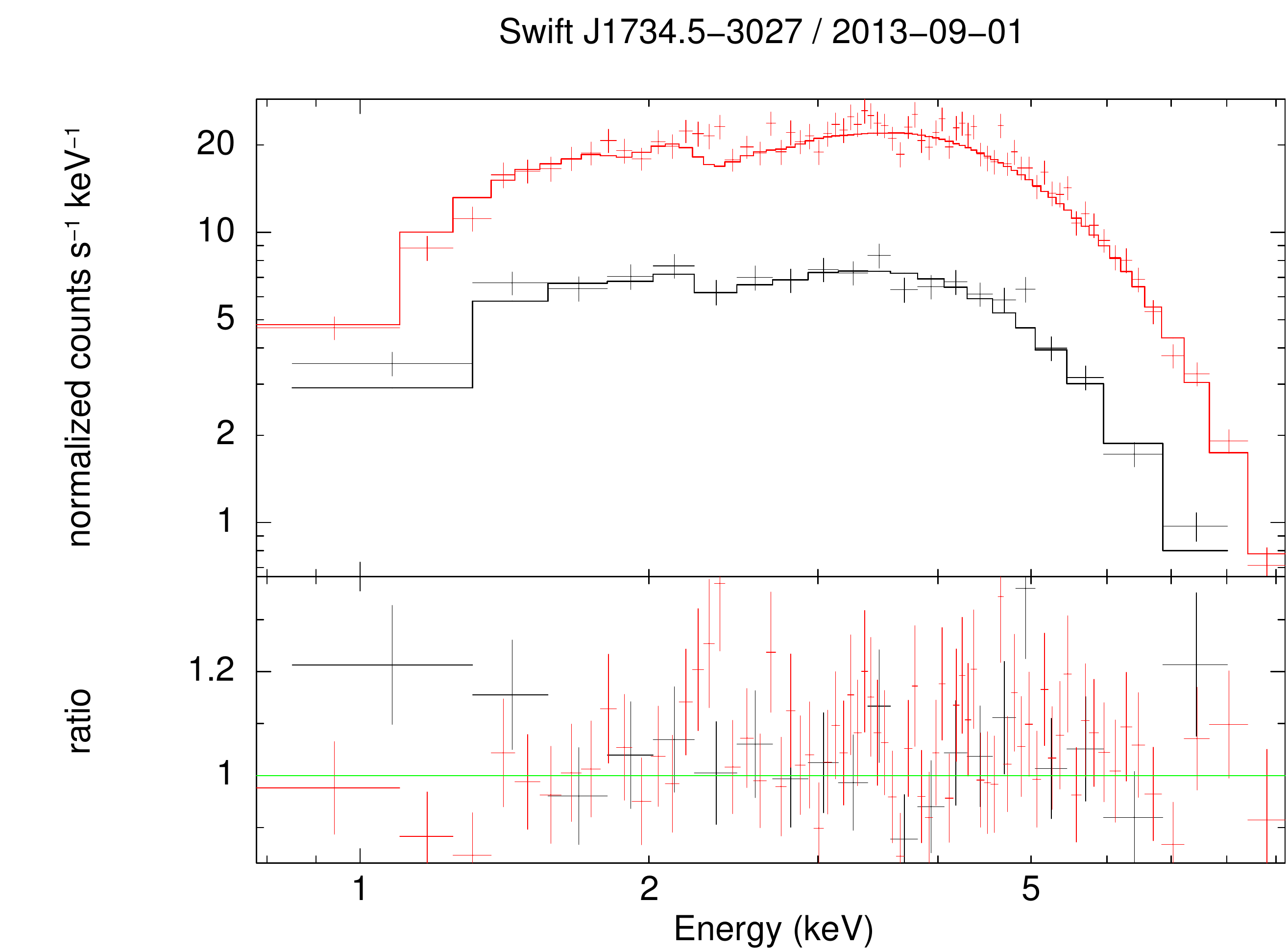}
\centering \includegraphics[height=0.33\textwidth,angle=270]{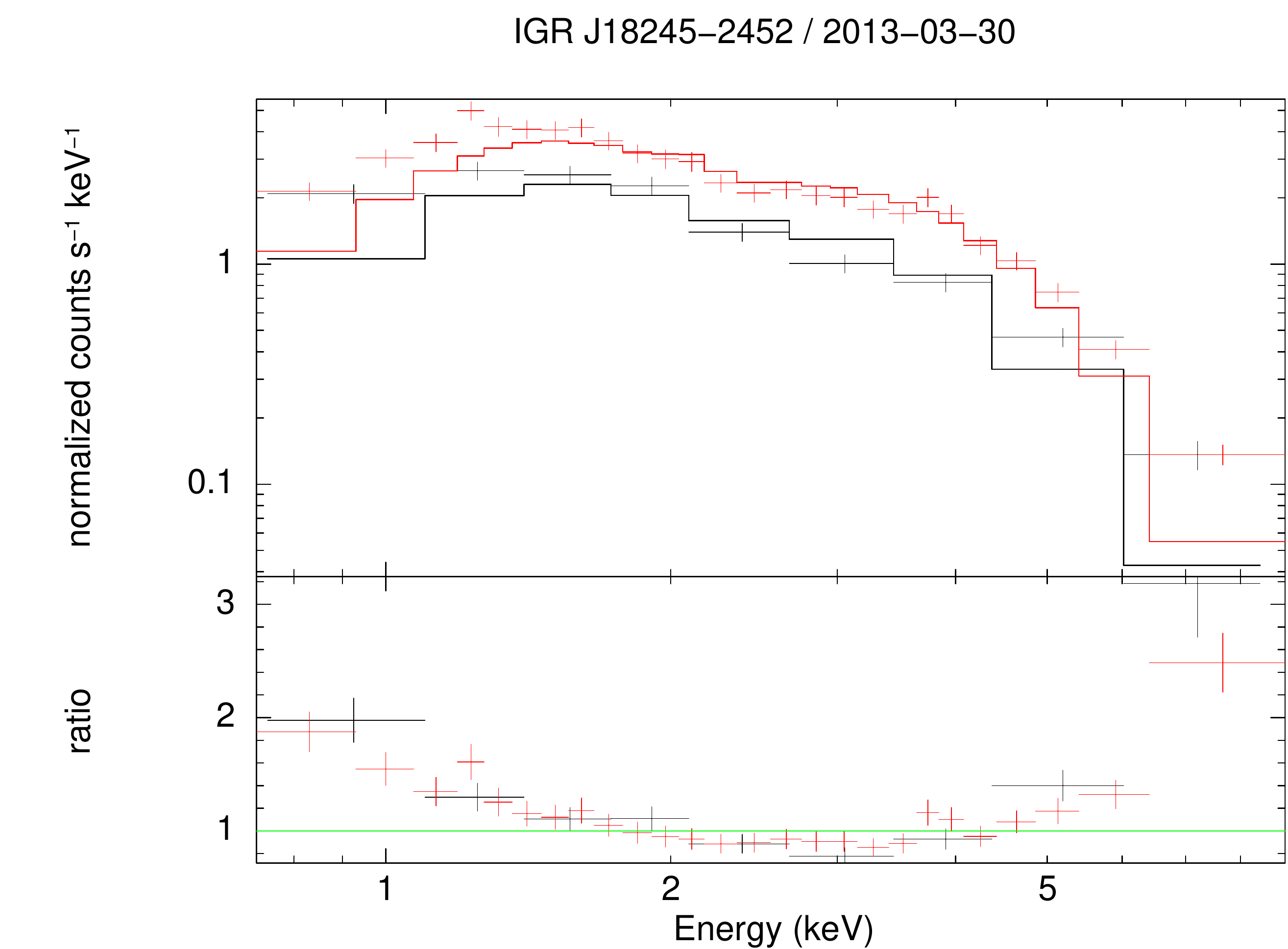}
\centering \includegraphics[height=0.33\textwidth,angle=270]{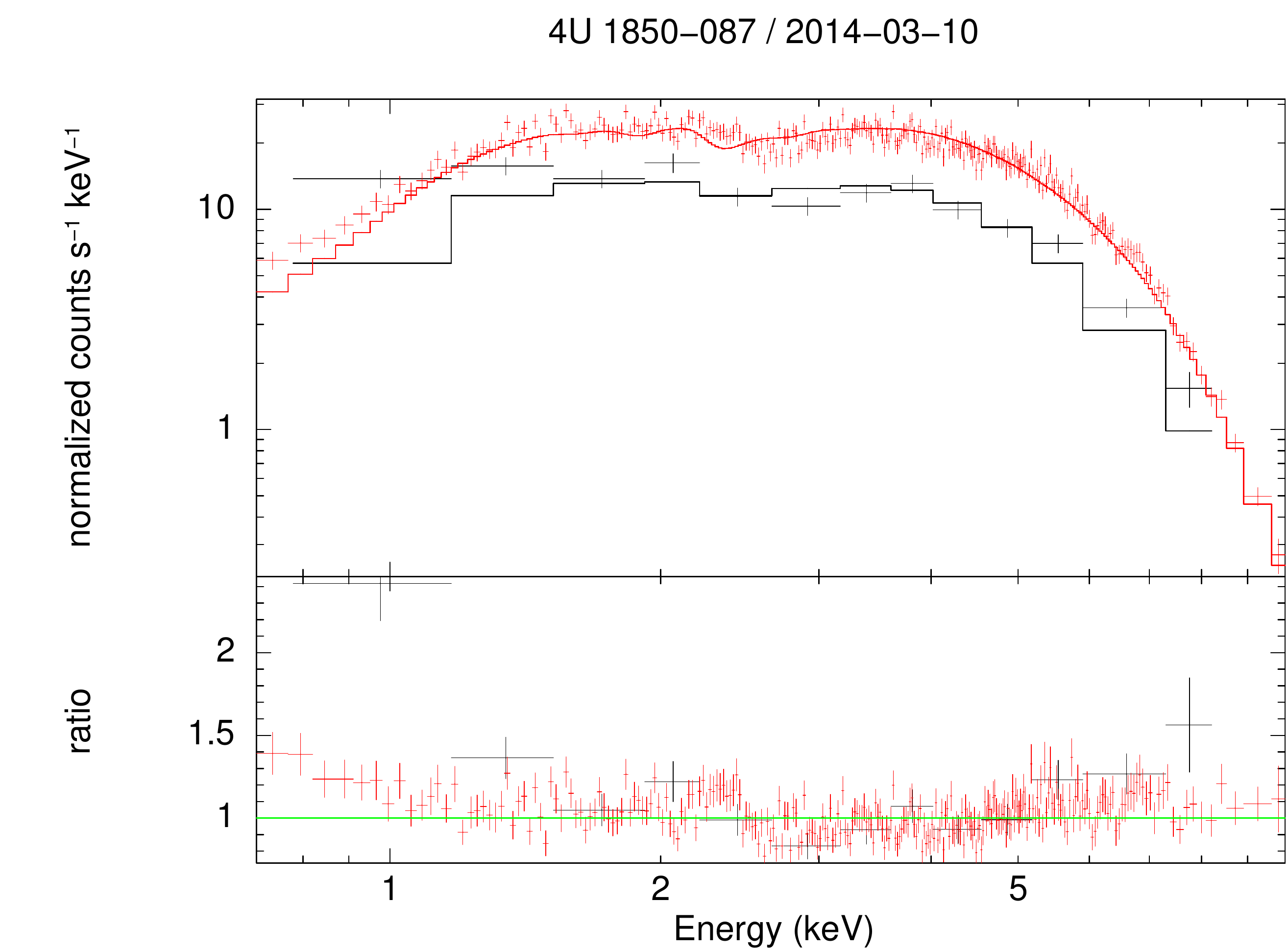}
\caption[]{Spectra of the upward and
  downward fluctuations for the six selected bursts. For IGR J17062-6143, only data above 2 keV
  were fitted. The lower panels show the ratio of the data to the
  applied model. Black data points are for `lo' data, red for the
  `hi'.
 \label{fig:sp}}
\end{figure*}

\begin{table*}
\caption[]{Spectral fits to six down/up spectra presented in
  Fig.~\ref{fig:sp}. For IGR J17062-6143, only data above 2 keV were
  fitted. $N_{\rm H}$ was not fitted (for literature values, see
  text). The last column refers to the chance probability that
  $\chi^2_\nu$ is higher than the observed value. \label{table:sp}}
\centering
\begin{tabular}{llllllll}
\hline
Burst id. & Exposures & $N_{\rm H}$ & k$T_{\rm lo}$  &  k$T_{\rm hi}$ & 0.7-10 keV flux & $\chi^2_\nu$ (dof) & Chance prob.\\
          &  (s)      & (10$^{21}$cm$^{-2}$)  &   (keV)        & (keV)            & ($10^{-9}$erg~s$^{-1}$cm$^{-2}$) lo/hi & & \\
\hline
IGR J17062-6143    / 2012-06-25 & 174.7 / 304.5& 1.2& 1.76$\pm$0.04 & 1.52$\pm$0.02 & 5.3(1)/7.8(1)   & 1.238(695) & $1.7\times10^{-5}$ \\
SAX J1712.6-3739   / 2011-09-26 & 158.7 / 25.0 &15.4& 2.11$\pm$0.08 & 2.26$\pm$0.08 & 3.1(2)/27.0(9)  & 1.237(247) & $6.5\times10^{-3}$ \\
SAX J1712.6-3739   / 2014-08-14 & 161.7 / 753.3&15.4& 1.85$\pm$0.12 & 1.64$\pm$0.02 & 0.90(8)/3.58(5) & 1.051(540) & $2.0\times10^{-1}$ \\
Swift J1734.5-3027 / 2013-09-01 & 68.9 / 79.9  & 5.7& 1.69$\pm$0.05 & 1.96$\pm$0.03 & 5.5(2)/19.3(4)  & 0.919(514) & $9.1\times10^{-1}$ \\
IGR J18245-2452    / 2013-03-30 & 139.2 / 265.6 & 2.2& 0.97$\pm$0.03 & 1.04$\pm$0.02 & 0.41(2)/0.73(2) & 1.713(208) & $6.8\times10^{-10}$ \\
4U 1850-087        / 2014-03-10 & 21.4 / 338.0 & 4.8& 1.64$\pm$0.07 & 1.82$\pm$0.02 & 7.9(5)/16.3(2)  & 1.700(786) & $3.6\times10^{-31}$ \\
\hline
\end{tabular}
\end{table*}

\subsection{General notes}
\label{sec:GenNot}

While the BAT light curves are continuous, the XRT light curves of the
six cases under investigation lack the burst onset as they are
automatic slew targets. Due to the high intensity, all interesting
data were acquired in Windowed Timing mode and we only investigate
these data.  In this mode, the detector is less susceptible to pile-up
at  high count rates. The downside is that imaging information is
lost in one direction and the background is accordingly higher.

XRT source data were extracted from a circular region of radius 90
arcsec around the source, the background data from all detector areas
at least 120 arcsec from the source. The background data were
normalized to the source extraction region area.

For spectral data, we only used grade-0 photons and excluded a central
region of 3.5 arcsec radius around the source position that may be
affected by pile-up. We believe that pile-up effects are minor for the
observed photon detection rates at the employed XRT Window Mode with a
readout time of only 1.7 ms per CCD column, but even minor pile-up
effects may be noticeable in spectral data. We expect pile-up
fractions of no more than 1\% given the maximum count rates of about
300 c~s$^{-1}$ \citep[as derived from HEASARC's {\tt WebPIMMS} tool,
  see also][]{romano2006}. When extracting light curves, we did not
exclude the central region because we wanted to use as many photons as
possible. Analysis of spectra was performed with {\tt XSPEC} version
12.9.1 \citep{arnaud1996}. Spectral data below 0.7 keV were ignored as
suggested by the XRT calibration
digest\footnote{\url{http://www.swift.ac.uk/analysis/xrt/digest_cal.php}}.
Spectra were background subtracted; light curves were not.  We binned
spectra with the {\tt grppha} tool. The minimum number of counts per
energy channel was set to 15 to allow use of chi-squared as a goodness
of fit. Spectra were sometimes further binned in {\tt XSPEC} for plot
aesthetics.

We calculated two hardness ratios from the XRT data: the ratio of the
flux in 2.5 to 10 keV to that in 0.3 to 2.5 keV and the ratio of the
flux in 4.0 to 10 keV to that in 2.0 to 4.0 keV. For this, we
generally use a bin time of 5 s. If the data are relatively noisy, we
experimented with a longer bin time to find a binning that decreases
the noise level but sustains some resolution to allow seeing evolution,
if present.

Time-resolved spectroscopy is done by extracting spectra in a series
of time intervals that are matched to the fluctuations in the light
curves. The spectra are modeled with an absorbed Planck function ({\tt
  wabs*bbodyrad} in {\tt XSPEC} jargon) in which $N_{\rm H}$ is fixed
to the literature value (see Table~\ref{table:sp}). This sometimes
implies small time intervals in which derived spectral parameters may
be ill constrained. The unabsorbed bolometric flux is calculated using
the {\tt cflux} model in {\tt XSPEC} between 0.1 and 30 keV.

\subsection{General analysis of all bursts}

For all six bursts, time profiles of full-bandpass XRT intensity and
both hardness ratios are drawn in Fig.~\ref{fig:lcs}. It should be noted
that 1) the light curves of the first burst from SAX J1712.6-3739 and
the burst from IGR J17062-6143 show fast fluctuations up and down,
while those of the others show only slow dips or partial eclipse-like
features (except IGR J18245-2452); 2) the fluctuations in the light
curves are traceable in the 2.5--10 keV/0.3--2.5 keV hardness ratio, but
not in  4--10 keV / 2--4 keV. In other words, the fluctuations in the
bursts are approximately achromatic above 2.5 keV and not below
\citep[cf.][]{zand2011,Degenaar2013IGR}. We note that only IGR
J17062-6143, Swift J1734.5-3027, IGR J18245-2452, and 4U 1850-87 have
low enough interstellar absorption to have ample signal below 2 keV;
3) the light curve of IGR J18245-2452 shows no decaying trend. Its
association with burst emission is uncertain (see
\S~\ref{sec:igrj1824}).

We carried out a standard spectral data reduction procedure on all six
bursts. Time intervals were determined from 1 s resolution intensity
time histories when the flux is below the anticipated decay trend and
when it is high. The latter condition applies  for times when
there are fast upward fluctuations (in SAX J1712.6-3739 and IGR
J17062-6143) and for times when there are no slow dips. Both `lo' and
`hi' spectra are fitted with a single absorbed blackbody, neglecting
non-burst emission from the same source. This non-burst emission is
not measurable because there is no pre-burst or immediate post-burst
data, but its flux is often 10$^2$ times smaller, as we  argue
below. The results are shown in Fig.~\ref{fig:sp} and
Table~\ref{table:sp}. We note that 1) IGR J17062-6143
exhibits a strong sub-2 keV component \citep[see
  also][]{Degenaar2017IGR,vandeneijnden2018}. This detection in our
sample may be a selection effect because it has the most photons and the
lowest $N_{\rm H}$; 2) the blackbody temperatures of the lo and hi
spectra often differ slightly (on a level of up to 5$\sigma$) and
there is no consistent behavior that the lowest or highest
temperatures coincide with the lo spectra; 3) spectral shapes for lo
and hi periods are similar above 2 keV; and 4) the data for the burst from
IGR J18245-2452 looks out of the ordinary, i.e.,  it is substantially
cooler.

\subsection{Burst from IGR J17062-6143}
\label{sec:IGR}

\begin{figure}[t]
\centering
    \includegraphics[width=0.9\columnwidth, trim=1.cm 0cm 10cm 12cm]{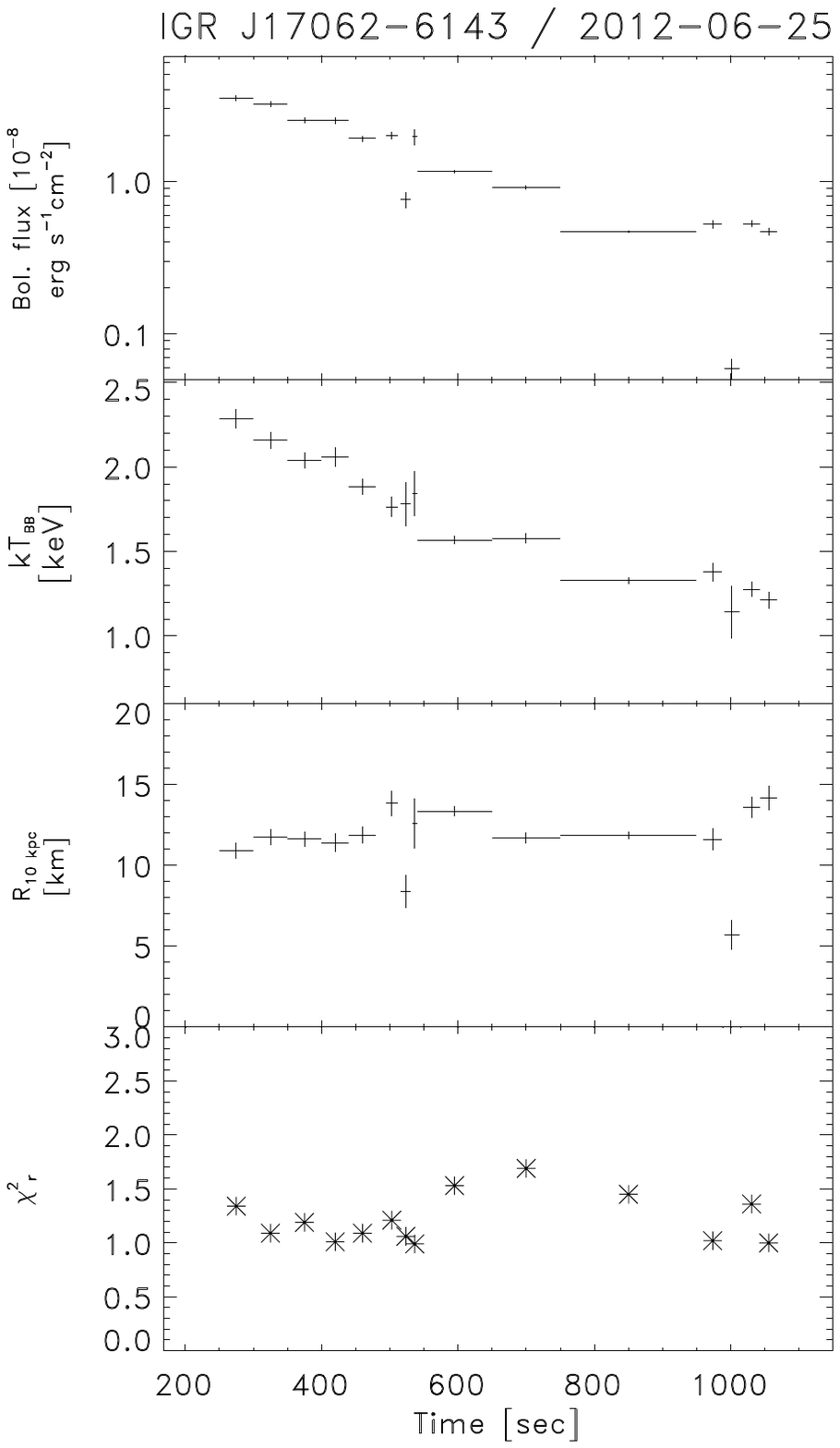}
    \caption{IGR J17062-6143: Time-resolved spectroscopy}
    \label{fig:trsIGR}
\end{figure}

IGR J17062-6143 is an accretion-powered millisecond pulsar with a spin
frequency of 164 Hz \citep{stroh17} that resides in an ultracompact
X-ray binary (UCXB) with an orbital period of 38 min \citep[Strohmayer
  et al. 2018; see also][]{Hernandez2018}. Such a small orbit can only
accommodate a small companion star, most likely a white dwarf that is
denuded from its upper hydrogen layers through mass transfer
\citep[e.g.,][]{Nelson1986,Savonije1986}. IGR J17062-6143 was
discovered in 2006 \citep{Churazov2007} and has been active ever since
with a low luminosity of $L$\textsubscript{X} $\simeq
(1-5)\times10^{35}\ (\frac{D}{5.0~{\rm kpc}})^2$ erg s$^{-1}$
\citep[][]{Ricci2008,Remillard2008,Degenaar2012,keek2017,vandeneijnden2018}. It
was identified as an accreting neutron star low-mass X-ray binary (NS
LMXB) through the X-ray burst discussed here that occurred on June 25,
2012, which was classified as a rare energetic intermediate-duration
X-ray burst by \citet{Degenaar2013IGR,Degenaar2017IGR}. A second X-ray
burst was detected in March 2015 with MAXI and was followed up with
XRT after 10$^4$~s \citep{keek2017}. The absorption column density
amounts to $N_{\rm H}=1.2\times10^{21}$~cm$^{-2}$
\citep{keek2017,vandeneijnden2018} which is the lowest in our sample.

This burst is a nice showcase for the kind of burst we present here
because it is bright, long, has small interstellar absorption, and is
well covered by BAT and XRT.  In the BAT light curve
(Fig.~\ref{fig:lc1}) there are a lot of fluctuations at the start of
the burst. These fluctuations are not significant, but
statistical in nature because the source is off-axis in the
beginning. They become smaller as Swift slews towards the source. The
 touchdown\ point is around 220 s after the start of the
burst. The XRT data, starting at 250 s after the burst start, show a
 plethora of fluctuation behavior starting at around 500 s with
initially fast fluctuations going up and down and later also longer
dips. The dip at the end is likely a long dip, which is below the
decay trend of the burst.

\cite{keek2017} have carefully assessed the bolometric flux due to
accretion and find an average of
$(0.92\pm0.07)\times10^{-10}$~\ecs\ between 2012 and 2015. This is
only 0.2\% of the peak luminosity of the burst.  We did not subtract
the persistent emission in the spectra.

For the fluctuation spectrum we used the same model as
\citet{Degenaar2013IGR} did, except we did not fit the absorption
lines and iron edges. In the fitted spectra we can clearly see that the
model is incorrect for the higher energies (Fig.~\ref{fig:sp}). Adding absorption lines and iron edges, as
\citet{Degenaar2013IGR} did, would improve this fit at higher
energies. An interesting feature of the fluctuation  spectrum is
that an emission line around 1 keV is observed. There seems to be a
much smaller drop in count rate between the up and down fluctuations
around 1 keV. The emission line energy is in the Fe-L
complex. \citet{Degenaar2013IGR} suggests that this emission line is
caused by the irradiation of relatively cold gas. By dividing the full
width at half maximum (FWHM) of the emission line by the line\qr s
energy, they derive a velocity of $\sim$0.16 c, where c is the speed
of light. This velocity implies a radial distance of $\sim8\times10^2$
km from the NS for a Keplerian orbit. \citet{Degenaar2013IGR} also
found absorption lines and edges at higher energies ($>7.5$ keV) which
we did not model. They conclude that the spectral features and the
fluctuations (on a timescale of 1--10 s) imply similar radial
distances from the NS, and are therefore likely caused by the same
material and mechanism.

Figure~\ref{fig:trsIGR} shows the time evolution of the parameters of
the blackbody spectral model. The strongest changes in bolometric
flux are attributed mostly to strong changes in the normalization of
the blackbody model and not in the temperature.

\subsection{Bursts from SAX J1712.6-3739}

\begin{figure}[t]
\centering
    \includegraphics[width=\columnwidth]{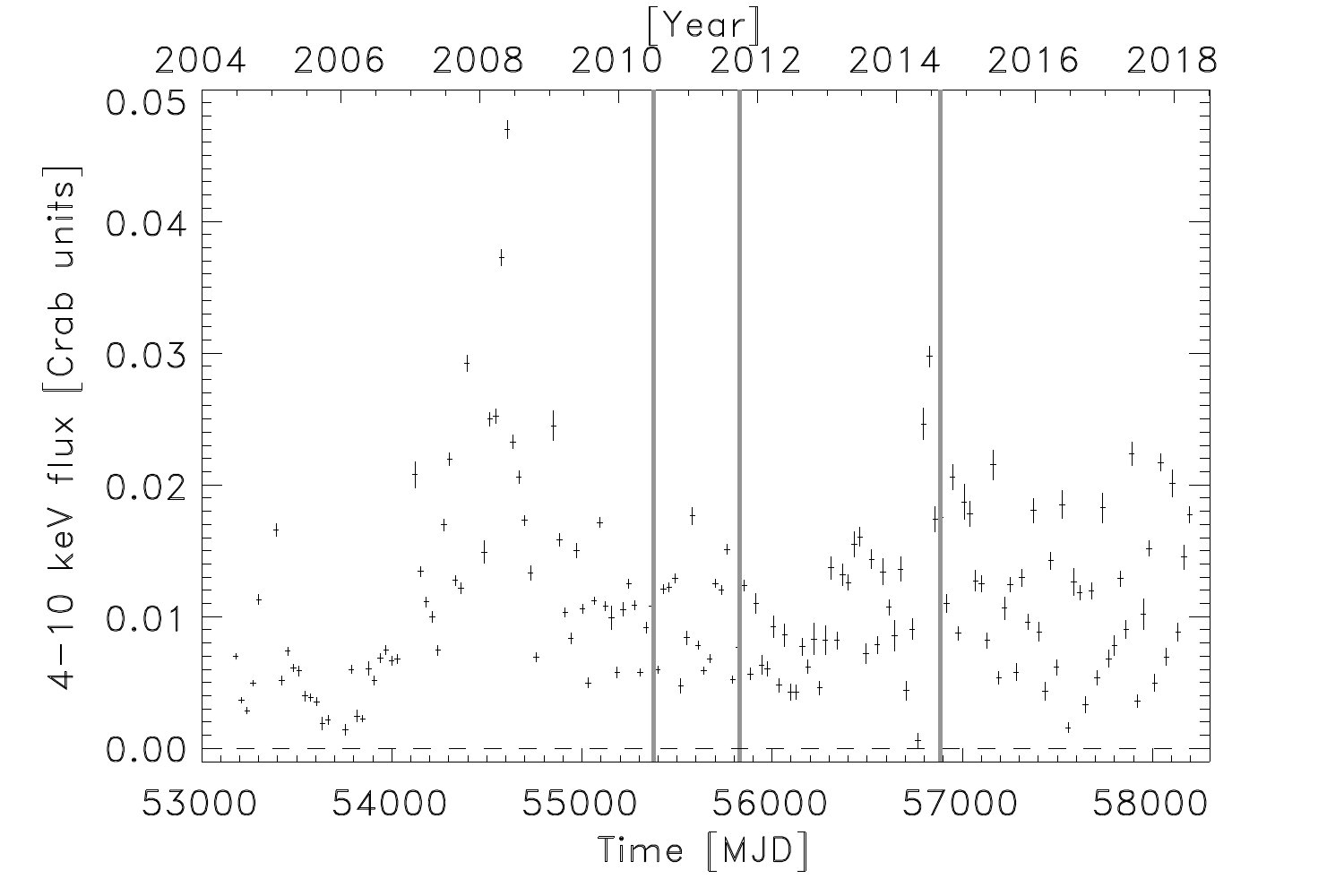}
    \includegraphics[width=\columnwidth]{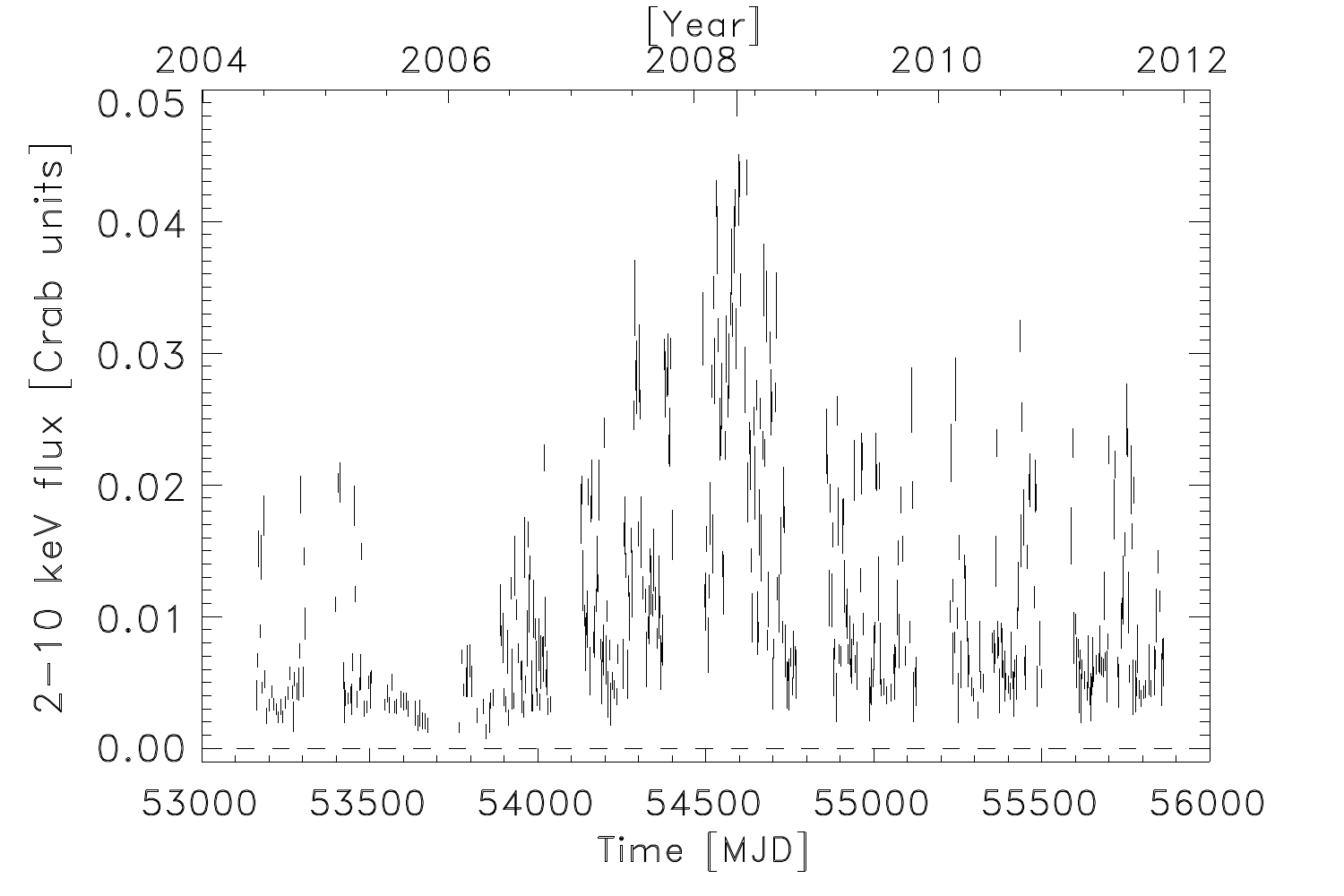}
    \caption{(Top) 4--10 keV light curve over 14 years for SAX
      J1712.6-3739 with a time resolution of 1 month. The data are
      from the PCA Bulge Scan program \citep{swank2001} between 2004
      and 2012, and MAXI observations between 2009 and 2018
      \citep[e.g.,][]{negoro2016}. A cross calibration has been
      performed for the two years that the programs overlap. Strong
      variability is obvious. Vertical bars indicate the times of the
      three bursts detected with Swift. (Bottom) 2--10 keV light curve
      measurements with PCA over 8 years for SAX
      J1712.6-3739. Typically, each data point represents 1 min of
      data and measurements were performed twice a week. Strong
      variability is also obvious at this resolution.}
    \label{fig:1712monitor}
\end{figure}

SAX J1712.6-3739 was discovered in 1999 \citep{Zand1999}, and since
then has always been detected whenever observed.  Immediately after
discovery, the source was seen to burst twice with a half-minute
duration \citep[e.g.,][]{Cocchi2001}. Figure~\ref{fig:1712monitor}
shows the long-term light curve between 2004 and 2018 with a one-month
resolution and a semi-week resolution. The data are from PCA Bulge
Scan \citep{swank2001} and MAXI observations \citep{negoro2016}. The
source shows wide variability between 0 and 48 mCrab on both
timescales with an average flux of $9.10\pm0.04$ mCrab, or about
$2\times10^{-10}$~\ecs.

When investigating the MAXI data at the highest 1.5 hr time
resolution, ten burst-like features show up in almost nine years of data,
when the 4--10 keV flux rises above 0.2 Crab units. Two of these show
high fluxes for multiple consecutive data points suggesting
intermediate-duration bursts. This is consistent with an earlier
finding of an intermediate-duration burst in RXTE-ASM data from 1999
\citep{kuulkers2009}.

In BAT Hard X-ray Transient Monitor daily-average data \citep{krimm},
the source averages 7 mCrab (15--50 keV) with an incidental excursion
of up to 36 mCrab, similar to values found with INTEGRAL-IBIS
\citep[4.7 mCrab in 20--40 keV during 2003--2006;][]{fiocchi2008} and
BeppoSAX \citep[6--32 mCrab in 1--10 keV in 1999;][]{Cocchi2001}. It
is clear from these measurements that the source is persistent but
wildly variable on timescales of 1 day and longer. However, it never
becomes brighter than about 50 mCrab. It is noted that the BAT Monitor
data recently showed a burst on May 8, 2018 \citep{lin2018}.

SAX J1712.6-3739 is classified as a NS LMXB \citep{Cocchi2001} which
implies an average luminosity of $2\times10^{36}$~erg~s$^{-1}$ or
roughly 1 \%\ of Eddington. \citet{zand2007} list this source as a
candidate UCXB. \citet{Wiersema2009SAX} find that the optical
magnitudes of the likely counterpart can accommodate a UCXB
nature. \citet{revnivtsev2013} come to the same conclusion on more
sensitive infrared measurements. Time-resolved optical spectroscopy of
this system would be able to reveal the binary period and therefore
the UCXB nature. According to \citet{Yoon2011}, SAX J1712.6-3739 is to
date the only known X-ray binary to display a prominent
H\textsubscript{$\alpha$} bow-shock nebula and \cite{Wiersema2009SAX}
speculate that it is due to a high velocity towards the Galactic plane
in a relatively dense medium and is powered by a jet. There are no
published papers on bursts observed from this source with Swift.

We found three burst detections with BAT and XRT from SAX
J1712.6-3739.  The first, on July 1, 2010, is the shortest of the
three and is unremarkable. The BAT rise time is 20 s. The second and
third, on 2011-09-26 and 2014-08-18, are very remarkable because they
are long and show strong fluctuations. This is the second source to
have shown fluctuations in multiple X-ray bursts, after 2S 0918-549
\citep{Zand20052S,zand2011}.

\begin{figure*}[t]
\centering
    \includegraphics[width=0.9\columnwidth, trim=1.cm 0cm 10cm 12cm]{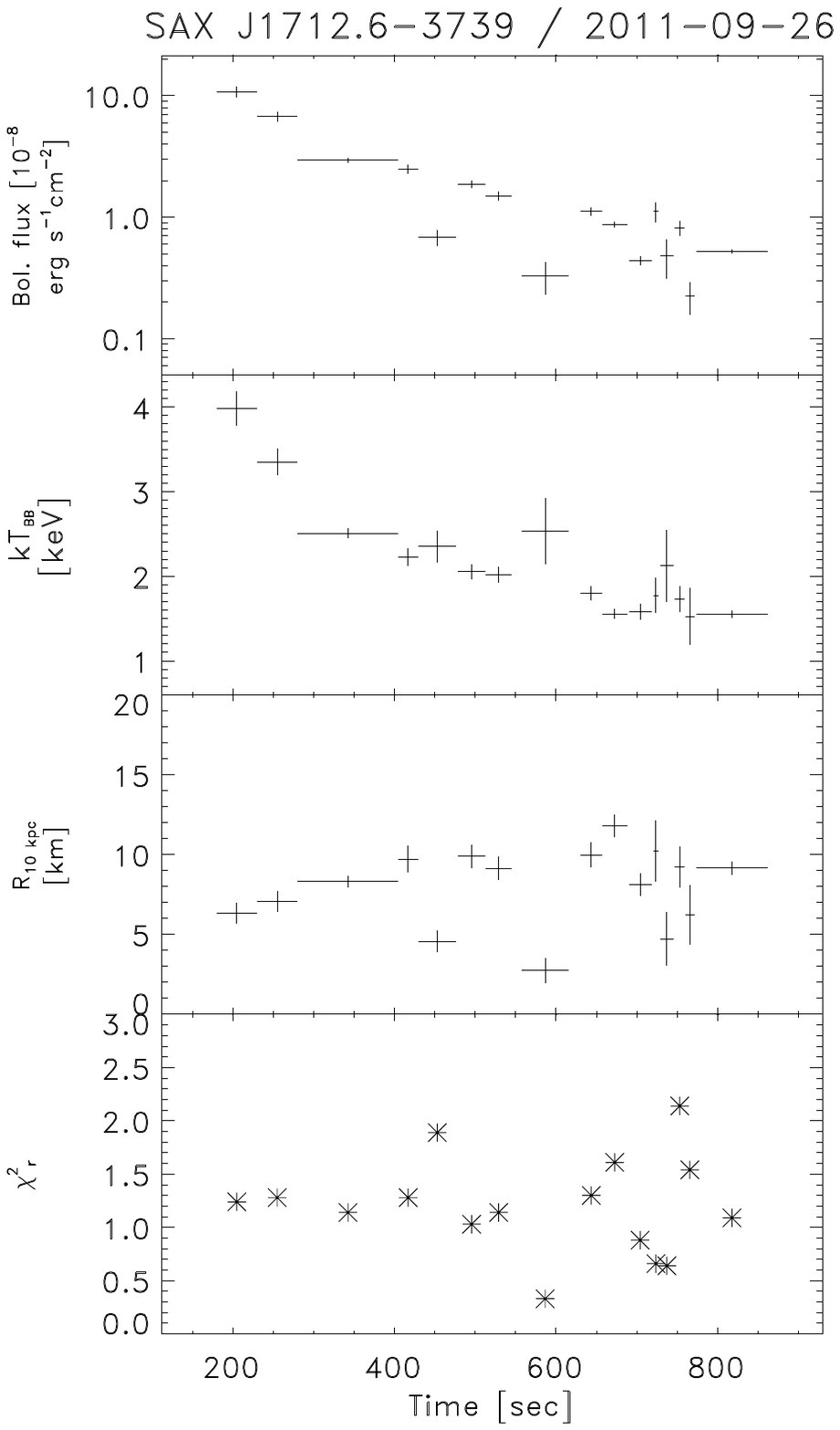}
    \includegraphics[width=0.9\columnwidth, trim=1.cm 0cm 10cm 12cm]{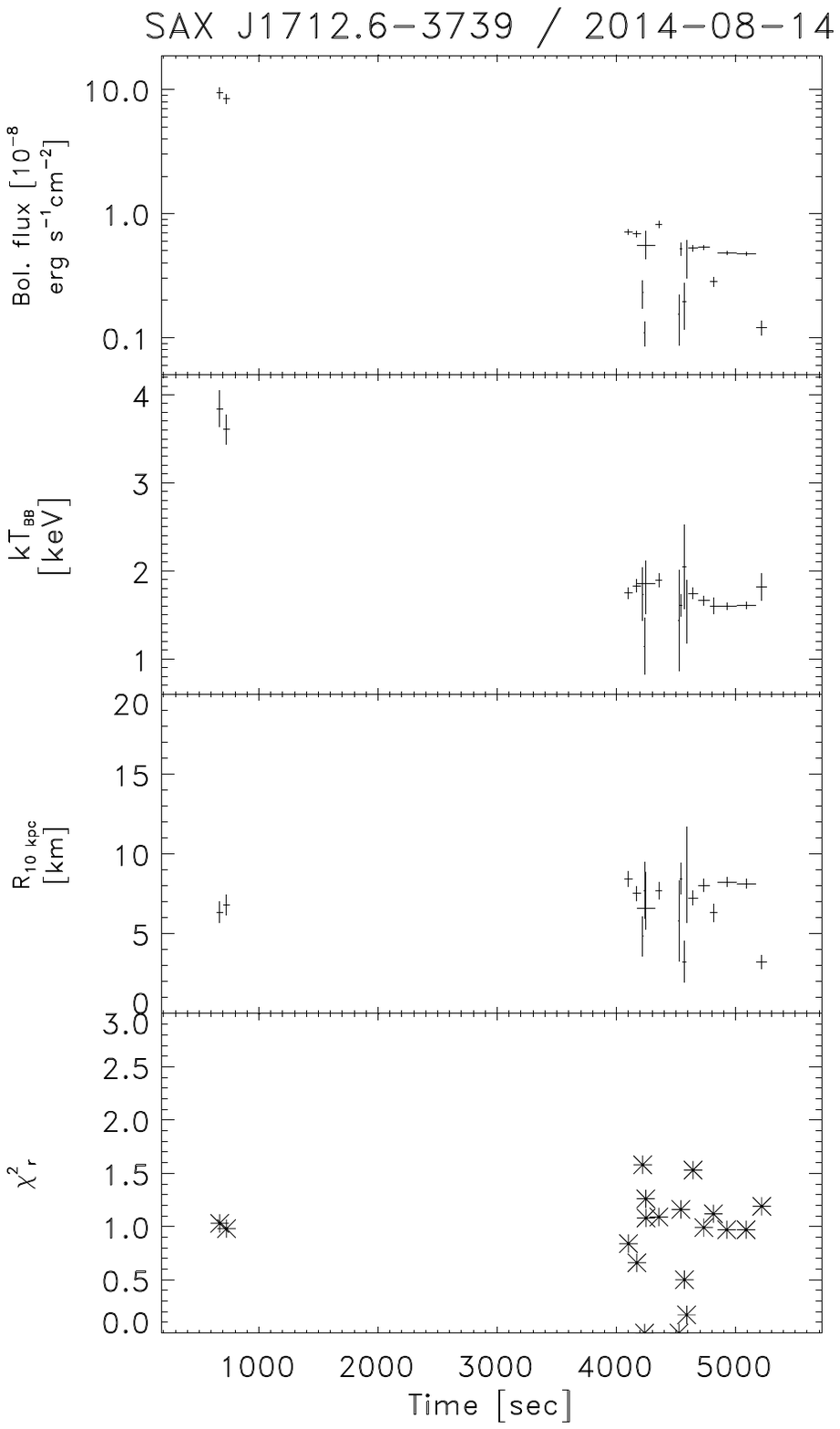}
    \caption{SAX J1712.6-3739: Time-resolved spectroscopy of both long bursts.}
    \label{fig:trsSAX}
\end{figure*}

\subsubsection{Burst on September 26, 2011}

In the BAT data (see Fig.~\ref{fig:lc1}), the end of the
Eddington-limited phase is around 190 s after the start of the
burst. The XRT light curve shows fast fluctuations between $\sim280$ s
and $\sim410$ s after the burst onset with both fluctuations up as
down. After the fast fluctuations have ceased, longer dips are
seen. The behavior of this burst is reminiscent of that of the burst
from IGR J17062-6143. The early dip at about 190 s after the burst
start could be the touchdown point.

For the  time-resolved spectroscopy (see Fig.~\ref{fig:trsSAX}) we
fitted an absorbed blackbody. The spectra are not subtracted with the
persistent flux of the source. The persistent flux measured with RXTE
PCA\footnote{\url{https://asd.gsfc.nasa.gov/Craig.Markwardt/galscan/html/SAXJ1712.6-3739.html}}
on MJD 55828.100 was 32.17 counts s$^{-1}$ 5PCU$^{-1}$ translating to
3.217 mCrab ($7.7208\times10^{-11}$ erg cm$^{-2}$ s$^{-1}$), which is
insignificant compared to the burst emission.  The maximum burst
emission of the XRT data is $4.1\times10^{-8}$ erg cm$^{-2}$
s$^{-1}$. After the burst on MJD 55835.219 the persistent flux was
66.95 counts s$^{-1}$ 5PCU$^{-1}$.

For the spectrum of the fluctuations (see Fig.~\ref{fig:sp}) we
selected the fluctuations between $\sim$340 and $\sim$480 s. There is
no noticeable difference between the shape of the lo and hi
spectrum. The $N_{\rm H}$ value for this source is much higher than
for the other sources. Therefore, less can be said about the lower
energies of this spectrum. Figure~\ref{fig:trsSAX} shows the
time-resolved spectroscopy in the left panel. The fluctuations are
represented by changes in the blackbody normalization.

\subsubsection{Burst on August 18, 2014}

\begin{figure}[t]
\centering
    \includegraphics[height=1.\columnwidth,angle=270]{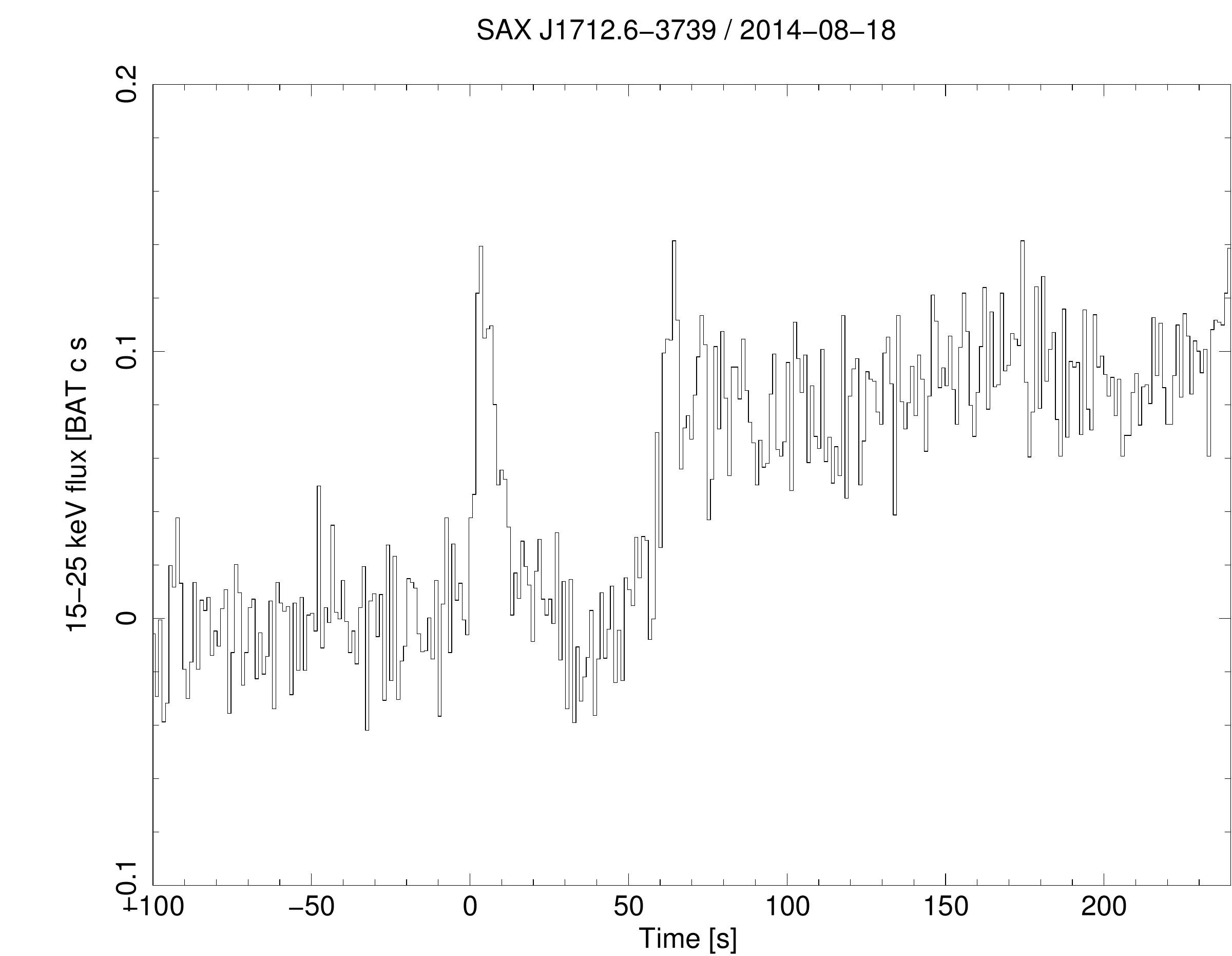}
    \caption{BAT light curve of the third burst from SAX~J1712.6-3739,
      reconstructed from rate meter data in 15--25 keV (pre-burst flux
      subtracted between -40 and 0 s from burst start and normalized
      to projected detector area as seen from source). During this
      data stretch BAT had a fixed pointing.}
    \label{fig:s1712rare}
\end{figure}

This burst is very unusual. The mask-tagged BAT data is,
unfortunately, incomplete because it is actually due to two triggers,
GRB140818A and a late image trigger from SAX J1712.6-3739.  The data
show a short-duration ($\sim15$ s) X-ray burst, then a period of 40 s
when no emission is detected, then a data gap of 450 s, followed by
data for about 320 s until 820 s after the burst start when the flux
decreases from about 100 to 70\% of the peak value of the initial
burst. A possible touchdown point can be identified at
$540\pm40$~s. To overcome the gap in the BAT data, we investigated the
rate data. The pointing of Swift was constant during that data
gap. The result is given in Fig.~\ref{fig:s1712rare}. It shows that
the BAT signal returns 50 s after the precursor ended.

The XRT data start 270 s after the burst start until 760 s and then
again from 4000 to 5500 s. We verified that there is no MAXI data
during the gap. At least four eclipse-like features with 70\% less
flux are visible in the second data stretch of duration 20--45 s except
the last one which lasts at least 90 s and is cut short by the
observation end.

This is the longest burst in our sample. The e-folding time of the XRT
data, simply estimated from the two observation periods, is 0.8
hr. This is similar to the shortest superbursts
\citep[cf.][]{zand2017}, but those superbursts were from a high
accretion-rate source. This brings us to an important question: is
this a superburst or not? We  discuss this in \S~\ref{sec:super}.

One could ask whether the late phase is really burst emission. We
tried fitting an absorbed power law and disk blackbody for the whole
period 4000--5300 s and found good fits that were, however,
inconsistent with earlier findings about $N_{\rm H}$.  Only a single
blackbody spectrum gave a consistent $N_{\rm H}$ value of
$(15.2\pm0.5)\times10^{21}$~cm$^{-2}$, compared to the literature
value of $15.4\times10^{21}$~cm$^{-2}$. We conclude that the emission
is consistent with that from the burst. The average temperature for
the whole period is $2.00\pm0.03$ keV.

For comparison, we also fitted the earlier XRT spectrum and find
$N_{\rm H}=(18.7\pm0.6)\times10^{21}$~cm$^{-2}$ and k$T$=$3.84\pm0.09$
keV $\chi^2_\nu=1.12$ ($\nu=746$). The temperature is higher than
expected even for an Eddington-limited flux \cite[cf.,][]{lewin1993}.
This is also visible in the time-resolved spectroscopy
(Fig.~\ref{fig:trsSAX}), also for the other burst from SAX
J1712.6-3739.  We are unaware of such high-temperature measurements in
other X-ray bursts and believe it points to a special circumstance in
SAX J1712.6-3739. It may be related to increased levels of the
persistent flux as is often seen, particularly in PRE bursts
\citep{zand13,Worpel2013}, but it is not possible to determine with
these data given the lack of pre-burst XRT data.

\subsection{Burst from Swift J1734.5-3027}
\label{sec:Swift1734}

Swift J1734.5-3027 is a NS LMXB transient that was discovered with
Swift through the burst on September 1, 2013,  discussed here
\citep{malesani2013}. The transient was active in November 2010, May
2013, and September--October 2013 \citep{negoro2016,Bozzo2015Swift} at
levels of $\sim$10 mCrab, which is two orders of magnitude smaller than
the burst peak. The low peak luminosity may indicate an UCXB origin
\citep{heinke2015}. An XMM observation \citep{Bozzo2015Swift}
exhibited cold interstellar absorption at a level of $N_{\rm
  H}=(5.7\pm0.8)\times10^{21}$ cm$^{-2}$. So far, only the one burst
was detected, testifying that the accretion rate never reaches values
in excess of about 1\% of Eddington.

\begin{figure}[t]
\centering
    \includegraphics[width=0.9\columnwidth, trim=1.cm 0cm 10cm 12cm]{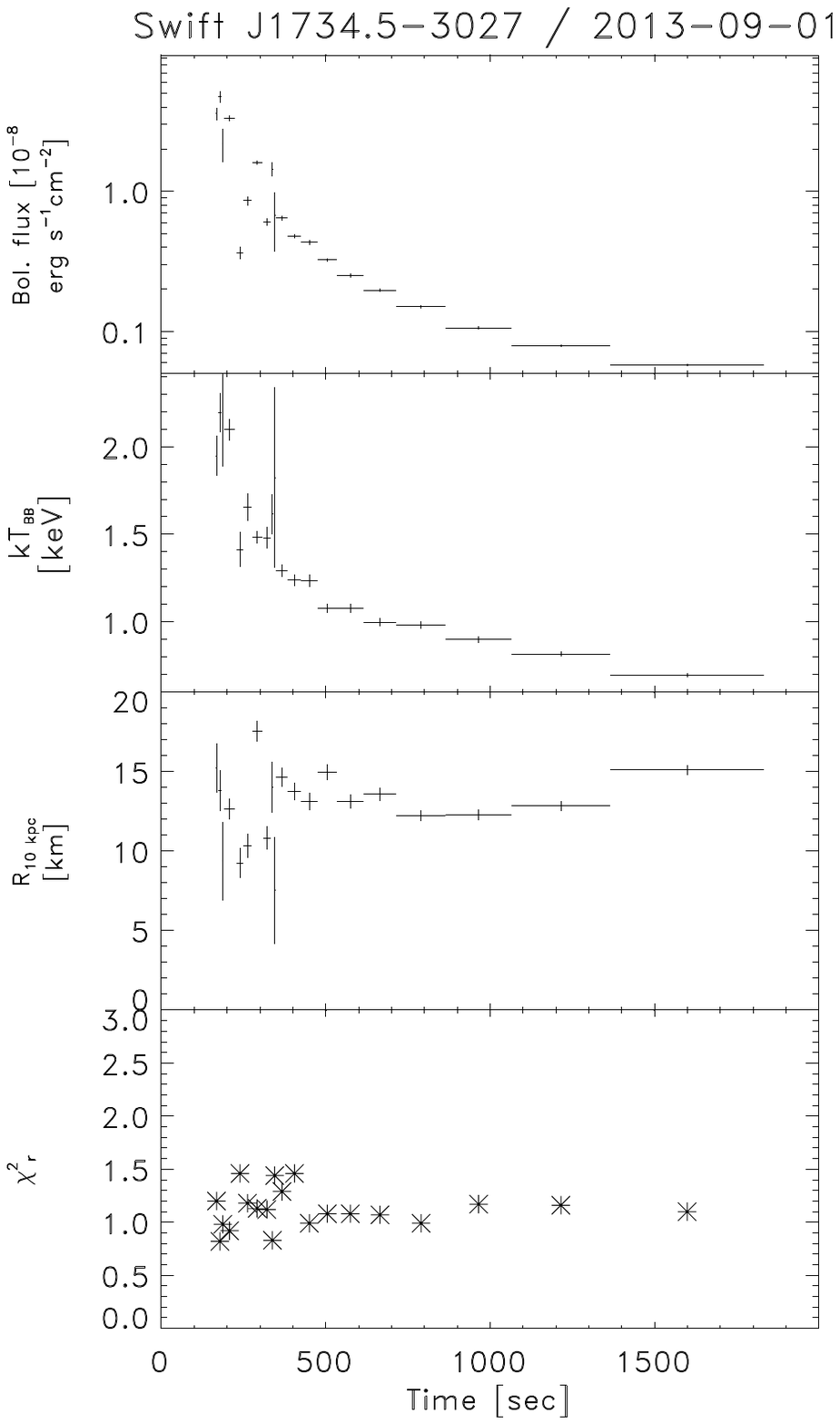}
    \caption{Swift J1734.5-3027: Time-resolved spectroscopy.}
    \label{fig:trs1734}
\end{figure}

The BAT light curve shows a fairly long rise time of $\approx 50$
s. \citet{Bozzo2015Swift} cannot firmly assess whether the burst
actually underwent PRE, but this light curve certainly suggests that
it did.  The XRT continuous measurement of the burst lasted for 1700
s. Some additional snapshots were taken 3000 s later. The XRT data
start with two short dips of about 50\% and 2 s and 5 s duration. Then
a 50 s long dip of depth 80\% is seen with a step upward of 20\%
halfway through the dip. Two final dips of 20 s and 6 s follow of
about 80\% depth.  \citet{Bozzo2015Swift} does not discuss the dips.

For the time-resolved spectroscopy, see Fig.~\ref{fig:trs1734}. It
shows that the dips are completely attributable to a decrease in the
blackbody normalization.

\subsection{Burst from IGR J18245-2452}
\label{sec:igrj1824}

Since the XRT light curve does not show a decreasing trend, we checked
whether this emission is due to the X-ray event detected with
BAT. First, we checked whether the spectrum is best described by a
blackbody. We tested the following prescriptions and found the
related $\chi^2_\nu$ values: a power law ($\chi^2_\nu=0.95$ with
$\nu=209$, $\Gamma=1.22\pm0.05$, $N_{\rm
  H}=(0.39\pm0.04)\times10^{22}$~cm$^{-2}$), a Comptonized spectrum
({\tt comptt} in {\tt XSPEC}; $\chi^2_\nu=0.96$ with $\nu=207$,
$N_{\rm H}=(0.40\pm0.13)\times10^{22}$~cm$^{-2}$), a disk blackbody
($\chi^2_\nu=0.86$ with $\nu=209$, k$T_{\rm in}=2.93\pm0.18$ keV,
$N_{\rm H}=(0.23\pm0.03)\times10^{22}$~cm$^{-2}$) and a blackbody
($\chi^2_\nu=1.24$ with $\nu=209$, k$T=1.11\pm0.02$, $N_{\rm
  H}<0.37\times10^{22}$~cm$^{-2}$). Although the data do not really
allow to make a distinction between models, a blackbody is the least
probable. IGR J18245-2452 is located in the globular cluster M28 with
a reddening of $E(B-V)=0.4$ \citep{Harris1996}, implying $N_{\rm
  H}=0.22\times10^{22}$~cm$^{-2}$ which is consistent with the value
found for the disk blackbody and simple blackbody. The absorbed
0.7-10 keV flux is $(8.6\pm0.3)\times10^{-10}$~erg~s$^{-1}$cm$^{-2}$.

\begin{figure}[t]
\centering
    \includegraphics[width=0.9\columnwidth, trim=1.cm 0cm 10cm 12cm]{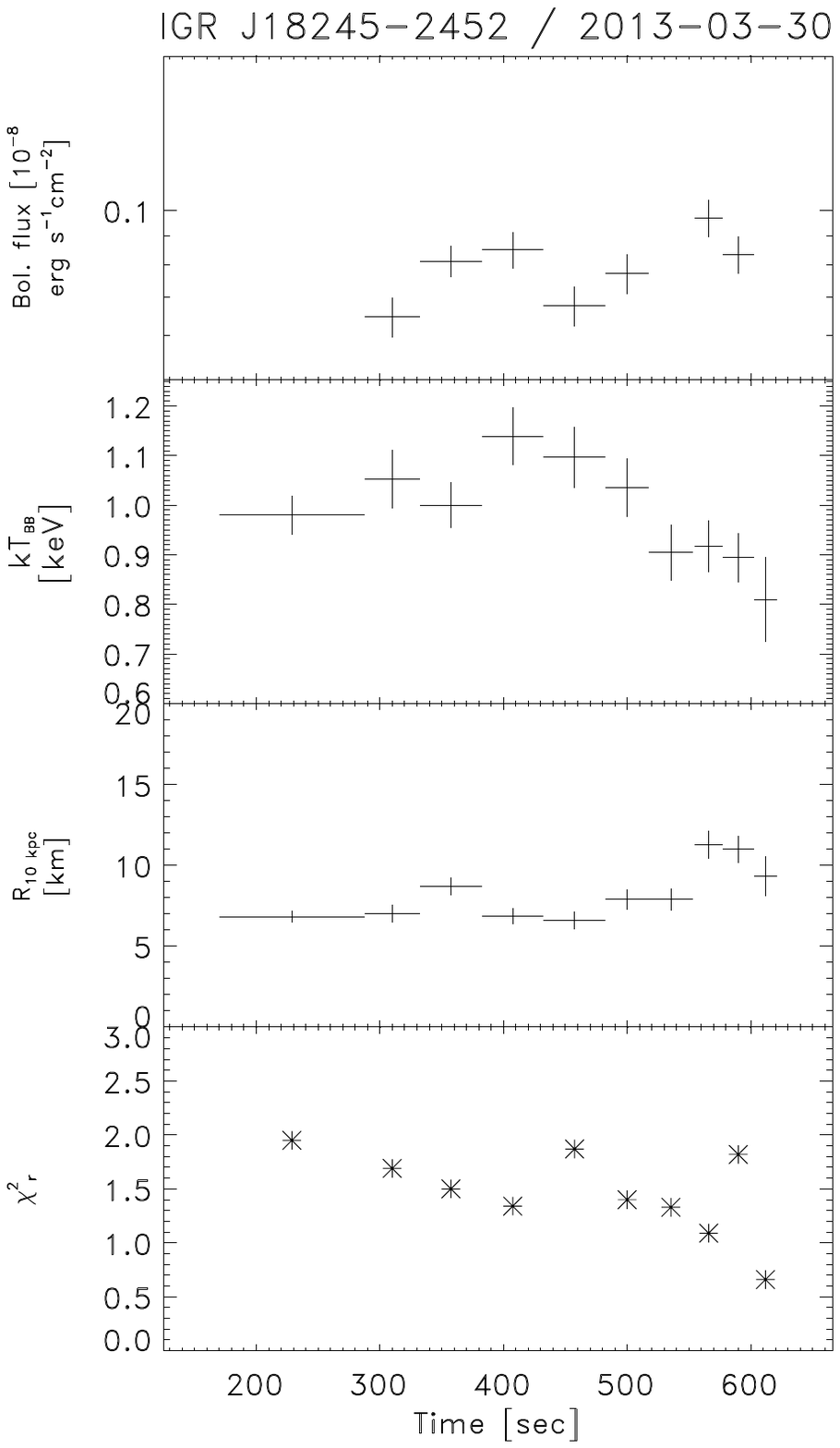}
    \caption{IGR J18245-2452: Time-resolved spectroscopy.}
    \label{fig:trs1824}
\end{figure}

Figure~\ref{fig:trs1824} shows the time-resolved spectroscopy of the
XRT data for this event. It shows a general cooling trend, although
only after the peak at 400 s at a rather low temperature which is
unexpected for a X-ray burst.

IGR J18245-2452 is a transient LMXB which was detected so far only
once. The outburst lasted from March 26 to April 17, 2013
\citep{falco2017}. It was highly variable with a peak bolometric flux
of $3\times10^{-8}$~erg~s$^{-1}$cm$^{-2}$. \cite{falco2017} report 3
X-ray bursts, one of which with Swift on 2013-04-07 \citep[][; see
  also Table~\ref{table:allbursts}]{papitto2013}. However, there may
have been a further detection with ASCA in 1995 \citep{gotthelf1997}.
IGR J18245-2452 is a transitional millisecond pulsar
\citep{papitto2013nature} with a very hard spectrum \citep{parikh2017}
and extreme variability \citep{wijnands2017}. The orbital period is
11.0 hr \citep{papitto2013}, so this is not a UCXB.

We here report a fourth burst-like event which was actually the first
in the outburst, 4.6 d after the outburst onset. \cite{falco2017}
study the broad-band persistent spectrum when the flux is
$4\times10^{-10}$~erg~s$^{-1}$cm$^{-2}$ and find that a power-law fit
yields a hard photon index $\Gamma=1.4$ which is similar to what we
find in the aftermath in XRT data of the burst detected with BAT. They
find $N_{\rm H}=0.24\times10^{22}$~cm$^{-2}$ which is also
consistent. This supports the notion that the emission in our XRT data
is not due to the burst and likely an expression of the strong
variability inferred by \cite{wijnands2017}. We, therefore, refrain
from further discussing this event.

\begin{figure}[t]
\centering
    \includegraphics[width=0.9\columnwidth, trim=1.cm 0cm 10cm 12cm]{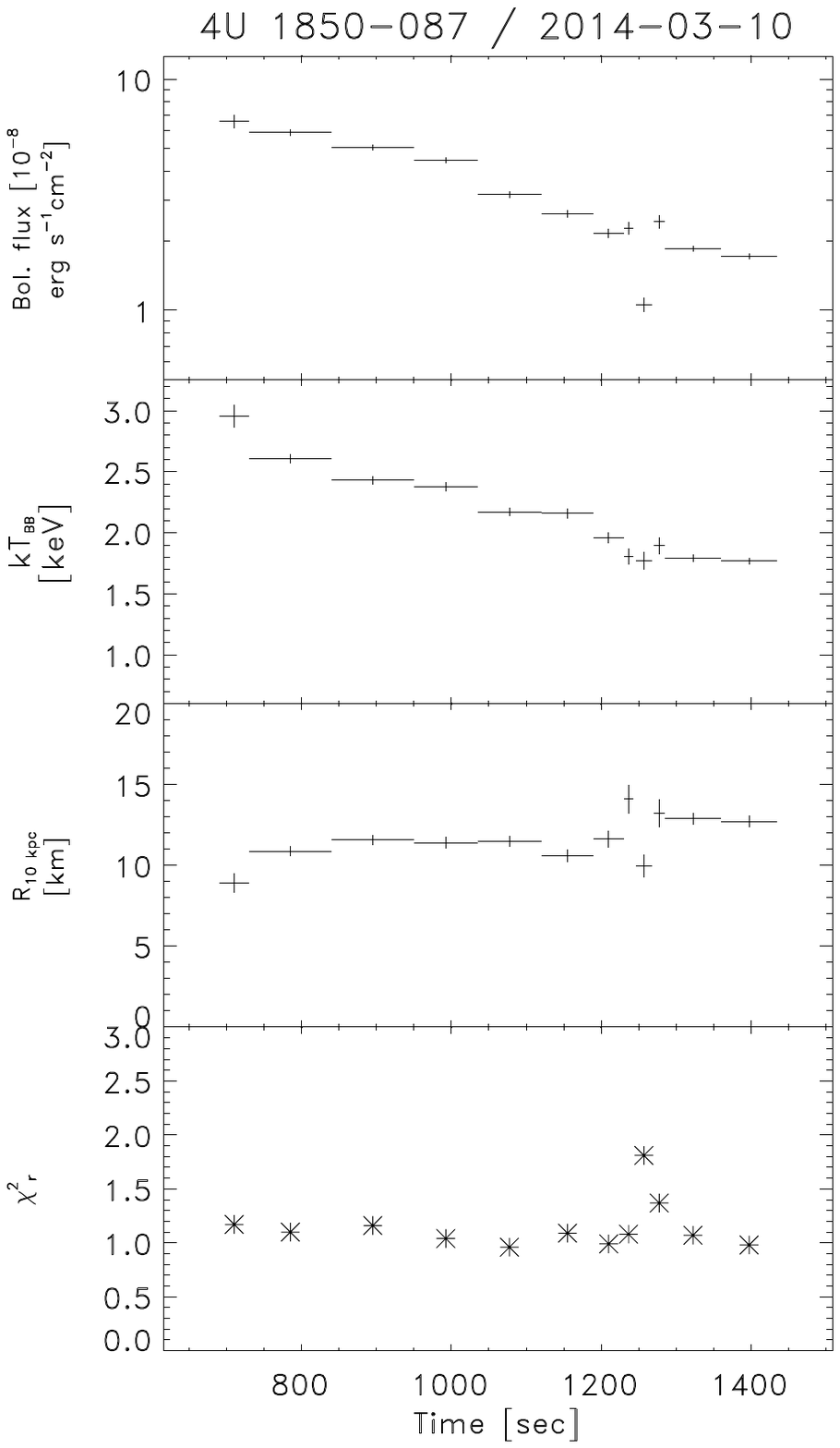}
    \caption{4U 1850-087: Time-resolved spectroscopy.}
    \label{fig:trs4U}
\end{figure}

\subsection{Burst from 4U 1850-087}
\label{sec:4U}

4U 1850-087 is an ultracompact X-ray binary (UCXB) with a tentative
orbital period of 20.6 min \citep{Homer1996}. 4U 1850-087 is located
in the globular cluster NGC 6712. From XMM-Newton observations,
\cite{sidoli2005} find $N_{\rm H}=(4-6.3)\times10^{21}$ cm$^{-2}$ with
evidence for extra absorption in the line of sight, since the best-fit
total $N_{\rm H}$ is always significantly higher than the optically
derived value in the direction of the host globular cluster of
$(1.8\pm0.2)\times10^{21}$ cm$^{-2}$. Therefore, the intrinsic
absorption ranges from 2 to $4.5\times 10^{21}$ cm$^{-2}$.

On March 10, 2014, 4U 1850-087 triggered Swift following an image
trigger over 392 s. After BAT completed the analysis in 26 s, the
observatory slewed to the source in 69 s. The first WT mode data is
available 691 s after the start of the burst \citep{zand2014b}. The
measurement lasted around 760 s. In the XRT light curve (see
Fig.~\ref{fig:lcs}) we see the decay of a thermonuclear X-ray burst. A
dip of $\sim20$ s long is clearly visible, which is $\sim660$ s after
the touchdown. Just before and after this dip, the count rate is above
the overall decay trend.

For the time-resolved spectroscopy, we fitted the spectra with an
absorbed blackbody (see Fig.~\ref{fig:trs4U}). The XRT data do not
show evidence for PRE, as expected given that the touch down point in
BAT data is prior to the XRT data. The dip is mostly due to a change
in the normalization of the blackbody and not to one in blackbody
temperature. An analogous remark can be made about the increased
fluxes before and after the dip.  The spectrum of the dip
(Fig.~\ref{fig:sp}) shows a small peak around 1 keV, although less
clear than in IGR J17062-6143.

The burst has a touchdown point which is about 550 s after the burst
onset according to BAT data. The implied duration of the
Eddington-limited phase rivals with the bursts from SLX 1735-269 and
the third burst from SAX J1712.6-3739 and is only superseded by the
1400~s seen in a superburst from 4U 1820-30 \citep{stroh02}.

The persistent emission of 4U 1850-087 as measured with MAXI from
56600 to 56730 MJD stays below approximately 0.02 counts cm$^{-2}$
s$^{-1}$ in both the 2--4 keV and 4--10 keV energy bands, except for the
days around the bursts when on average the count rate was $\sim0.01$
counts cm$^{-2}$ s$^{-1}$. Using the {\tt
  WebPIMMS}\footnote{\url{http://heasarc.gsfc.nasa.gov/cgi-bin/Tools/w3pimms/w3pimms.pl}}
tool we estimate the persistent flux (2.0--4.0 keV), which is
$\sim10^{-10}$ erg cm$^{-2}$ s$^{-1}$ for a power-law  index of 2. The
bolometric flux may be a factor of $\sim$2 larger. Since we did not model
the persistent spectrum we assumed the power-law index to be 2. The
persistent bolometric flux should therefore be taken with a large
uncertainty of 50\%. The persistent emission is, however, much smaller
than the burst emission which was $5.0\times10^{-8}$ erg cm$^{-2}$
s$^{-1}$ at the start of the XRT data. The peak bolometric flux might
be a bit higher. We did not subtract the persistent emission in the
spectra.

\begin{figure*}[t]
\centering
    \includegraphics[width=0.9\columnwidth, trim=1.5cm 1cm 4.5cm 2.2cm,clip=true]{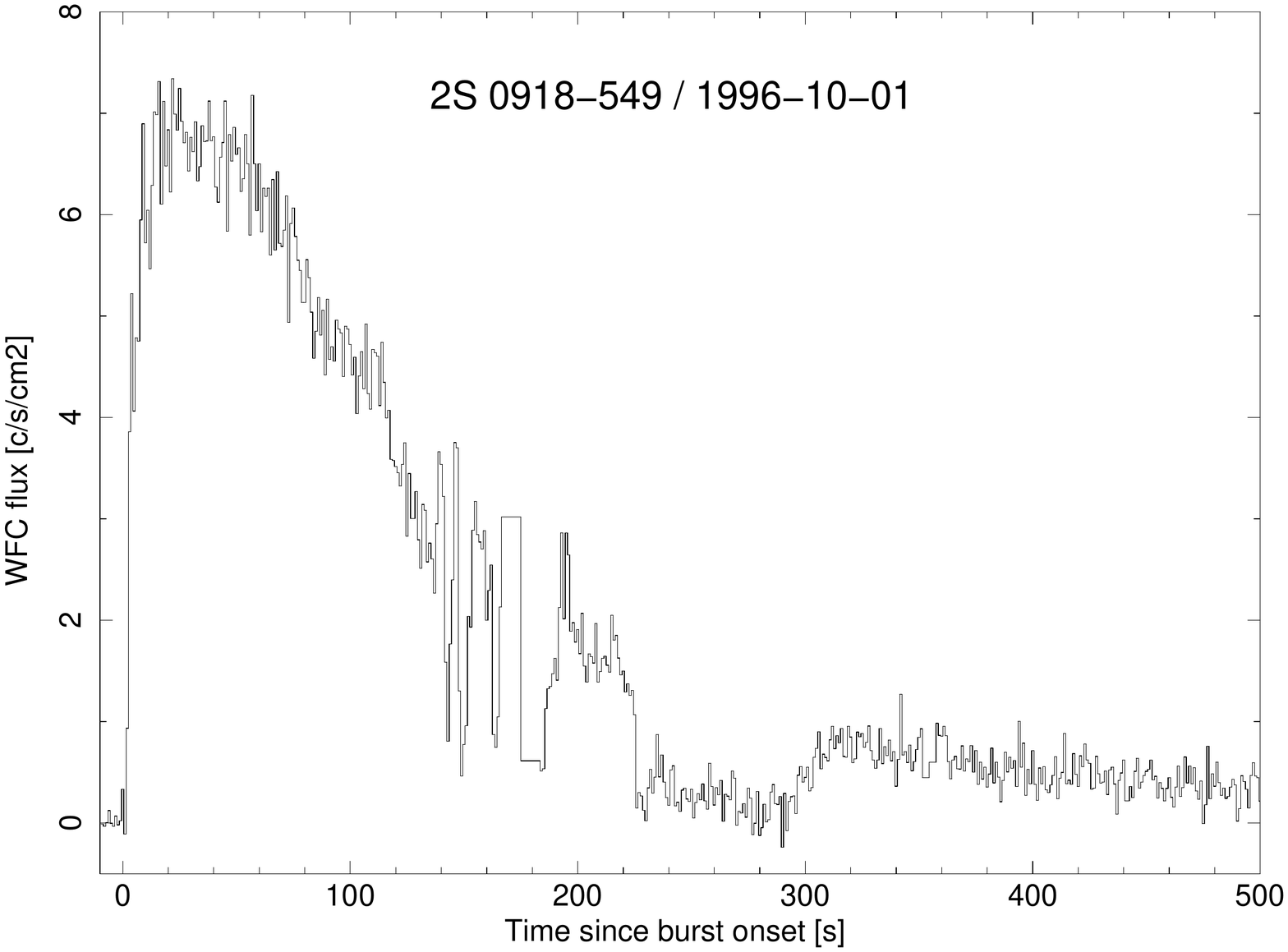}\hspace{4mm}
    \includegraphics[width=0.9\columnwidth, trim=1.5cm 1cm 4.5cm 2.2cm,clip=true]{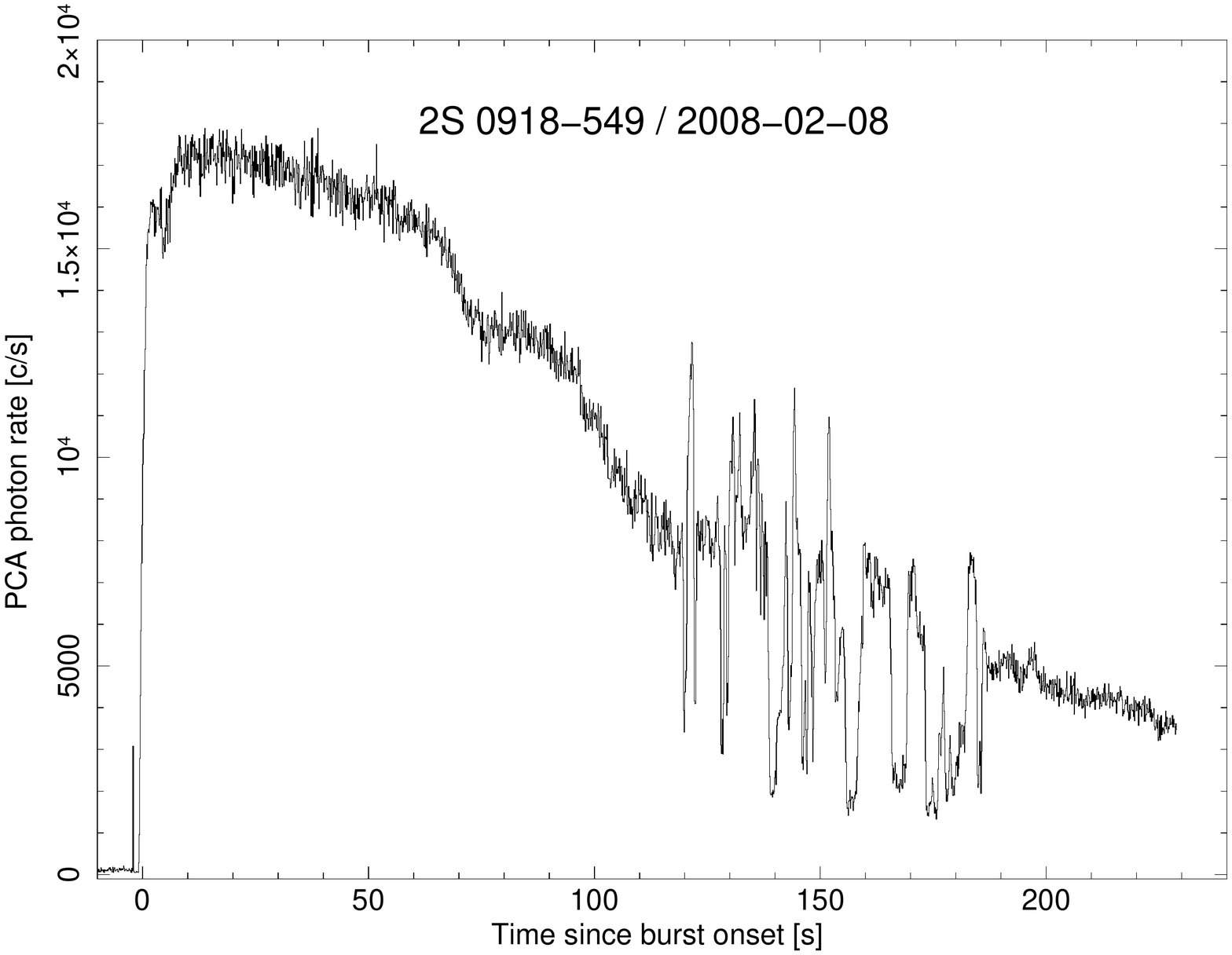}
    \includegraphics[width=0.9\columnwidth, trim=1.5cm 1cm 4.5cm 2.2cm,clip=true]{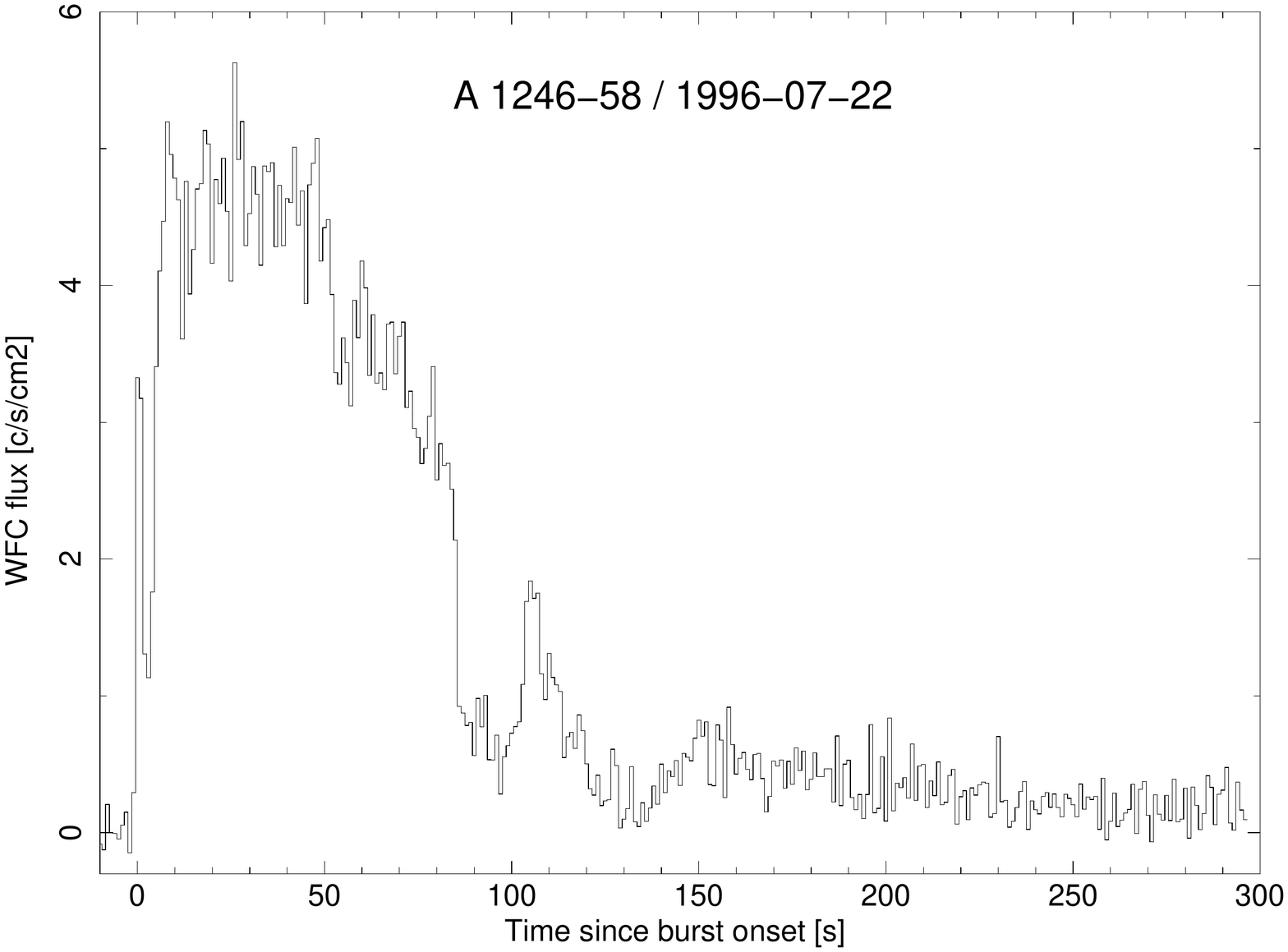}\hspace{4mm}
    \includegraphics[width=0.9\columnwidth, trim=1.5cm 1cm 4.5cm 2.2cm,clip=true]{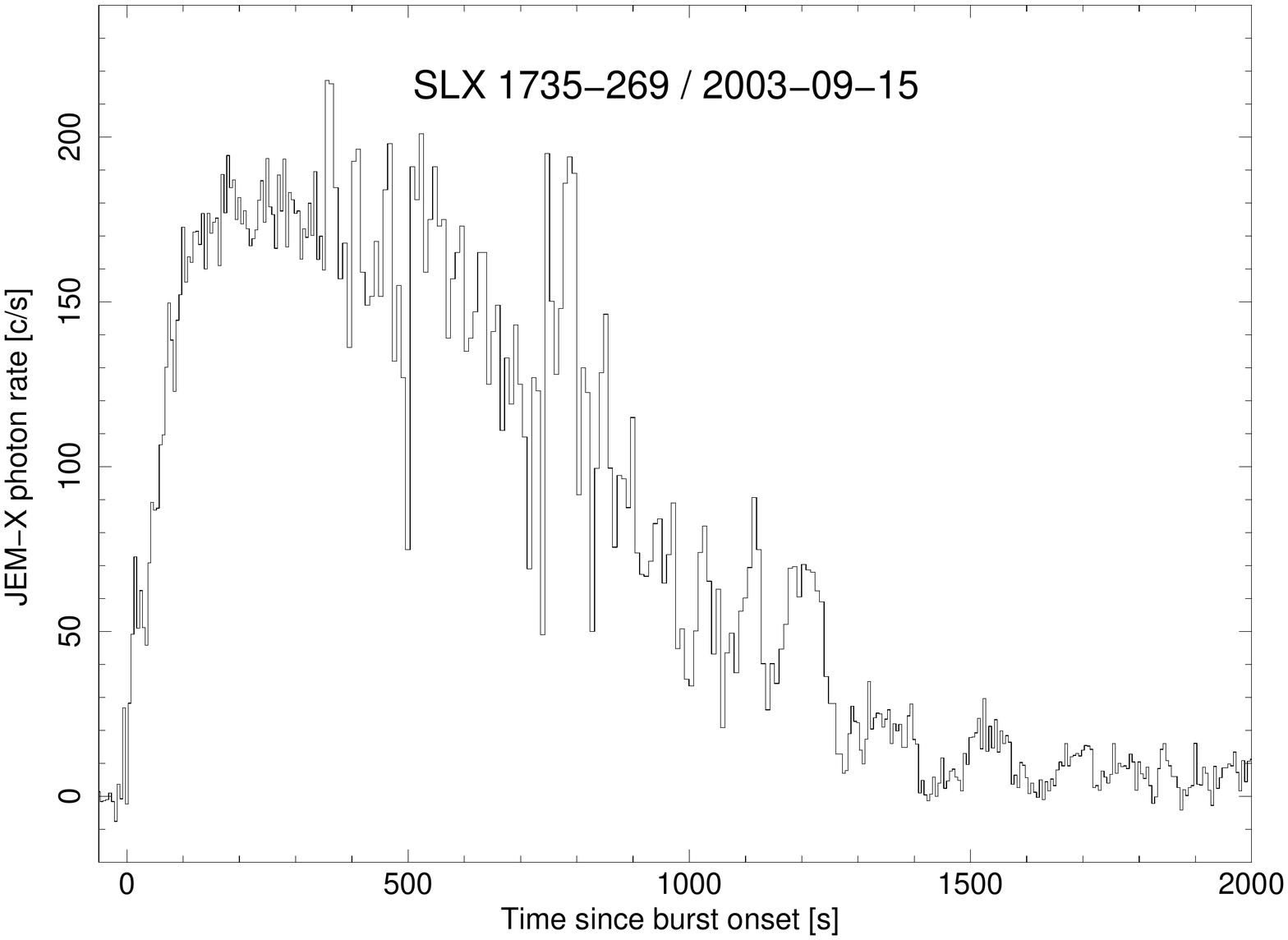}
    \includegraphics[width=0.9\columnwidth, trim=1.5cm 1cm 4.5cm 2.2cm,clip=true]{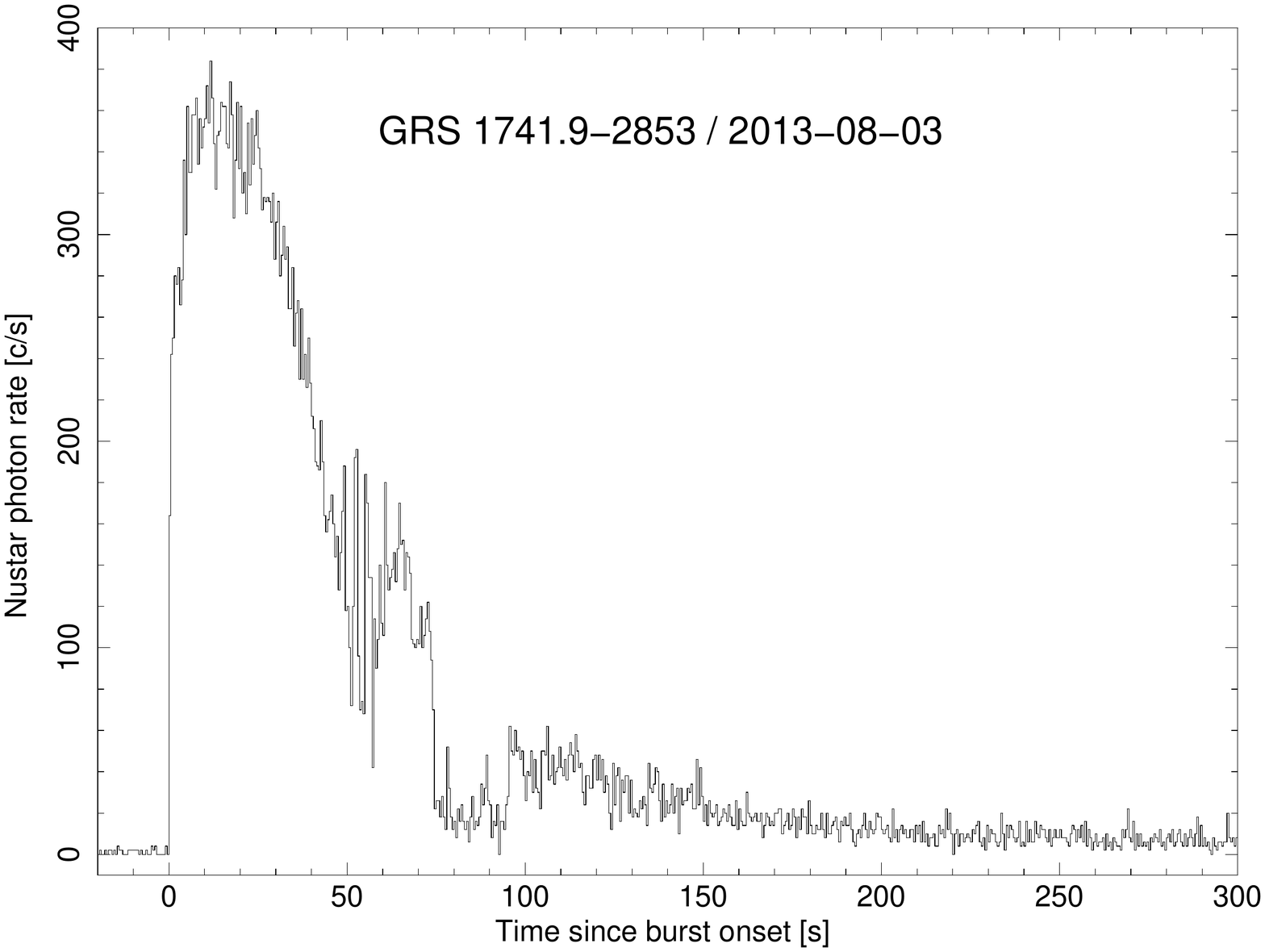}\hspace{4mm}
    \includegraphics[width=0.9\columnwidth, trim=1.5cm 1cm 4.5cm 2.2cm,clip=true]{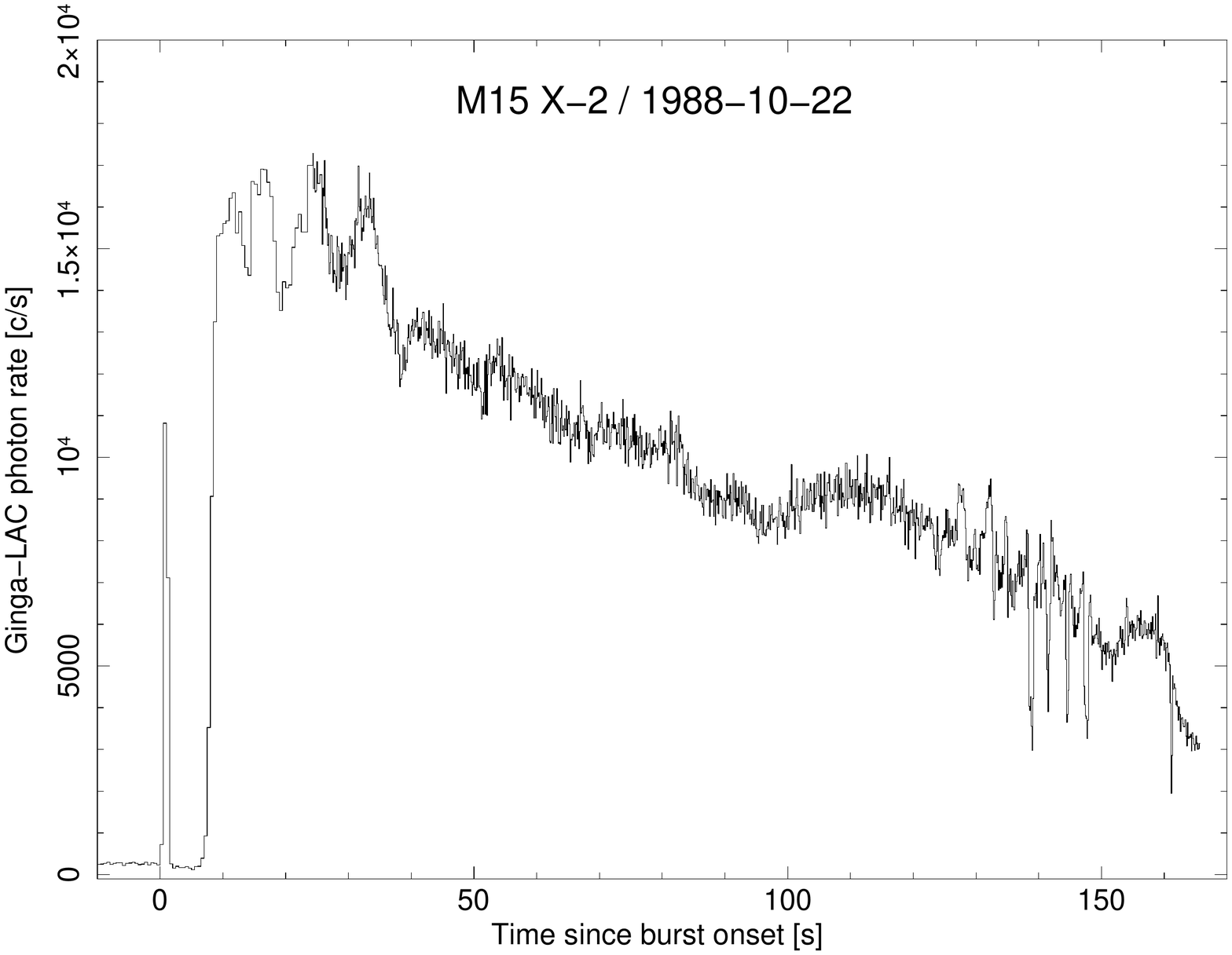}
    \caption{Light curves of literature bursts with fluctuations.}
    \label{fig:LClit}
\end{figure*}

\section{Literature bursts}
\label{sec:literature}

Nine  literature cases are known (eight are listed in \citealt{degenaar2018}; an   additional case is from  A1246-58).  We  provide
here some details of seven of these cases that are not among the Swift
cases. Light curves are provided in Fig. \ref{fig:LClit} and details
in Tables \ref{tab:thetableGEN}--\ref{tab:thetableDIP}.

\subsection{2S 0918-549}
2S 0918-549 is a very likely UCXB \citep{juett2003,zand2007} with a
tentative orbital period of 17.4 min \citep{Zhong2011}. Three bursts
were detected with the WFC \citep{Zand20052S}, one of which was much
longer than the others, lasting over 2500 s. The first 1000 s of this
long burst are displayed in Figure~\ref{fig:LClit}. The long burst
showed a strong photospheric radius expansion. There were two data
drop outs: from 165 to 180 s and from 350 to 354 s. After $\sim$2
minutes a short period of strong fluctuations is observed. After these
fluctuations a long dip starts quickly and then rises in $\sim$70 s
back to the decay trend of the burst.

Five X-ray bursts were detected with RXTE-PCA \citep{zand2011}. The
last one (on February 8, 2008) is at least seven times longer than the
others. This burst (see Fig.~\ref{fig:LClit}) has a very short
precursor of 40 ms \citep[see also][]{zand2014c} with the main burst
starting 1.2 s later. This burst is Eddington limited and has PRE. The
interesting feature of this burst is the strong fluctuation starting
122 s after burst onset and lasting 66 s.

\subsection{A 1246-58}

A 1246-58 is a UCXB candidate \citep{Bassa2006}. \citet{zand2008}
investigated the UCXB nature of A 1246-58 further and reported four
X-ray bursts, detected with the BeppoSAX WFC. All four bursts are
Eddington limited, one of which (Fig.~\ref{fig:LClit}) shows two deep
dips in the tail lasting tens of seconds. The dips are between 85 and
150 s while the PRE phase lasts until 55 s after burst onset. The
burst exhibits superexpansion.

\subsection{SLX 1735-269}

SLX 1735-269 \citep{skinner1987} is a UCXB candidate
\citep{zand2007}. During observations of the Galactic center region in
2003, INTEGRAL detected six X-ray bursts \citep{Molkov2005SLX}. One of
them (see Fig.~\ref{fig:LClit}) lasted over 2000 s and starts with a 2
s precursor. Eight seconds after the precursor, the main burst
started. This burst has an Eddington-limited phase of about 450
s. This burst also has superexpansion based on the fact that a
precursor is seen. \citet{Molkov2005SLX} conclude that the long burst
was most likely a result of the unstable burning of hydrogen and
helium, but we contend that this looks like an intermediate-duration
burst. The burst has many fluctuations, both up and down in count
rate. These fluctuations start before the touchdown ($\sim$500 s),
although the early fluctuations should be considered with caution
since they were measured in a non-standard observation mode (Chenevez,
priv. comm.).

\subsection{GRS 1741.9-2853}

GRS 1741.9-2853 is a transient LMXB with peak accretion luminosities
$<10^{37}$~\lum\ \citep{muno2003,trap2009,degenaar2015}. Whether GRS
1741.9-2853 is a UCXB is an unsettled matter. None of the previously
detected X-ray bursts from this source
\citep[][]{muno2003,galloway2008,trap2009} clearly resolves an
rp-component in the light curve, which would identify hydrogen and
suggest a non-UCXB nature \citep[e.g.,][]{zand2017}. The burst series
and repetitive buildup of burst ignition conditions indicates that
pure helium is burned during the bursts \citep{trap2009}. The sole
presence of pure helium bursts at apparently low accretion rates would
be typical of UCXBs.

In August 2013, NuSTAR detected an intermediate-duration PRE burst
lasting at least 800 s and with a fairly short exponential decay time
of 21 s \citep{barriere2015}. The LMXB was on the rise of a transient
outburst and was at a persistent bolometric luminosity about 445 times
smaller than the burst peak luminosity. The burst has two periods of
fluctuations: up and down fluctuations from 48 to 65 s on a timescale of a few seconds and with a dip between 75 and 95 s. The PRE
touchdown point is at about 30 s.

\subsection{4U 1820-30}

4U 1820-30 is the prototypical UCXB with the shortest orbital period
of 11 min \citep{stella1987}. In 1999 it exhibited a superburst in
RXTE/PCA observations \citep{stroh02}. These data show fluctuations
happening \citep{zand2011} on timescales between 1 and 10 s and
amplitudes up to 30\% between 3365 and 7200 s with a PRE phase until
1400 s. No eclipse-like features were seen.

\subsection{M15 X-2}

This burst was detected in 1988  from M15 X-2 with the Large Area
Counter on the Japanese X-ray observatory Ginga \citep{jvpm15}.  M15
X-2 is a UCXB with an orbital period of 22.6 min in the globular
cluster M15 \citep{dieball2005}. The burst is an intermediate-duration
burst visible for 160 s. It has superexpansion and a PRE stage lasting
about 95 s. During the post-PRE stages the general e-folding decay
time is about 60 s. The source shows fluctuations on a typical timescale of 1 s between 127 and 152 s.  No long timescale dips have been
detected, but that could be because the observation stopped
prematurely.

\section{Comparative analysis}
\label{sec:comparison}

\subsection{Data}
\label{subsec:TableInfo}

Light curves of all bursts, except the superburst of 4U 1820-30, are
provided in Figs. \ref{fig:lcs} and \ref{fig:LClit}.
Table~\ref{tab:thetableGEN} contains general information about the ten
sources including the type of binary, the hydrogen absorption column
density,  the type of accretion (persistent or transient), and
references. Table~\ref{tab:thetableFAST} presents the parameters
describing each burst, such as the e-folding decay time and the timing
properties of the fluctuations. All decay times are longer than
$\sim$100 s, except for the bursts from 1A 1246-588, which is 55 s,
and GRS 1741.9-2853, which is 46 s. The $t_{5\%}$ duration is often
constrained by  a lower limit alone because the data are
incomplete. However, we note that all are consistent with those of
intermediate-duration bursts ($>150$~s).

For the bursts detected with Swift, we determined the touchdown time
by examining the peaks in the BAT light curves. Unless the touchdown
time was reported in the reference of the literature burst, the
touchdown time was estimated from the peak in the plot of the  blackbody
temperature (k$T$, to be more exact).  We also determined the
timing properties of the fast fluctuations seen in the light curves
(see Figs.~\ref{fig:lcs} and \ref{fig:LClit}). The last column of this
table presents how much the bolometric flux has dropped, as a percentage
of the peak flux, at the start of the fluctuations. For the
calculation of the drop in bolometric flux we used the HEASARC {\tt
  WebPIMMS} tool. This tool allows us to calculate count rates from
fluxes or vice versa for different satellites. As input for the five
Swift bursts we use the BAT peak count rate. The input energy range of
the BAT is 15--150 keV and as output energy range we use 0.01--40 keV as
bolometric flux for a blackbody. The tool then calculates the flux of
a blackbody with a peak temperature. The accuracy  of these
calculations is estimated to be on the  order of ten percent.

In the last table (Table~\ref{tab:thetableDIP}), we present timing
properties of the eclipse-like features called \ql dips\qr . When more
than one dip is seen in the light curve, the dip duration column
presents the average duration of the dips seen.

\subsection{Findings}
\label{subsec:TableFind}

Of the ten sources, eight are  UCXBs (or candidates), which implies a small
orbit and hydrogen-deficient accretion. The nature of the two remaining
sources (Swift J1734.5-3027 and GRS 1741.9-2853) is undetermined, but
a UCXB nature is not excluded.

We distinguish two general types of variability: fast fluctuations
above and below the interpolated burst decay trend, and slow dips or
partial eclipse-like features. The fast fluctuations have a typical
timescale of a few to several tens of seconds (see
Table~\ref{tab:thetableFAST}) and an amplitude of $\sim$70\% above or
below the average trend. In the eclipse-like features the flux only
goes down, sometimes by as much as $\sim$90\%. In most cases the burst
decay remains visible during the eclipse (see
Table~\ref{tab:thetableDIP}). This is not the case for the burst from
Swift J1734.5-3027 where halfway through the two dips the flux
suddenly rises to a level between the first part of the dip and the
out-of-dip decay (see Fig.~\ref{fig:lcs}).  Seven of the 12 bursts
discussed here have both types of variabilities, 2 only have
eclipse-like features and 3 only have only fast fluctuations. The lack
of any of the two types of variability may be due to lack of full
burst coverage and it could well be true that all bursts exhibit both
kinds of variability.

The variability always occurs after the PRE-phase (i.e., after the
touchdown) except perhaps for SLX 1735-269 where the variability seems
to occur $\sim$40 s before the touchdown point, but this measurement
may be affected by instrumental behavior (the observational mode was
not standard; J. Chenevez, priv. comm.). Disregarding SLX 1735-269,
the fluctuations start at 1.5$\pm0.4$ times the PRE duration for the
bursts for which this ratio could be determined. The fluctuations and
dips are visible in the hardness ratio of 2.5--10/0.3--2.5 keV for all
the burst, except for the one from SAX J1712.6-3739 (probably because
$N_{\rm H}$ is relatively high), but are never seen in the hardness
ratio of 4.0--10/2.0--4.0 keV. This suggests that the fluctuations are
not completely achromatic as earlier suggested by \citet{zand2011}, but
only above about 2 keV, ergo above the spectral regions where most
line and edge features are situated.

The long-known and widely studied sources 4U 1850-087, 2S 0918-549, A
1246-58, SAX J1712.6-3739, SLX 1735-269, and GRS 1741.9-2853 are not
known to be dippers or eclipsers \citep{zand2007,trap2009}. Dips are
due to a bulge of the accretion disk, where the accretion stream
impacts the accretion disk. Eclipses are due to the donor star. Both
features periodically or quasi-periodically obscure the line of sight
to the inner X-ray bright part of the disk. Dips and/or eclipses from
the sources Swift J1734.5-3027 and IGR J17062-6143 have also not been
seen, although it should be noted that they have not been observed
much yet. This indicates that the sources are all viewed from an
inclination angle of less than 60$^{\rm o}$ \citep{horne1985}.

The bursts are all intermediate-duration bursts, lasting longer than
normal X-ray bursts (tens of seconds) but shorter than superbursts
(several hours, see also \S \ref{sec:super}). It is tempting to
attribute the fluctuations that we see to a selection effect because
it is, of course, more likely to see fluctuations in longer bursts
than in shorter ones. However, short bursts have been seen $\sim$100
times as often as intermediate-duration bursts, which counteracts this
selection effect.

The bursts all have photospheric radius expansion and SLX 1735-269
\citep{Molkov2005SLX}, 2S 0918-549 \citep{zand2011}, and A 1246-58
\citep{zand2008} show superexpansion, reaching an emission radius of
over 1000 km. The first burst from 2S 0918-549 \citep{Zand20052S} does
not have superexpansion unless the precursor lasts shorter than
$\sim$0.1 s as the second burst does, when it would not be
detectable. The burst from GRS 1741.9-2853 for certain has no
superexpansion \citep{barriere2015}. For the five bursts detected with
Swift, no XRT data are available for the first 100 s. Therefore, it is
difficult to check whether superexpansion
happened. \citet{Degenaar2013IGR} suggest that it is plausible that
superexpansion happened in the burst from IGR J17062-6143. The BAT
data of the last burst from SAX J1712.6-3739 show a clear long (15 s)
precursor and an unprecedented long (50 s) intermission to the main
burst phase. The 2--10 keV X-ray flux may, however, remain above zero.

We cannot exclude that the fluctuations are more likely to be related
to the intermediate-duration nature of the bursts in combination with
long PRE phases than with the UCXB nature of the hosts. The reason for
the dominance UCXBs may well be due to intermediate-duration bursts
occurring predominantly, but not exclusively, on UCXBs.

\begin{table*}[t]
\caption{General properties of sources and data coverage of
  fluctuating bursts (literature bursts below the dividing line).  All
  times are in seconds after the start of the burst, unless noted
  otherwise.\label{tab:thetableGEN}}
\begin{tabular}{rlllllll}
    \hline
Burst &    Object       & Ultracompact?                   & $N_{\rm H}/10^{22}$ & P/T         & Date & Instrument & Data coverage [s] \\ 
    \hline
1& IGR J17062-6143      & confirmed$^{[3,4,13]}$      & 0.12$^{[17]}$       & P$^{[4]}$  & 2012-06-25   & Swift/BAT+XRT & $-89/+1321$ \\
2& SAX J1712.6-3739     & candidate$^{[1],[5]}$       & 1.54$^{[6]}$        & P$^{[1]}$  & 2011-09-26   & Swift/BAT+XRT & $-119/+924^a$ \\ 
3& SAX J1712.6-3739     & ``                         & ``                 & ``         & 2014-08-18   & Swift/BAT+XRT & $0/+5200$ \\
4& Swift J1734.5-3027   & possibly                   & 0.57$^{[2]}$   & T$^{[2]}$  & 2013-09-01    & Swift/BAT+XRT & $-215/+1691^b$ \\
5& 4U 1850-087          & confirmed$^{[1]}$           & 0.48$^{[9]}$       & P$^{[1]}$  & 2014-03-10    & Swift/BAT+XRT & $-39/+1444$ \\
\hline
6& 2S 0918-549          & semi-confirmed$^{[10]}$     & 0.35$^{[9]}$       & P$^{[1]}$   & 1996-10-01    & BeppoSAX/WFC & $-13/+1002$ \\
7& 2S 0918-549          & ``                         &    ``              &            & 2008-02-08    & RXTE/PCA     & $-768/+315$ \\ 
8& A 1246-58            & candidate$^{[1]}$           & 0.50$^{[8]}$       & P$^{[1]}$   & 1996-07-22    & BeppoSAX/WFC & $-31/+199$ \\ 
9& SLX 1735-269         & candidate$^{[1]}$           & 1.50$^{[7]}$       & P$^{[1]}$   & 2003-09-15    & INTEGRAL/JEM-X & $-1200/+2235$ \\
10&GRS 1741.9-2853      & probably not$^{[11,14]}$       & 11.4$^{[11,14]}$       & T          & 2013-08-03    & NuSTAR         & $<-100/+800$ \\
11&4U 1820-30           & confirmed$^{[15]}$          & 0.16               & P          & 1999-09-09    & RXTE/PCA     & $-152/+9200$ \\
12&M15 X-2              & confirmed$^{[16]}$          & 0.07$^{[12]}$       & P          & 1988-10-20    & Ginga/LAC    & $<-100/+160$ \\
    \hline
    \multicolumn{8}{p{18cm}}{\tiny
        $^a$ This data coverage is continuous. The rest of the XRT data beyond 5713 s are short intervals with count rates below 9.5 c/s; 
        $^b$ This data coverage is continuous. The rest of the XRT data beyond 5007 s are short intervals with count rates below 30 c/s; 
        References:
        [1]: \citet{zand2007},
        [2]: \citet{Bozzo2015Swift},
        [3]: \citet{keek2017,Hernandez2018},
        [4]: \citet{Degenaar2013IGR,Degenaar2017IGR},
        [5]: \citet{Wiersema2009SAX},
        [6]: \citet{Fiocci2008ucxb}, 
        [7]: \citet{David1997SLX}, 
        [8]: \citet{zand2008}, 
        [9]: \citet{sidoli2005},
        [10]: \citet{Zhong2011},
        [11]: \citet{barriere2015},
        [12]: \citet{jvpm15},
        [13]: \citet{strohmayer2018},
        [14]: \citet{trap2009},
        [15]: \citet{stella1987},
        [16]: \citet{dieball2005} and
        [17]: \citet{vandeneijnden2018}
        }
    
\end{tabular}
\end{table*}

\begin{table*}[h]
\caption{Fast fluctuation properties. All times are in seconds after the start of the burst, unless noted otherwise.}
\label{tab:thetableFAST}
\centering
\begin{tabular}{rllrlllrrr}
    \hline
    Burst  & Object    & E-folding                   & $t_{5\%}$ & Touch   & Start time     & Start time            & End             & Time-  & Drop bol.  \\ 
           &           & decay time                  & [s]      & down    & fluct. (after  & fluct.                & time            & scale & flux [\%] \\ 
           &           & [s]                         &          & time    & touchdown)     & (Eddington            & fluct.          & fluct.            & until  \\ 
           &           &                             &          & [s]     & [s]            & phase duration)       & [s]             & [s]               & fluct.       \\ 
    \hline
    1&IGR J17062-6143    & 743        & 1300     & 220     & 540(240)       & $\frac{540}{300}=1.80$ & 1125            & 19   & 78 \\
    2&SAX J1712.6-3739   & 449        & 1050     & 190     & 330(110)       & $\frac{330}{200}=1.38$ & 480             & 11   & 74 \\ 
    3&SAX J1712.6-3739   & 3000       & 6000     & 500     & -              & -                      & -               & -    & -  \\ 
    4&Swift J1734.5-3027 & 174        & 600      &  75     & Before start XRT&       -               & 210             & 20   & 71 \\
    5&4U 1850-087        & 1017$^a$   & 1400     & 480     & -              & -                      & -               & -    & - \\
    \hline
    6&2S 0918-549        & 130        & 420      &85$^{[3]}$& 135(50)        & $\frac{135}{85}=1.59$  & 195            & 9    & 68 \\
    7&2S 0918-549        & 97         &  500     &77$^{[4]}$& $122(45)^{[13]}$& $\frac{122}{77}=1.58$  & 188            & 6    & 75 \\
    8&A 1246-58          & 55         & 200      &55 $^{[2]}$& -             & -                      & -              & -    & - \\
    9&SLX 1735-269       & 425        & 2200     &420$^{[1]}$& 380(-40)       & $\frac{450}{420}=1.07$ & 1230           & 49   & 0 \\
   10&GRS 1741.9-2853    & 46         & 240      &30        & 48(18)         & $\frac{48}{30}=1.6$   &  65            &  60  & 25\\
   11&4U 1820-30         & 3950       & $>$9200  &1400      & 2640           & $\frac{2640}{1400}=1.9$& 7170           & 3    & 50\\
   12&M15 X-2            & 98         & $>$165   &95        & 125(95)        & $\frac{125}{95}=1.3$ & 147            &  70  & 55\\
    \hline
    \multicolumn{10}{p{18cm}}{\tiny
        $^a$From $\sim20$ s to $\sim300$ s the light curve is flattened. If the exponential function is fitted to the light curve after this flat phase, the e-folding time decreases significantly to $130.37\pm31.35$ s. More information in \S~\ref{sec:4U};
        References:
        [1]: \citet{Molkov2005SLX};
        [2]: \citet{zand2008}; 
        [3]: \citet{Zand20052S};
        [4]: \citet{zand2011};
        }\\
    
\end{tabular}
\end{table*}

\begin{table*}[h]
\caption{Dip properties. All times are in seconds after the start of the burst, unless noted otherwise. When  multiple dips are seen, an average duration of the dips is given.}
\label{tab:thetableDIP}
\centering
\begin{tabular}{rllllll}
    \hline
    Burst & Object             & Start dips                  & End dips                   & Ingress \& egress                     & Dip                        & Decay            \\
          &                    &                    [s]      &                  [s]       &                   time         [s]    &     duration\newline [s]   &       during dip?\\
    \hline
    1&IGR J17062-6143    & 1125                        & Beyond data coverage?      & 3 \& 3                                & $\geq107$                  & Yes\\
    2&SAX J1712.6-3739   & 480                         & 840                        & 8 \& 3 (or 9)                         & 60                         & Yes \\ 
    3&SAX J1712.6-3739   &                             &                            & 8 \& 3 (or 9)                         & 200                        & No \\ 
    4&Swift J1734.5-3027 & 210                         & 320                        & 3 \& 3                                & 38                         & No \\
    5&4U 1850-087        & 1020 or 1240                & Beyond data or 1270        & 1 \& 2                                & 30                         & Yes\\
    \hline
    6&2S 0918-549        & 220                         & 310                        & 2 \& 70                               & 90                         & Yes\\
    7&2S 0918-549        & -                           & -                          & -                                     & -                          & -\\
    8&A 1246-58          & 85                          & 150                        & 2 \& 4; 4 \& 8                        & 30                         & Yes\\
    9&SLX 1735-269       & 1230                        & 1652                       & 5 \& 10                               & 80                         & No\\
   10&GRS 1741.9-2853    & 72                          & $>$165                     & 5 \& 5                                & 5                          & Yes \\
   11&4U 1820-30         & -                           & -                          & -                                     & -                          & - \\
   12&M15 X-2            & 145                         & 95                         & 2 \& 2                                & 23                         & Yes \\
    \hline
\end{tabular}
\end{table*}



Comparing the spectra from 4U 1850-087 (Fig.~\ref{fig:sp}) and IGR
J17062-6143 (Fig.~\ref{fig:sp}), we see that both have an interesting
feature below 2 keV. Although we are looking at two different parts of
the burst (at the dip for 4U 1850-087 and at the fluctuations for IGR
J17062-6143), they both have excesses there. In IGR J17062-6143 it is
a broad emission line, while in 4U 18050-087 it is a broader
excess. Logically, the count rate is lower for the downward
fluctuations than the count rate for the upward fluctuations, but
around 1 keV the difference in count rate is much smaller. In the case
of 4U 1850-087 the count rate is approximately equal around 1 keV for
both the dip ($\sim$30 s) and the times around the dip (from 30 s before
 the dip start and from the end of the dip to 30 s after the
dip). A small peak in the downward fluctuations of Swift J1734.5-3027
(Fig.~\ref{fig:sp}) is also visible around $\sim$1.4 keV. The model in
that spectrum is also underestimated in the 1 keV area as is the case
with 4U 1850-087 and IGR J17062-6143. A 1 keV emission feature is most
prominent in the burst from IGR J17062-6143. SAX J1712.6-3739 has a
$N_{\rm H}$-value which is much higher than for the other three bursts
which makes it difficult to detect any feature in the spectrum below 2
keV (see Fig.~\ref{fig:sp}).

\section{Discussion}
\label{ch:Discussion}

\subsection{Fluctuations}
\label{subsec:flucs}

\cite{zand2011} interpret burst fluctuations on the basis of six
bursts.  The sample has now doubled, but most observational
parameters of this phenomenon have not changed, in particular the
amplitude, timescale, and delay time with respect to the
Eddington-limited phase. One change is that the fluctuations appear
achromatic above 2 keV, but not at lower energies. There sometimes is
an excess, which in one case is consistent with a broad emission
line. Another change is that the UCXB nature may not be a prerequisite
for fluctuations.

The fluctuations may be associated with the presence of superexpansion
in the following way \citep[for details, see][]{zand2011}. When
superexpansion is observed, an optically thick shell is thought to
expand from the neutron star and skim the accretion disk within
seconds. The shell can disturb the disk. The inner 10$^2$~km of the
disk may be swept up due to the large ram pressure of the shell
(we note that the expansion is fast,  up to a few tens of percent of the
speed of light), as might be the disk surface of inner
$10^{3-4}$~km. The disk will resettle on a viscous timescale, but not
before the Eddington-limited phase and the associated radiation
pressure has ended. It is possible that  the wind and radiation pressure of the
regular photospheric expansion magnifies the disturbances imposed by
the passing shell, and it will do so more if the expansion phase lasts
longer.

The disturbance will be characterized by a vertical structure on the
disk. This may be envisaged as clouds. Such highly ionized clouds
may backscatter burst photons into the line of sight, when at the far
side of the neutron star as seen by the observer, yielding upward
fluctuations. Conversely, the cloud may scatter photons out of the
line sight when it is in the line of sight. Since the observed
amplitudes of the fluctuations are so high, only one cloud can be
active at a time. This model \citep[some simple calculations were
  performed in][]{zand2011} does not explain the occurrence of two
types of variability. The fast fluctuations must originate closer to
the neutron star than the slow dips, nor does it explain why these two
types of variability happen after each other (slow after fast).

The observational data do not change this picture. The fact that the
low-energy flux does not show such a high amplitude and that the best
data from IGR J17062-6143 indicates a 1 keV emission feature as the
culprit suggests that this may be due to a different emission region
for the low-energy flux than for the high-energy flux. That would be
consistent with the general idea that the accretion flow around
neutron stars consists of multiple components, namely accretion disk,
accretion disk corona, disk wind, and/or jet \citep[for a recent review
  of similar accretion flows around black holes and further
  references, see, e.g.,][]{blaes2014}, all with different regions and
sizes.

\subsection{Is the burst from SAX J1712.6-3739 on August 14, 2014, a superburst?}
\label{sec:super}

The e-folding decay time of the event from SAX J1712.6-3739 on
August 14, 2014, of $\sim$0.8 hr is comparable to that of the shortest of 27
detected superbursts \citep{zand2017,iwakiri2018}. The four shortest
superbursts have e-folding decay times of 0.7 and 1.0 hr for
superbursts from GX 17+2, 0.5 hr for one from 4U 1820-30 and 1.2 hr
for one from Ser X-1). The particular event from 4U 1820-30 is not thought 
to be a superburst \citep{serino2016}.  GX 17+2 and Ser X-1 have, at
the times of the superbursts, accretion rates in excess of 20\% of
Eddington. This is unlike SAX J1712.6-3739.  Considering that some
characteristics of the event from SAX J1712.6-3739 are consistent with
the longest intermediate-duration bursts (superexpansion, a long PRE
phase, low accretion rate; see, e.g., Table~\ref{tab:thetableFAST}), we
think it more likely that it is a long intermediate-duration bursts, in
other words a flash of a thick helium pile instead of a thick carbon
pile. This conclusion is identical to that drawn for other long bursts from
SAX J1712.6-3739 \citep[e.g.,][]{kuulkers2009,iwakiri2018}.

An increasing number of   bursts have been identified, not only in this study, that have
durations  on the short end of the superburst distribution and
on the long end of the distribution for intermediate-duration
bursts. This gives rise to uncertainty in superburst
identifications, not only for the burst from SAX J1712.6-3739, but
also  for IGR J1706-6143 \citep{iwakiri2015,keek2017}, 4U
1820-30, and SLX 1735-269 \citep{serino2016,serino2017}, among others. They can either
be superbursts in high accretion rate systems or intermediate-duration
bursts on cold neutron stars. It depends on the temperature of the
neutron star and the characteristics of the bursts. The data sets on
these diagnostics are, unfortunately, scarce and progress in their
understanding requires better data sets that are hard to obtain
because of their rarity.

\section{Summary and conclusion}

We found five bursts in the Swift data with eclipse-like features and/or
strong fluctuations during their decay. The five bursts are discussed in
detail and are from 4U 1850-087, Swift J1734.5-3027, IGR J17062-6143,
and SAX J1712.6-3739. Seven additional bursts from the literature show
similar features. We provide a table with information about all twelve
bursts, their variabilities, and general properties of the source.

We find that the typical timescale of the fluctuations is 1--50 s and
the amplitude $\sim$70\% above and below the burst decay trend. The
variabilities (fluctuations and eclipse-like features) almost always
occur after the PRE-phase.

The fluctuations show similar spectral characteristics such as an
emission line around $\sim$1 keV. The spectrum does not change, except
in total emission, between upward and downward fluctuations. The
emission from this line is  stronger for the downward
fluctuations.

As previously suggested by \citet{zand2011}, the fluctuations are
possibly due to a disturbed accretion disk, which is likely caused by
a combination of an expanding shell and a near-Eddington flux. These
causes provide an explanation for the variabilities seen in these 12
bursts. The validity of this suggestion needs to be verified through
modeling the interaction of shells, winds and radiation pressure and
the accretion disk, using  as constraints the superexpansive character of
the bursts and the long Eddington-limited flux.  We suggest that such
a theoretical study should investigate the similarity with the instabilities
seen in flags driven by laminar wind flows.

\begin{acknowledgements}
We thank the Swift Team for their long-lasting support of a wonderful
mission for burst science, not only gamma-ray bursts, but also X-ray
bursts. We are grateful to Neil Gehrels and his team for their
willingness to open up Swift automatic slewing to X-ray bursts, which
resulted in the interesting findings presented here. May Neil Gehrels
rest in peace. We thank Motoko Serino and Wataru Iwakiri for providing
the MAXI light curve for SAX J1712.6-3739 at our request, and Tadayasu
Dotani, Ken Ebisawa, and Moto Kokubun for providing the light curve of
M15 X-2.  ND is supported by a Vidi grant from the Netherlands
Organization for Scientific Research (NWO). This work benefited from
discussions at the BERN18 Workshop supported by the National Science
Foundation under Grant No. PHY-1430152 (JINA Center for the Evolution
of the Elements).
\end{acknowledgements}

\bibliographystyle{aa} 
\bibliography{references}

\appendix
\section{All Swift X-ray burst detections with event data}
\label{appendix:1}

{ \normalsize

On March 31, 2018, we downloaded all XRT data concerning the 111
Galactic X-ray bursters known at the time, extracted light curves at 1
s resolution of all data, and searched for X-ray bursts. The data set
consists of 15.2 Msec exposure. Particularly the four bursters that
are in the field of view of the Galactic center (XMM J174457-2850.3,
GRS 1741.9-2853, AX J1745.6-2901, and 1A 1742-289) were widely covered
with about 2.1--2.2 Msec each. We found 90 X-ray bursts from 45
bursters. The numbers per burster are listed in
Table~\ref{table:xrtdata}. Swift discovered 13 new bursters: the 6
Swift sources in the list plus IGR J00291+5934 \citep{kuin2015}, MAXI
J1647-227 \citep{kennea2012}, XTE J1701-407 \citep{markwardt2008}, IGR
J17062-6143 \citep{Degenaar2012}, 1RXH J173523.7-354013
\citep{degenaar2010}, IGR J17511-3057 \citep{bozzo2009}, and IGR
J18245-2452 \citep{papitto2013}.

We provide a list of the 90 bursts detected with XRT in
Table~\ref{table:allbursts}, plus the 45 bursts that were detected
with BAT and for which no XRT signal was found. This table includes
the trigger numbers if appropriate, the trigger times if appropriate,
the burst start times (estimated with an accuracy of a few seconds),
and indications of the e-folding decay times in both BAT and XRT data,
if appropriate. Furthermore, we list 41 BAT triggers on burster
locations for which there is no clear signal in BAT. This includes
image triggers that are less certain to involve X-ray bursts.

Figure~\ref{fig:allbursts1} shows the light curves for all 90 bursts that
have XRT coverage.

}

\begin{table*}
\begin{center}
\caption[]{Summary of all XRT observations of all 111 X-ray bursters until March 31, 2018. These observations yielded 90 detections of X-ray bursts.\label{table:xrtdata}}
\small
\begin{tabular}{lrrlrr}
Source         &   exposure    & No.   & Source & exposure    & No. \\
               &   time (ksec) & bursts&        & time (ksec) & bursts\\
\hline
IGR J00291+5934      &     50.8 & 1 &IGR J17464-2811      &     63.9 & 0 \\ 
4U 0513-40           &     38.8 & 0 &IGR J17473-2721        &    7.0 & 1 \\ 
4U 0614+09           &    463.5 & 2 &SLX 1744-300           &    2.4 & 0 \\ 
EXO 0748-676         &    311.2 & 2 &SLX 1744-299           &    2.2 & 0 \\ 
4U 0836-429          &      0   & 0 &GX 3+1                 &    5.2 & 0 \\ 
2S 0918-549          &     25.3 & 0 &IGR J17480-2446        &  210.8 & 0 \\ 
4U 1246-588          &     13.7 & 1 &EXO 1745-248           &  210.8 & 1 \\ 
4U 1254-69           &     37.4 & 0 &Swift J174805.3-244637 &  210.8 & 2 \\ 
SAX J1324.5-6313     &     38.0 & 0 &1A 1744-361            &   10.9 & 0 \\ 
4U 1323-62           &     46.3 & 0 &SAX J1748.9-2021       &  101.2 & 5 \\ 
MAXI J1421-613       &     21.9 & 1 &Swift J1749.4-2807     &   86.5 & 1 \\ 
Cen X-4              &    142.6 & 0 &IGR J17498-2921        &   87.3 & 2 \\ 
Cir X-1              &    180.6 & 1 &4U 1746-37             &   54.8 & 3 \\ 
UW CrB               &     36.2 & 0 &SAX J1750.8-2900       &   76.8 & 0 \\ 
4U 1608-522          &    107.9 & 0 &EXO 1747-214           &    0   & 0 \\
MAXI J1621-501       &     23.7 & 0 &GRS 1747-312           &   16.3 & 0 \\
4U 1636-536          &    125.3 &11 & IGR J17511-3057        &  80.5 & 3 \\ 
MAXI J1647-227       &     10.6 & 2 &SAX J1752.3-3138       &    1.9 & 0 \\ 
XTE J1701-462        &    177.2 & 0 &SAX J1753.5-2349       &   20.5 & 0 \\ 
XTE J1701-407        &     52.6 & 2 &AX J1754.2-2754        &   52.1 & 1 \\ 
MXB 1658-298         &     78.0 & 2 &IGR J17597-2201        &   11.2 & 0 \\ 
4U 1702-429          &      7.4 & 0 &1RXS J180408.9-342058  &  115.3 & 6 \\ 
IGR J17062-6143      &    137.9 & 1 &SAX J1806.5-2215       &  105.9 & 1 \\ 
4U 1708-23           &     10.4 & 0 &2S 1803-245            &    0   & 0 \\ 
4U 1705-32           &     19.8 & 0 &SAX J1808.4-3658       &  353.7 & 1 \\ 
4U 1705-44           &      8.7 & 0 &XTE J1810-189          &   19.1 & 2 \\ 
XTE J1709-267        &     32.4 & 0 &SAX J1810.8-2609       &   13.8 & 1 \\ 
XTE J1710-281        &      5.3 & 1 &XTE J1812-182          &    2.2 & 0 \\ 
4U 1708-40           &      0.5 & 0 &XTE J1814-338          &    2.0 & 0 \\ 
SAX J1712.6-3739     &     21.7 & 3 &GX 13+1                &   25.7 & 0 \\ 
2S 1711-339          &      2.8 & 0 &4U 1812-12             &   13.2 & 0 \\ 
RX J1718.4-4029      &    104.1 & 0 &GX 17+2                &  121.7 & 0 \\ 
1H 1715-321          &      1.1 & 0 &Swift J181723.1-164300 &   21.9 & 0 \\ 
IGR J17191-2821      &      9.9 & 1 &SAX J1818.7+1424       &   33.1 & 0 \\ 
XTE J1723-376        &      0.7 & 0 &4U 1820-303            &   50.0 & 1 \\ 
IGR J17254-3257      &     10.3 & 0 &IGR J18245-2452        &  131.1 & 2 \\ 
4U 1722-30           &      2.0 & 0 &4U 1822-000            &    0   & 0 \\ 
4U 1728-34           &     21.8 & 1 &SAX J1828.5-1037       &   41.8 & 0 \\ 
MXB 1730-335         &    111.3 & 3 &GS 1826-24             &   51.7 & 2 \\ 
KS 1731-260          &      0   & 0 &XB 1832-330            &    5.1 & 0 \\ 
Swift J1734.5-3027   &     60.3 & 1 &Ser X-1                &    6.0 & 0 \\ 
1RXH J173523.7-354013&     39.7 & 1 &Swift J185003.2-005627 &   27.0 & 1 \\ 
SLX 1732-304         &      3.6 & 0 &4U 1850-086            &   11.8 & 1 \\ 
IGR J17380-3747      &      3.1 & 0 &HETE J1900.1-2455      &   74.7 & 0 \\ 
SLX 1735-269         &     14.7 & 1 &XB 1905+000            &    0   & 0 \\ 
4U 1735-444          &    114.1 & 5 &Aql X-1                &  612.8 & 2 \\ 
XTE J1739-285        &     16.6 & 1 &XB 1916-053            &   54.7 & 0 \\ 
SLX 1737-282         &      8.5 & 0 &Swift J1922.7-1716     &   55.9 & 1 \\ 
IGR J17445-2747      &     23.8 & 0 &XB 1940-04             &    0   & 0 \\ 
KS 1741-293          &    102.7 & 0 &XTE J2123-058          &    0   & 0 \\ 
XMM J174457-2850.3   &   2149.2 & 1 &4U 2129+12             &   57.9 & 0 \\ 
GRS 1741.9-2853      &   2149.2 & 2 &XB 2129+47             &    0   & 0 \\ 
AX J1745.6-2901      &   2190.7 & 2 &Cyg X-2                &  144.0 & 1 \\ 
1A 1742-289          &   2191.2 & 0 &SAX J2224.9+5421       &   26.5 & 0 \\ 
1A 1742-294          &      8.6 & 0 &4U 1543-624            &    7.4 & 0 \\ 
SAX J1747.0-2853     &     65.3 & 0 &                       &        &   \\

\hline\hline
\end{tabular}
\end{center}
\end{table*}

\clearpage
\onecolumn
\begin{longtable}[c]{llcclccrr}
\caption{List of 134 X-ray bursts detections with Swift
  until March 31, 2018, for which event data is available. At the
  bottom, an additional 41 BAT triggers (without slew) on burst source
  locations are listed that reveal no clear burst signal in
  BAT. \label{table:allbursts}}\\ 
\hline
Source   & Date and time (UTC) & Det.$^{\rm a}$ & Trigger & Time (MET)$^{\rm b}$ & MJD(UTC)$^{\rm b}$ & BAT & BAT & XRT   \\
         &                     &              & id.     & \hspace{5mm}(s)        &          & double & dur.$^{\rm c}$ & decay$^{\rm d}$  \\
         &                     &              &         &            &          &  peaked? &    (s) & (s) \\
\hline
\endfirsthead
\multicolumn{8}{l}
{\tablename\ \thetable\ -- \textit{Continued from previous page}} \\
\hline
Source   & Date and time (UTC) & Det.$^{\rm a}$ & Trigger & Time (MET)$^{\rm b}$ & MJD(UTC)$^{\rm b}$ & BAT & BAT & XRT   \\
         &                     &              & id.     & \hspace{5mm}(s)        &          & double & dur.$^{\rm c}$ & decay$^{\rm d}$ \\
         &                     &              &         &            &          & peaked? &     (s) & (s) \\
\hline
\endhead
\hline \multicolumn{8}{r}{\textit{Continued on next page}} \\
\endfoot
\hline
\endlastfoot
4U 1702-429            & 2005-01-29 17:04:14 & b  & 104039 &  128711053.70 & 53399.711277 & y &  14 &    \\ 
4U 1812-12             & 2005-01-30 17:26:09 & b  & 104090 &  128798768.38 & 53400.726493 & y &  13 &    \\ 
EXO 0748-676           & 2005-02-12 16:56:54 & x  &        &  129920221.71 & 53413.706192 &   &     & 26 \\
EXO 0748-676           & 2005-02-12 20:00:28 & x  &        &  129931234.71 & 53413.833658 &   &     & 17 \\
4U 1636-536            & 2005-02-13 22:11:40 & b  & 106034 &  130025499.90 & 53414.924774 &   &  14 &    \\ 
4U 1728-34             & 2005-02-14 09:08:32 & b  & 106070 &  130064911.49 & 53415.380927 &   &   4 &    \\ 
4U 1702-429            & 2005-02-17 14:57:15 & b  & 106276 &  130345033.98 & 53418.623086 & y &  11 &    \\
4U 1636-536            & 2005-02-22 13:48:58 & b  & 106666 &  130772936.96 & 53423.575666 &   &   6 &    \\
4U 1702-429            & 2005-02-23 08:30:40 & b  & 106724 &  130840239.36 & 53424.354629 &   &  14 &    \\
4U 1812-12             & 2005-02-24 12:40:46 & sl & 106799 &  130941645.31 & 53425.528309 & y &  21 &    \\ 
4U 1812-12             & 2005-03-14 03:18:38 & b  & 110772 &  132463117.31 & 53443.137938 & y &  16 &    \\
4U 1636-536            & 2005-07-02 09:01:17 & b  & 143840 &  141987676.42 & 53553.375889 &   &   8 &    \\ 
4U 1702-429            & 2005-07-03 20:29:23 & b  & 144067 &  142115362.24 & 53554.853734 &   &  11 &    \\
4U 1728-34             & 2005-07-21 11:14:30 & b  & 147029 &  143637270.40 & 53572.468412 & y &  10 &    \\
SLX 1735-269           & 2005-07-28 21:36:40 & b  & 147919 &  144279400.00 & 53579.900467 &   &  11 &    \\
HETE J1900.1-2455      & 2005-08-17 12:19:58 & b  & 150823 &  145973997.70 & 53599.513866 & y &  19 &    \\
HETE J1900.1-2455      & 2005-08-28 15:09:37 & b  & 152451 &  146934576.32 & 53610.631674 & y &  17 &    \\
XTE J1739-285          & 2005-10-31 20:39:19 & x  &        &  152483980.62 & 53674.860648 &   &     &  3 \\
XTE J1723-376          & 2006-02-12 23:42:39 & b  & 181323 &  161480560    & 53778.987951 &   &   5 &    \\
SAX J1747.0-2853       & 2006-03-25 00:53:03 & b  & 202662 &  164940784    & 53819.036839 &   &     &    \\
2S 0918-549            & 2006-04-15 03:38:09 & b  & 205373 &  166765090.62 & 53840.151499 &   &  12 &    \\
Swift J1749.4-2807     & 2006-06-02 23:54:34 & sl & 213190 &  170985275.14 & 53888.996226 &   &  20 &   7\\ 
GRS 1741.9-2853        & 2006-06-04 09:40:27 & x  &        &  171106844.23 & 53890.403100 &   &     &  11\\
4U 1636-536            & 2006-06-11 15:09:47 & bx &        &  171731387.10 & 53897.6231802&   &   8 &  11\\
4U 1636-536            & 2006-06-12 05:24:08 & bx &        &  171782648.10 & 53898.225101 &   &  10 &   9\\
4U 1246-588            & 2006-08-11 02:59:55 & sl & 223918 &  176957996.86 & 53958.124948 &   &  70 &    \\ 
Aql X-1                & 2006-08-15 11:07:18 & bx  &       &  177332850.70 & 53962.463408 & y &  10 &  16\\ 
SAX J1747.0-2853       & 2006-08-19 14:43:20 & b  & 225393 &  177691401.86 & 53966.613431 &   &  11 &    \\
4U 0614+09             & 2006-10-21 09:02:00 & b  & 234849 &  183114121.98 & 54029.376394 &   &  40 &    \\
GRS 1741.9-2853        & 2007-01-22 06:12:58 & b  & 257213 &  191139180.16 & 54122.259010 &   &  13 &    \\
AX J1745.6-2901        & 2007-02-16 22:59:18 & bx  &       &  193359575.23 & 54147.957857 &   &  10 &  11\\
GRS 1741.9-2853        & 2007-03-05 20:19:00 & bx &        &  194818757.23 & 54164.846537 & y &  10 &  15\\ 
4U 0614+09             & 2007-03-30 08:53:21 & b  & 273106 &  196937603.14 & 54189.370385 &   &  23 &    \\
SLX 1744-299           & 2007-04-26 06:03:50 & b  & 277191 &  199260232    & 54216.252663 &   &  20 &    \\
IGR J17191-2821        & 2007-05-01 13:20:53 & x  &        &  199718455.33 & 54221.556173 &   &     &  17\\
4U 1702-429            & 2007-05-16 21:00:25 & b  & 279418 &  201042026.75 & 54236.875287 &   &  10 &    \\
XB 1832-330            & 2007-05-30 07:44:15 & b  & 280846 &  202203857.47 & 54250.322401 &   &   7 &    \\
AX J1754.2-2754        & 2007-07-12 09:12:28 & x  &        &  205924368.81 & 54293.383660 &   &     & -- \\ 
SAX J1810.8-2609       & 2007-08-05 11:27:26 & sl & 287042 &  208006048.51 & 54317.477389 & y &  23 &  84\\ 
SAX J1810.8-2609       & 2007-09-01 11:19:26 & b  & 288384 &  210338367.94 & 54344.471826 & y &  22 &    \\
SAX J1810.8-2609       & 2007-09-16 15:54:17 & b  & 291218 &  211650858.75 & 54359.662691 & y &  12 &    \\
Aql X-1                & 2007-09-19 07:56:35 & b  & 291524 &  211881397.50 & 54362.330964 & y &   5 &    \\
SAX J1810.8-2609       & 2007-09-27 15:09:44 & b  & 292421 &  212598586.50 & 54370.631762 &   &   5 &    \\
XTE J1810-189          & 2008-03-18 22:32:53 & sl & 306737 &  227572375.17 & 54543.939497 &   &  10 &  --\\ 
XTE J1810-189          & 2008-03-21 14:56:33 & bx &        &  227804200.86 & 54546.622613 &   &  20 &  31\\ 
IGR J17473-2721        & 2008-03-31 09:03:33 & sl & 308196 &  228647016.13 & 54556.377470 & y &  25 &  17\\
SAX J1750.8-2900       & 2008-04-27 18:35:45 & b  & 310319 &  231014147.33 & 54583.774822 &   &   8 &    \\
1RXH J173523.7-354013  & 2008-05-14 10:32:37 & sl & 311603 &  232453960.   & 54600.439320 &   & 240 & 289\\
4U 1636-536            & 2008-05-19 22:21:23 & x  &        &  232928483.10 & 54605.931524 &   &     &  14\\
XTE J1701-407          & 2008-07-17 13:30:00 & sl & 317205 &  237994202.69 & 54664.562497 &   & 125 & 119\\
XTE J1701-407          & 2008-07-27 22:31:20 & sl & 318166 &  238890682.82 & 54674.938424 &   &  15 & $<40$\\
SAX J1808.4-3658       & 2008-09-24 20:14:24 & b  & 325827 &  243980067.33 & 54733.843336 & y &  15 & \\
MXB 1730-335           & 2009-03-05 13:28:15 & x  &        &  257952505.27 & 54895.561292 &   &     &  11\\
4U 1820-303            & 2009-06-06 09:28:19 & b  & 354224 &  265973303.55 & 54988.394661 & y &   6 &    \\
4U 1820-303            & 2009-06-06 12:43:33 & bx &        &  265985018.58 & 54988.530252 &   &   4 &   7\\
IGR J17511-3057        & 2009-09-14 00:50:31 & bx &        &  274582236.37 & 55088.035084 &   &  10 &  10\\ 
IGR J17511-3057        & 2009-09-15 17:17:13 & bx &        &  274727838.37 & 55089.720292 &   &  20 &  18\\
IGR J17511-3057        & 2009-09-30 18:31:57 & sl & 371210 &  276028321.66 & 55104.772182 &   &  20 &  15\\
GS 1826-24             & 2009-10-21 17:44:55 & x  &        &  277839902.24 & 55125.739529 &   &     &  82\\ 
4U 1728-34             & 2009-10-30 10:19:42 & bx &        &  278590782.29 & 55134.430355 &   &  35 &  33\\ 
Aql X-1                & 2009-11-16 02:03:24 & bx &        &  280029816.70 & 55151.085699 &   &  15 &  22\\ 
4U 1850-086            & 2010-02-16 06:07:42 & b  & 412503 &  287993267.26 & 55243.255344 &   &   6 &    \\
SAX J1712.6-3739       & 2010-07-01 14:55:41 & sl & 426405 &  299688946.69 & 55378.621999 &   &  30 & 119\\
IGR J17464-2811        & 2010-08-13 21:03:30 & b  & 431582 &  303426216    & 55421.877430 &   & 110 &    \\
1E 1145.1-6141         & 2011-04-03 02:38:09 & b  & 450610 &  323491096    & 55654.109829 &   &   5 &  \\
XTE J1810-189          & 2011-06-19 00:59:37 & sl & 455640 &  330137984.   & 55731.041400 &   & 400 &  53\\
Swift J185003.2-005627 & 2011-06-25 00:06:08 & sl & 456014 &  330653174.72 & 55737.004255 &   &  40 &  80\\
IGR J17498-2921        & 2011-08-18 11:22:34 & x  &        &  335359374.87 & 55791.474009 &   &     &   8\\
IGR J17498-2921        & 2011-08-28 10:12:04 & x  &        &  336219144.87 & 55801.425051 &   &     &  14\\
KS 1741-293            & 2011-09-01 12:07:22 & b  & 502024 &  336571649.66 & 55805.505120 &   &  10 &    \\
SAX J1712.6-3739       & 2011-09-26 20:11:29 & sl & 504101 &  338760696.   & 55830.841303 &   & 315 & 200\\
Swift J1922.7-1716     & 2011-11-03 14:12:13 & sl & 506913 &  342022340.67 & 55868.591819 & y &  50 & 106\\        
Swift J1922.7-1716     & 2011-12-02 09:44:42 & b  & 508855 &  344511889.47 & 55897.406040 & y &  18 &    \\
MAXI J1647-227         & 2012-06-19 09:49:38 & x  &        &  361792199.29 & 56097.409479 &   &     &  10\\
IGR J17062-6143        & 2012-06-25 22:42:32 & sl & 525148 &  362356960.   & 56103.946200 &   & 420 &1186\\
MAXI J1647-227         & 2012-06-27 08:10:17 & x  &        &  362477438.29 & 56105.340485 &   &     &  12\\
Swift J174805.3-244637 & 2012-07-17 21:05:29 & bx &        &  364251943.28 & 56125.878818 &   &  10 &  20\\ 
XMM J174457-2850.3     & 2012-08-11 04:43:54 & sl & 530588 &  366353043.33 & 56150.197151 & y &  75 &  --\\
Swift J174805.3-244637 & 2012-08-13:09:13:34 & sl & 530808 &  366542023.81 & 56152.384424 & y &  35 &$\sim100$\\
Cyg X-2                & 2013-02-23 14:14:37 & x  &        &  383321688.64 & 56346.593491 &   &     &   7\\
IGR J18245-2452        & 2013-03-30 15:10:38 & sl & 552369 &  386349048.   & 56381.632380 &   & 130 &  --\\ 
IGR J18245-2452        & 2013-04-07 22:15:21 & bx &        &  387065731.77 & 56389.927331 &   &  20 &  43\\ 
Swift J1734.5-3027     & 2013-09-01 09:13:17 & sl & 569022 &  399719608.   & 56536.384225 &   & 155 & 188\\
MAXI J1421-613         & 2014-01-18 08:39:20 & sl & 584155 &  411727171.2  & 56675.360664 & y &  15 &   9\\
4U 1746-37             & 2014-02-16 19:18:41 & x  &        &  414271134.68 & 56704.804644 &   &     &  12\\
4U 1746-37             & 2014-02-22 19:07:59 & x  &        &  414788892.68 & 56710.797213 &   &     &   9\\
4U 1746-37             & 2014-02-22 20:40:29 & x  &        &  414794442.68 & 56710.861450 &   &     &  13\\
4U 1735-444            & 2014-02-23 01:46:51 & bx &        &  414812824.26 & 56711.074210 &   &   5 &  12\\
4U 1735-444            & 2014-02-23 14:36:45 & x  &        &  414859018.26 & 56711.608863 &   &     &   9\\
4U 1735-444            & 2014-02-23 16:16:39 & bx &        &  414865012.26 & 56711.678238 &   &   5 &   7\\ 
4U 1735-444            & 2014-02-24 08:10:48 & x  &        &  414922261.26 & 56712.340842 &   &     &  10\\
4U 1735-444            & 2014-02-24 14:19:45 & x  &        &  414944398.26 & 56712.597058 &   &     &   8\\
4U 1850-086            & 2014-03-10 21:05:00 & sl & 591237 &  416178312.   & 56726.878475 &   & 840 &1999\\
GS 1826-24             & 2014-06-24 23:46:16 & x  &        &  425346383.24 & 56832.990467 &   &     &  10\\
SAX J1712.6-3739       & 2014-08-18 17:10:04 & sl & 609878 &  430074616.   & 56887.715319 &   &$>800$&  42\\
4U 0614+09             & 2014-12-31 08:52:27 & bx &        &  441708760.80 & 57022.369764 &   &  15 &  26\\
1RXS J180408.9-342058  & 2015-02-06 21:24:14 & x  &        &  444950668.12 & 57059.891840 &   &     &  58\\
4U 0614+09             & 2015-02-19 16:42:24 & sl & 631747 &  446056956.93 & 57072.696108 &   &   6 &  11\\
1RXS J180408.9-342058  & 2015-02-21 20:36:20 & x  &        &  446243794.12 & 57074.858576 &   &     &  33\\
1RXS J180408.9-342058  & 2015-02-22 01:10:45 & bx &        &  446260258.12 & 57075.049132 &   &     &  36\\
1RXS J180408.9-342058  & 2015-02-26 02:48:18 & x  &        &  446611711.12 & 57079.116875 &   &     &  37\\
SAX J1748.9-2021       & 2015-02-26 21:47:19 & x  &        &  446680052.25 & 57079.907859 &   &     &  14\\ 
1RXS J180408.9-342058  & 2015-03-06 02:12:47 & x  &        &  447300781.12 & 57087.092222 &   &     &  48\\
SAX J1748.9-2021       & 2015-03-09 01:57:46 & bx &        &  447559079.25 & 57090.081783 &   &     &  16\\
SAX J1748.9-2021       & 2015-03-16 05:26:37 & x  &        &  448176410.25 & 57097.226818 &   &     &  27\\
EXO 1745-248           & 2015-03-25 05:06:02 & x  &        &  448952776.28 & 57106.212533 &   &     &  10\\
1RXS J180408.9-342058  & 2015-03-31 23:44:56 & bx &        &  449538310.12 & 57112.989548 &   &     &  59\\ 
SAX J1748.9-2021       & 2015-04-06 10:26:22 & bx &        &  450008795.25 & 57118.434977 &   &     &  30\\
SAX J1808.4-3658       & 2015-04-11 19:36:25 & sl & 637765 &  450473798.66 & 57123.816959 &   &  25 &  46\\
SLX 1735-269           & 2015-04-18 00:20:55 & x  &        &  451009273.70 & 57130.014526 &   &     &   3\\
MXB 1730-335           & 2015-05-30 01:45:51 & x  &        &  454643161.27 & 57172.073515 &   &     &  21\\
SAX J1748.9-2021       & 2015-06-23 22:42:09 & b  & 645944 &  456792142.4  & 57196.945933 &   &  10 &  \\
IGR J00291+5934        & 2015-07-25 02:12:05 & sl & 650221 &  459483139.84 & 57228.091724 & y &  20 & 128\\
4U 1636-536            & 2015-09-05 18:35:23 & x  &        &  463170923.10 & 57270.774580 &   &     &  16\\
MXB 1658-298           & 2015-09-09 10:07:54 & x  &        &  463486090.72 & 57274.422159 &   &     &  14\\
SAX J1750.8-2900       & 2015-10-16 01:25:59 & b  & 659734 &  466651574.53 & 57311.059714 &   &  12 &    \\
AX J1745.6-2901        & 2016-02-10 21:07:15 & x  &        &  476831252.23 & 57428.880044 &   &     &  --\\
MXB 1658-298           & 2016-04-18 19:09:33 & x  &        &  482699389.72 & 57496.798305 &   &     &  41\\
XTE J2123-058          & 2016-05-19 06:15:59 & b  & 686845 &  485331376    & 57527.261108 &   &  10 &    \\
Cir X-1                & 2016-09-18 14:05:27 & x  &        &  495900327.20 & 57649.587128 &   &     &   4\\
XTE J1710-281          & 2017-02-24 17:01:09 & x  &        &  509648488.62 & 57808.709143 &   &     &  14\\
SAX J1806.5-2215       & 2017-04-01 19:00:53 & sl & 745022 &  512766072.   & 57844.792281 &   & 250 & 107\\
Swift J181723.1-164300 & 2017-07-28 16:57:58 & sl & 765081 &  522953897.41 & 57962.706921 & y &  15 &    \\ 
Swift J181723.1-164300 & 2017-07-30 21:32:05 & b  & 765422 &  523143144.45 & 57964.897280 &   &   8 &    \\
4U 1636-536            & 2017-08-03 05:25:59 & x  &        &  523430759.10 & 57968.226385 &   &     &  21\\
MXB 1730-335           & 2017-08-26 21:15:15 & x  &        &  525474925.27 & 57991.885598 &   &     &  8\\
4U 1636-536            & 2017-09-08 08:53:02 & x  &        &  526553582.10 & 58004.370170 &   &     &  21\\
4U 1636-536            & 2017-09-14 17:44:23 & bx &        &  527103863.10 & 58010.739163 &   &  20 &  27\\
4U 1636-536            & 2017-09-17 04:52:32 & x  &        &  527316752.10 & 58013.203156 &   &     &  16\\
4U 1636-536            & 2017-09-23 08:56:56 & bx &        &  527849816.10 & 58019.372878 &   &  20 &  13\\
4U 1636-536            & 2017-09-26 15:08:56 & x  &        &  528131336.10 & 58022.631212 &   &     &   9\\
SAX J1748.9-2021       & 2017-10-05 19:32:22 & bx &        &  528924755.25 & 58031.814144 &   &     &  20\\
4U 1636-536            & 2017-10-23 09:26:11 & bx &        &  530443571.10 & 58049.393191 &   &  15 &  11\\ 
XTE J701-407           & 2018-03-09 18:20:35 & b  & 813449 &  542312456.   & 58186.764299 &   & 150 &    \\
\hline 
4U 0836-429            & 2005-01-06 16:59:00 &  n & 101731 &               &               &   &     & \\
Cyg X-2                & 2005-02-05 07:08:18 &  n & 104695 &               &              &   &     & \\
XB 1916-053            & 2005-02-26 01:29:30 &  n & 106912 &               &               &   &     & \\
4U 1735-44             & 2005-03-17 17:57:54 &  n & 111378 &               &               &   &     & \\
4U 1636-536            & 2005-03-20 19:28:56 &  n & 111765 &               &               &   &     & \\
IGR J17473-2721        & 2005-06-23 20:56:56 &  n & 142554 &               &               &   &     & \\
GS 1826-24             & 2005-06-24 21:03:28 &  n & 142667 &               &               &   &     & \\
GX 13+1                & 2005-07-09 09:35:20 &  n & 145011 &               &               &   &     & \\
4U 1746-37             & 2005-07-26 19:33:36 &  n & 147857 &               &               &   &     & \\
1E 1145.1-6141         & 2005-08-13 04:40:48 &  n & 150131 &               &              &   &     & \\
EXO 0748-676           & 2005-10-06 13:52:40 &  n & 158539 &               &               &   &     & \\
SLX 1737-282           & 2006-02-01 06:06:39 &  n & 179811 &               &               &   &     & \\
XTE J1701-462          & 2006-03-15 18:50:22 &  n & 201721 &               &               &   &     & \\
SAX J1747.0-2853       & 2006-03-28 13:50:14 &  n & 203045 &               &               &   &     & \\
XB 1832-330            & 2006-04-08 13:15:42 &  n & 204415 &               &               &   &     & \\
XTE J1701-462          & 2006-06-02 06:23:10 &  n & 213056 &               &               &   &     & \\
XTE J1701-462          & 2006-06-25 10:07:26 &  n & 215987 &               &               &   &     & \\
MXB 1730-335           & 2006-07-01 18:45:33 &  n & 216664 &               &               &   &     & \\
SAX J1808.4-3658       & 2008-09-24 00:37:16 &  n & 325730 &               &              &   &     & \\
IGR J17480-2446        & 2010-10-13 19:27:21 &  n & 436241 &               &               &   &     & \\
IGR J17480-2446        & 2010-10-28 22:35:21 &  n & 437313 &               &               &   &     & \\
IGR J17480-2446        & 2010-10-31 09:52:01 &  n & 437466 &               &               &   &     & \\
4U 1608-522            & 2011-08-12 03:24:40 &  n & 500082 &               &               &   &     & \\
GX 17+2                & 2011-08-12 05:40:32 &  n & 500093 &               &               &   &     & \\
GS 1826-24             & 2011-08-12 07:08:16 &  n & 500108 &               &               &   &     & \\
SAX J1808.4-3658       & 2011-11-04 02:55:12 &  n & 506961 &               &               &   &     & \\
1E 1145.1-6141         & 2012-04-10 17:59:52 &  n & 519850 &               &               &   &     & \\
IGR J17480-2446        & 2012-07-13 14:16:54 &  n & 526511 &               &               &   &     & \\
IGR J17480-2446        & 2012-07-16 20:59:50 &  n & 526892 &               &               &   &     & \\
SAX J1828.5-1037       & 2012-10-10 23:57:42 &  n & 535733 &               &               &   &     & \\
GX 17+2                & 2013-02-26 22:17:01 &  n & 549824 &               &               &   &     & \\
SAX J1806.5-2215       & 2013-06-20 02:21:57 &  n & 558631 &               &               &   &     & \\
Cyg X-2                & 2013-09-08 16:39:48 &  n & 570069 &               &               &   &     & \\
2S 1803-245            & 2014-11-14 07:32:03 &  n & 618556 &               &               &   &     & \\
1RXS J180408.9-342058  & 2015-01-31 21:14:18 &  n & 629206 &               &               &   &     & \\
1RXS J180408.9-342058  & 2015-02-06 20:45:14 &  n & 630047 &               &               &   &     & \\
SAX J1748.9-2021       & 2015-02-20 23:38:50 &  n & 631946 &               &               &   &     & \\
IGR J17480-2446        & 2015-04-05 18:15:06 &  n & 637212 &               &               &   &     & \\
IGR J00291+5934        & 2015-07-24 05:23:37 &  n & 650140 &               &               &   &     & \\
Ser X-1                & 2016-03-05 20:42:32 &  n & 677890 &               &               &   &     & \\
Swift J1922.7-1716     & 2017-08-01 17:03:16 &  n & 765783 &               &               &   &     & \\
\end{longtable}
$^{\rm a}$``b''=BAT detection, ``x''=XRT detection, ``sl''=BAT trigger and AT, ``n''=BAT nor XRT detection.
$^{\rm b}$The times refer to the trigger time for those that triggered BAT (i.e., with a trigger ID) or
to the burst onset time for the others. The trigger times were extracted from gcn.gsfc.nasa.gov/swift\_grbs.html.
$^{\rm c}$The duration of the BAT signal, judged by eye, only if detected.
$^{\rm d}$The exponential decay time of the XRT signal, fitted, only if detected.
\twocolumn

\begin{figure*}[p]
\centering \includegraphics[width=0.9\textwidth]{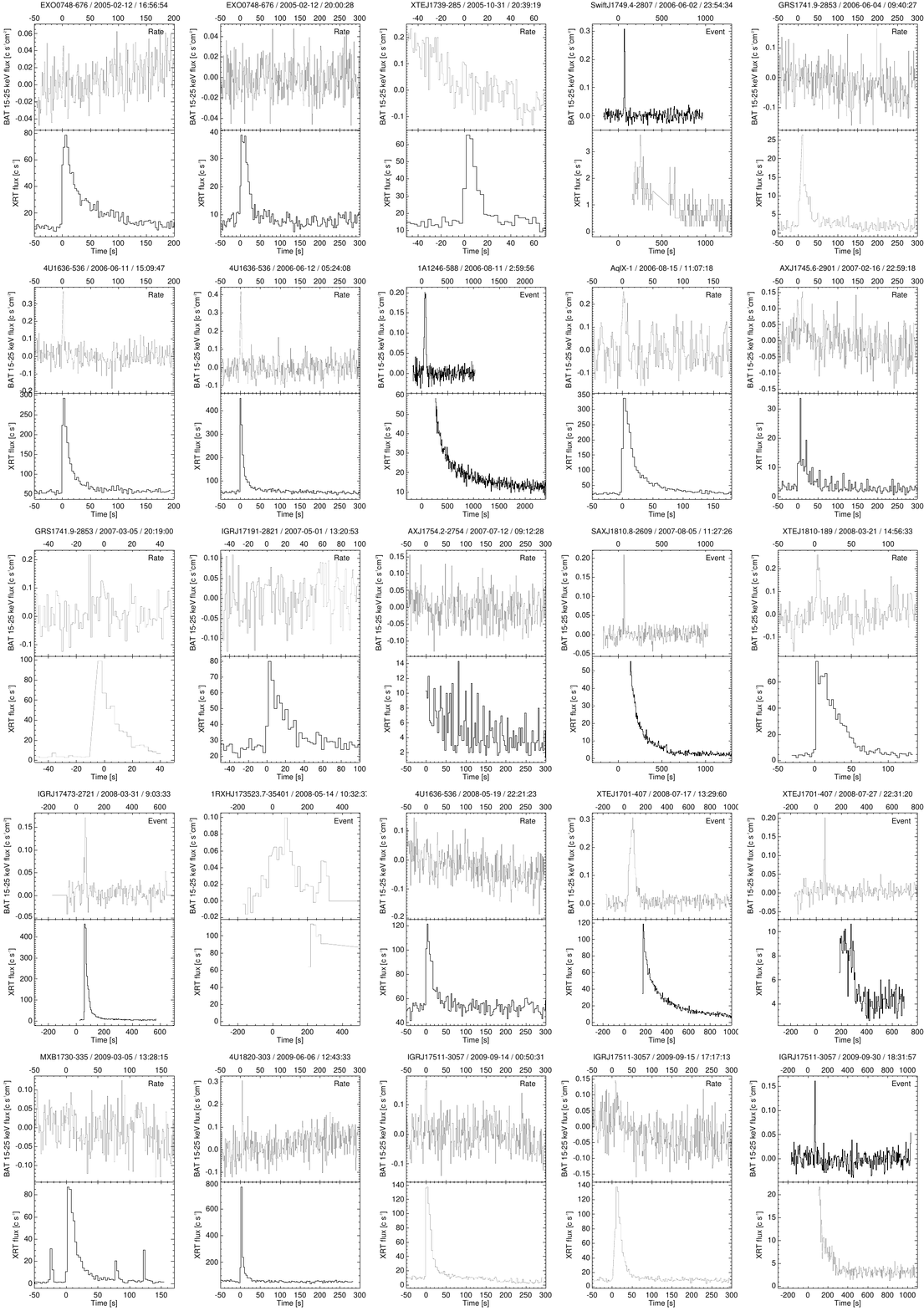}
    \caption{BAT (upper panels) and XRT light curves (lower panels) of
      all 90 X-ray bursts that have XRT coverage. The BAT light curves
      are labeled `Event' if they are from mask-tagged event data or
      `Rate' if they are inferred from rate meter data. In the latter
      case, the rate meter data have been pre-burst subtracted and
      divided by the photon-collecting area illuminated by the source
      that was identified on board.}
    \label{fig:allbursts1}
\end{figure*}
\setcounter{figure}{0}
\begin{figure*}[!t]
\centering \includegraphics[width=0.95\textwidth, trim=0cm 0cm 0cm 0cm,clip=true]{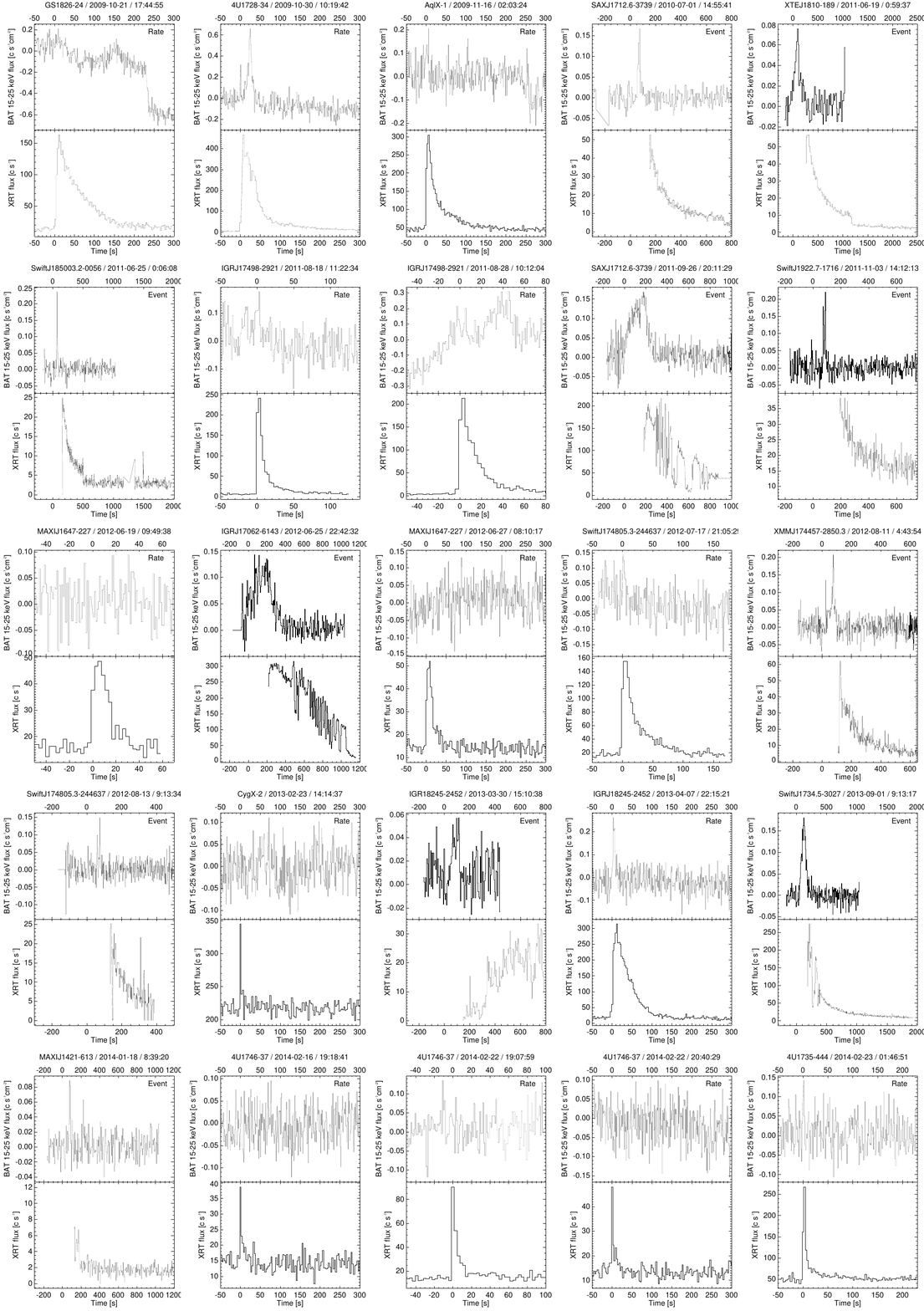}
    \caption{{\it (continued)}}
    \label{fig:allbursts2}
\end{figure*}
\setcounter{figure}{0}
\begin{figure*}[!t]
\centering \includegraphics[width=0.95\textwidth, trim=0cm 0cm 0cm 0cm,clip=true]{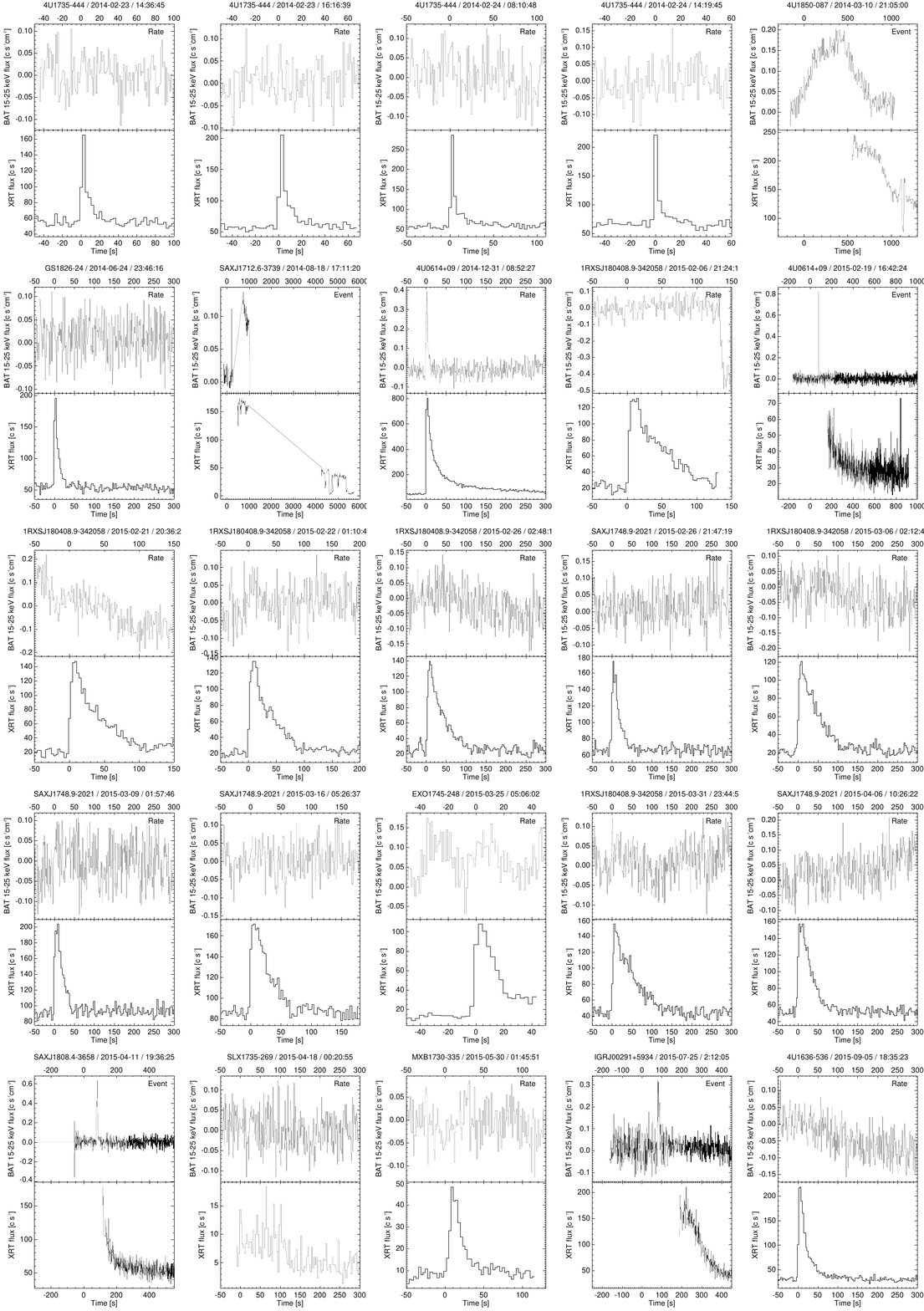}
    \caption{{\it (continued)}}
    \label{fig:allbursts3}
\end{figure*}
\setcounter{figure}{0}
\begin{figure*}[!t]
\centering \includegraphics[width=0.95\textwidth, trim=0cm 0cm 0cm 0cm,clip=true]{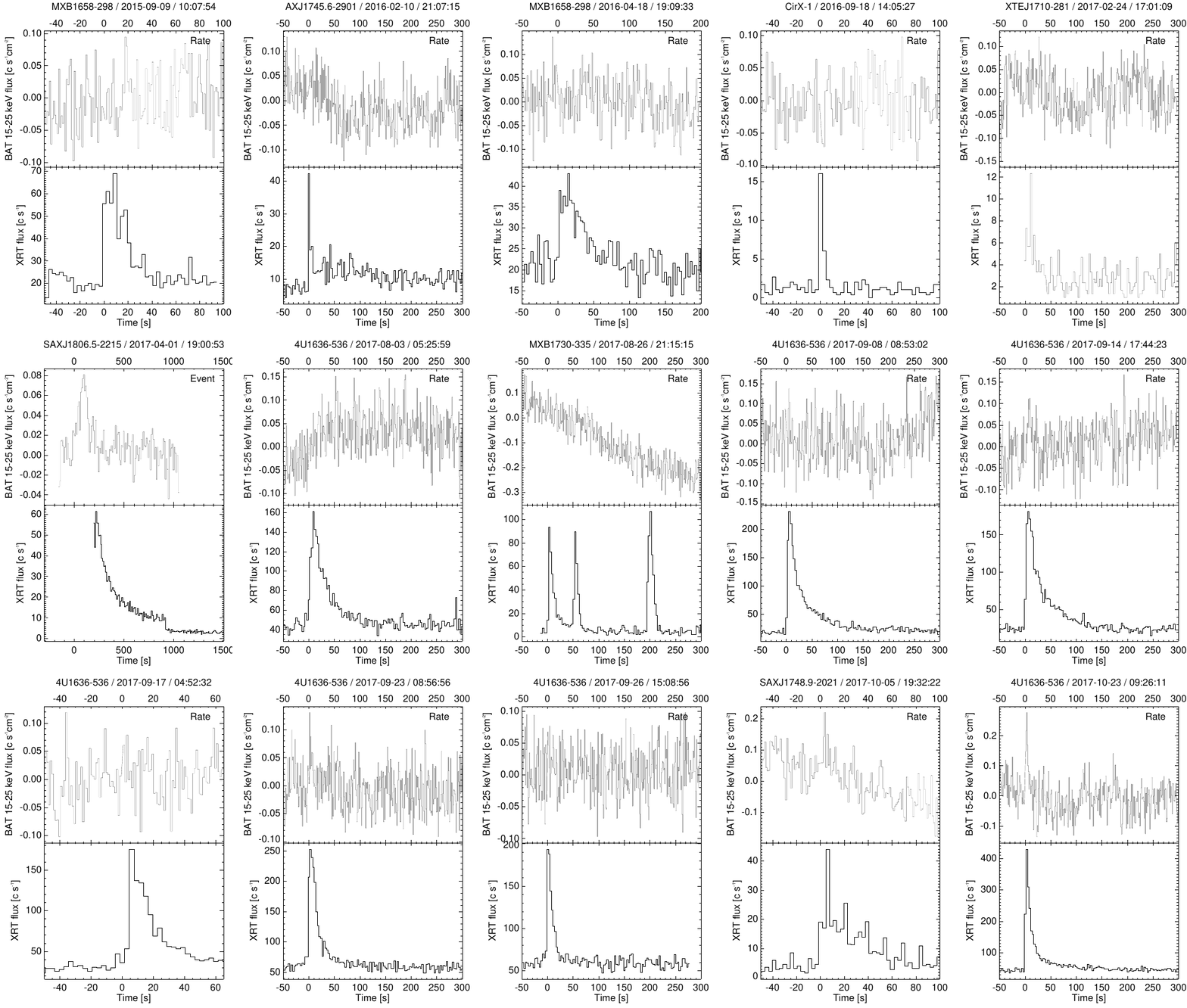}
    \caption{{\it (continued)}}
    \label{fig:allbursts4}
\end{figure*}

\end{document}